\documentclass[11pt]{article}

\usepackage[top = 1.in, bottom = 1.3in, left   = 1.25in, right  = 1.25in]{geometry}

\usepackage{endnotes}
\usepackage{amssymb}
\usepackage{amsfonts}
\usepackage{amsmath} 
\usepackage{amsthm}
\usepackage{url}
\usepackage{graphicx,color}

\usepackage{comment}
\usepackage{chicago}
\usepackage{array}

\usepackage{url} 

\usepackage{setspace}
\definecolor{amber}{rgb}{1.0, 0.75, 0.0}
\definecolor{ao(english)}{rgb}{0.0, 0.5, 0.0}
\definecolor{armygreen}{rgb}{0.29, 0.33, 0.13}

\def\R{\mathbb{R}}
\def\endproof{\hfill\diamondsuit}

\def\sF{{\mathcal F}}
\def\sN{{\mathcal N}}

\def\ta{\tilde{a}}

\def\E{\mathbb{E}}
\def\V{\mathbb{V}}
\def\sF{\mathcal{F}}
\def\P{\mathbb{P}}

\numberwithin{equation}{section}
\theoremstyle{plain}                
\newtheorem{theorem}{Theorem}[section]

\newtheorem{lemma}[theorem]{Lemma}

\theoremstyle{definition}           
\newtheorem{definition}[theorem]{Definition}

\theoremstyle{remark}               

\begin{document}
\pagestyle{empty}

\begin{center}
\large{\bf Learning about latent dynamic trading demand}$^\ast$
\makeatletter{\renewcommand*{\@makefnmark}{}\footnotetext{\hspace{-.35in} $^\ast${The authors have benefited from helpful comments from Dan Bernhardt and participants at the SIAM math finance conference (2021) and at Tepper (Carnegie Mellon). Jin Hyuk Choi is supported by the National Research Foundation of Korea (NRF) grant funded by the Korea government (MSIT) (No. 2020R1C1C1A01014142 and No. 2021R1A4A1032924). Kasper Larsen has been supported by the National Science Foundation under Grant No. DMS 1812679 (2018 - 2021). Any opinions, findings, and conclusions or recommendations expressed in this material are those of the author(s) and do not necessarily reflect the views of the National Science Foundation (NSF). The corresponding author is Kasper Larsen. Xiao Chen has email: xc206@scarletmail.rutgers.edu, Jin Hyuk Choi has email: jchoi@unist.ac.kr,  Kasper Larsen has email: KL756@math.rutgers.edu, and Duane J. Seppi has email: ds64@andrew.cmu.edu. The datasets generated during and/or analysed during the current study are available from the corresponding author on reasonable request.}\makeatother}}
\end{center}

\begin{center}

\ \\
{ \bf Xiao Chen}\\
Department of Mathematics

Rutgers University

110 Frelinghuysen Road

Piscataway, NJ 08854-8019, USA

\ \\

{ \bf Jin Hyuk Choi}\\ 

Department of Mathematical Sciences

Ulsan National Institute of Science and Technology (UNIST) 

UNIST-gil 50

Ulsan 689-798, Republic of Korea

\ \\

{ \bf Kasper Larsen}\\
Department of Mathematics

Rutgers University

110 Frelinghuysen Road

Piscataway, NJ 08854-8019, USA

\ \\

{ \bf Duane J. Seppi}\\
Tepper School of Business

Carnegie Mellon University

5000 Forbes Avenue

Pittsburgh PA 15213, USA

\ \\ \ \\

{\normalsize \today }
\end{center}

\newpage

\begin{center}

\end{center}

\begin{verse}
{\sc Abstract}: We present an equilibrium model of dynamic trading, learning, and pricing by strategic investors with trading targets and price impact. Since trading targets are private, rebalancers and liquidity providers filter the child order flow dynamically over time to estimate the latent underlying parent trading demand imbalance and to forecast its impact on subsequent price pressure dynamics. We prove existence of an  equilibrium and solve for equilibrium trading strategies and prices in terms of the solution to a system of coupled ODEs. Trading strategies are combinations of trading towards investor targets, liquidity provision for other investors’ demands, and front-running based on learning about latent underlying trading demand imbalances.

\end{verse}

\begin{verse}
\end{verse}

\begin{verse}

\ \\
\end{verse}

\newpage

\pagestyle{plain}
\addtocounter{page}{-2}

\section{Introduction}

The price formation process in financial markets involves equating supply and demand for securities over time for arriving investors with heterogeneous trading preferences.  In present day markets, large investors act on their underlying trading preferences, sometimes called {\em parent demands}, by splitting their trading into dynamic sequences of smaller orders, called {\em child orders} (see O’Hara  (2015)) to minimize their price impact.  Since the parent demands driving child-order trading are private information, investors use information from arriving child orders to form inferences over time about the dynamically evolving fundamental state of the market.  In particular, investors learn about imbalances in the underlying aggregate parent  demands and the associated pressure on future market-clearing prices and incorporate this information in their current child orders.  Given the widespread prevalence of optimized order-splitting of parent orders into flows of child orders, dynamic learning about aggregate parent demands is a critical part of market dynamics.\footnote{See van Kervel and Menkveld (2019), Korajczyk and Murphy (2019), and van Kervel, Kwan, and Westerholm (2020).}

This paper is the first to provide an analytically tractable equilibrium model of dynamic learning, trading, and pricing with parent trading demands.  We consider a continuous-time model with high-frequency   trading at times $t\in[0,1]$ over short time-horizons with $[0,1]$ being a day or an hour.  Trading occurs between price-sensitive optimizing traders with two different types of parent trading targets:  One group has fixed individual targets, and the other group wants to track a stochastically evolving  target over time. Since parent targets are initially not public, information about parent demand imbalances is partially revealed through market-clearing stock prices. Our analysis models the dynamic learning process and the equilibrium holdings and  stock-price processes. 

Our main results are:

\begin{itemize}

\item We construct and solve  two different equilibrium models: A simpler price-impact equilibrium and a subgame perfect Nash financial-market equilibrium. In the subgame perfect Nash equilibrium, price impact is partially endogenous.  We find that these two equilibria are numerically similar.  

\item Intraday price drifts due to price pressure change over the trading day and are path-dependent. This leads to time-varying incentives for investors to provide liquidity to the child orders of other investors.

\item A practical application of our model is that we can compute total trading costs for investors given the effects of dynamic learning and front-running by other investors. We show  these costs are quadratic in the rebalancers' trading targets.

\item  Trading in our model reflects a combination of liquidity provision and front-running but not predatory trading. We conjecture that the absence of predatory trading is because our model replaces the exogenous price-elastic residual supply used in both Brunnermeier and Pedersen (2005) and Carlin, Lobo, and Viswanathan (2007) with endogenous demands coming from rational profit-maximizing investors.

\end{itemize}

Our paper advances several strands of research on market microstructure. First, dynamic learning and trading have been extensively studied in the context of markets with strategic investors with long-lived asymmetric information as in Kyle (1985). However, equilibrium  trading, learning, and pricing with optimized dynamic order-splitting by large uninformed investors are less well understood. Thus, we model price pressure to equate supply and demand rather than adverse selection. Second, Grossman and Miller (1988) model pricing and liquidity provision with impatient traders who submit single orders equal to their parent demands with symmetric payoff information.  In contrast, we model liquidity provision with optimized order-splitting of parent demands into child order flows. Third, Choi, Larsen, and Seppi (2019) construct an equilibrium with optimized dynamic trading and learning in a market with a strategic rebalancer with an end-of-day trading target and an informed investor who trades on private long-lived asset-payoff information. By filtering the order flow over time, the rebalancer learns about the underlying asset payoff, the informed investor learns about the rebalancer’s trading target, and market makers learn about both when setting prices.  That earlier paper provides a characterization result for equilibrium and gives numerical examples but does not have an existence proof or analytic solutions. In contrast, our model is solved analytically and gives the equilibrium in closed form. Fourth, Brunnermeier and Pedersen (2005) and Carlin, Lobo, and Viswanathan (2007) show how dynamic rebalancing by a large investor can lead to predatory trading.  However, these papers abstract from the learning problem by assuming the parent trading needs are publicly observable.  They also make an ad hoc assumption about the price sensitivity of the residual market-maker trading demand in  the form of exogenous price-elastic noise traders. In contrast, our model assumes the underlying parent trading demands are private information, which leads to a learning problem. In addition, our prices are rationally set with no ad hoc residual demand. Fifth, a large body of research models optimal order-splitting strategies for a single strategic investor given an exogenous pricing rule with no learning about latent trading demands of other investors (see, e.g.,  Almgren and  Chriss (1999, 2000), Almgren (2003), and  Schied and Sch\"oneborn (2009)). In contrast, we solve for optimal trades, learning, and pricing jointly. Van Kervel, Kwan, and Westerholm (2020) solve for optimal trading strategies for two dynamic rebalancers with learning over time about each other’s latent trading demands. This leads to predictions about the effect of aggregate parent demand on individual investor child orders, which are then verified empirically. However, they assume an ad hoc linear pricing rule, and there are no existence proofs or analytic solutions.  In contrast, price pressure in our model is endogenously determined in equilibrium, and we solve our model analytically.  As in van Kervel, Kwan, and Westerholm (2020),   trading in our model is a combination of front-running along with trading-demand accommodation.

The mathematics of our model is tractable because we use a modeling approach from the asset-pricing literature for non-dividend paying stocks.  The simplification involves finding  equilibrium price drifts that clear the market without determining the levels of market-clearing prices as discounted future cash flows. Karatzas and Shreve (1998, Chap. 4) use this approach in complete market settings, and Cuoco and He (1994) consider an extension to incomplete markets. Atmaz and Basak (2021) show that non-dividend paying stocks are relevant for asset pricing. However, models using the non-dividend paying stock approach are new in the mainstream microstructure literature. G\^arleanu and Pedersen (2016), Bouchard, Fukasawa, Herdegen, and Muhle-Karbe (2018), and Noh and Weston (2020) use the zero-dividend stock approach to model prices given exogenous transaction costs.  Our model uses this approach with endogenous price impact.

\section{Model}

We model equilibrium trading, learning, and pricing in a market with a risky stock and a riskless bank account over a short time horizon $[0,1]$ (e.g., a trading day).  For simplicity, the net supply of both the stock and bank account are set to zero. Since the time horizon is short, the risk-free interest rate on the bank account is set to zero.  Stock differs from the bank account in two ways: First, investors have individual parent demands for the stock.  Second, stock prices are stochastic over time.  Stock valuation can be viewed as the sum of two components:  One component is a fundamental valuation of future dividends absent price pressure from trading targets. The other component is incremental price pressure for markets to clear given parent trading demand imbalances.  It is the price pressure component that is the focus of our analysis. Our analysis treats these two components as being orthogonal and, for simplicity,  normalizes the dividend valuation component to zero. Thus, hereafter, when we refer to the ``stock price’’, this is shorthand, for brevity, for the ``price pressure valuation component of stock prices.’’ Thus, prices are random here due to random price pressure due to random trading demand imbalances. In a more complicated model, a separate fundamental dividend valuation component could be added to our stock price pressure valuation to get the full stock price.

Two different groups of investors trade in our equilibrium model. 
\begin{itemize}

\item[(i)]  Price-sensitive rebalancers. Rebalancer $i\in\{1,...,M\}$ maximizes her expected profit subject to a parent trading target $\ta_i$  where $\ta_i$ is private  information for $i$. The targets $(\ta_1,...,\ta_M)$ are assumed independent and homogeneously distributed $\ta_i \sim \mathcal{N}(0,\sigma_{\ta}^2)$ for all rebalancers $i\in\{1,...,M\}$ with identical zero means and  standard deviations $\sigma_{\ta}$. The aggregate target is
\begin{align}\label{aS}
\ta_\Sigma := \sum_{i=1}^M \ta_i.
\end{align}

Rebalancer $i$'s  control is her stock holdings, which are denoted by $(\theta_{i,t})_{t\in[0,1]}$ for $i\in\{1,...,M\}$. For simplicity, the initial endowed holdings of both the bank account and the stock are normalized to zero for all rebalancers.

When $\ta_i= 0$, rebalancer $i$ is a ``high-frequency" liquidity provider with inventory penalties. Because $\ta_i$ is private information for $i$, other traders $k$, $k\neq i$, do not know whether rebalancer $i$ has an active latent trading demand $(\ta_i\neq 0)$ or is a pure liquidity provider $(\ta_i=0)$. 

\item[(ii)] Price-sensitive trackers. Trackers $j\in\{M+1,...,M+\bar{M}\}$ all track a dynamic target given by a common exogenous Brownian motion process $w_t$ over time $t\in[0,1]$ 
\begin{align}\label{w_t}
w_t := w_0 + w^\circ_t,\quad t\in(0,1],
\end{align} 
where the initial target is $w_0 \sim \sN(0,\sigma^2_{w_0})$, the drift is zero,  and  $w^\circ_t$ is a standard Brownian motion that starts at zero, has a zero drift, and a unit volatility.\footnote{ Adding a volatility coefficient $\sigma_w$ in front of  $w^\circ_t$ in \eqref{w_t} does not increase model flexibility because --- as we shall see ---  the stock volatility $\gamma$ is a free model parameter and $\gamma$ and $\sigma_w$ would play identical roles. Moreover, our model can be extended to include a drift term $\mu_w t$ for a constant $\mu_w$ in \eqref{w_t}.} While trackers observe the same $w_t$ at time $t\in[0,1]$, rebalancers   do not and instead filter $w_t$ over time $t\in[0,1]$. Tracker $j$'s  control is her stock holdings, which are denoted by $(\theta_{j,t})_{t\in[0,1]}$ for $j\in\{M+1,...,M+\bar{M}\}$. Their initial stock and money market holdings are also normalized to zero.

We assume the random variables $(\ta_1,...,\ta_M)$, $w_0$, and $(w^\circ_t)_{t\in[0,1]}$ are all independent.  

\end{itemize}

In the following, index $k\in \{1,...,M+\bar M\}$ denotes any generic trader, index $i\in \{1,...,M\}$ denotes a rebalancer, and index $j\in \{M+1,...,M+\bar M\}$ denotes a tracker. This allows us to express the stock-market clearing condition as
\begin{align}\label{xSigmainitial}
0=  \sum_{k=1}^{M+\bar{M}} \theta_{k,t}=  \underbrace{\sum_{i=1}^M \theta_{i,t}}_\text{rebalancer demand}+\underbrace{\sum_{j=M+1}^{M+\bar{M}} \theta_{j,t}}_\text{tracker demand},\quad t\in[0,1].
\end{align}
\noindent Investor stock demands change over time due to stochastic shocks to the tracker target $w_t$ and due to randomness in imperfect learning about the rebalancer targets. As a result, the stock-price process that clears the market as in \eqref{xSigmainitial} changes randomly over time. 
Thus, stock randomness in our model --- given that the fundamental dividend valuation is normalized to zero --- comes from learning about traders’ parent targets (which are initially private information of the individual rebalancers and the trackers) and from random changes over time in the trackers’ target $w_t$. \footnote{Our model features asymmetric information and learning about parent demands. However, because there are no stock dividends, there can be no asymmetric information related to future dividends.}

Investor information is represented as generic filtrations ${\cal F}_{i,t}$ and ${\cal F}_{j,t}$ for rebalancers and trackers.  These filtrations are constructed explicitly in the equilibria considered below. In Section \ref{sec:PI}, the filtrations $\sF_{i,t}$ and $\sF_{j,t}$ are
\begin{align}\label{Rfiltration}
\begin{split}
&\sigma(\tilde{a}_i,S_{i,u})_{u\in[0,t]},\quad t\in[0,1],\quad i\in\{1,...,M\},\\
&\sigma(w_u,S_{j,u})_{u\in[0,t]},\quad  t\in[0,1],\quad j\in \{M+1,...,M+\bar{M}\},
\end{split}
\end{align}
where $S_{i,t}$ and $S_{j,t}$ denote perceived stock-price processes for a rebalancer $i$ and a tracker $j$. However, in the Nash equilibrium in Section \ref{sec:eq1}, more complicated filtrations are needed to derive traders' optimal off-equilibrium response functions.

Our model is a model of dynamic learning.  As we shall see, trackers will be able to infer the aggregate target $\ta_\Sigma$ in \eqref{aS} from the initial stock price, and so trackers have no need to filter the rebalancers’ individual targets $(\ta_1, ..., \ta_M)$. The situation is different for each rebalancer $i\in \{1,...,M\}$, who only observes her own target $\ta_i$ and past and current stock prices. When $\sigma^2_{w_0} >0$, these observations are insufficient to infer $\ta_\Sigma$ and $w_t$ separately, so rebalancer $i$ filters based on $\ta_i$ and on past and current stock price observations to learn about the underlying latent parent demands $\ta_\Sigma$ and $w_t$.  In contrast, when $\sigma_{w_0}:=0$, the model only has static learning about $\ta_\Sigma$ at time $t=0$ from the initial stock price. At later times $t\in (0,1]$, the rebalancers can infer $w_t$ from their stock-price observations. The static learning model with  $\sigma_{w_0}:=0$ was developed in Choi, Larsen, and Seppi (2021).

\subsection{Individual maximization problems}

This section introduces the individual maximization problems. A generic trader $k$'s optimal stock holdings are determined  in terms of a trade-off between expected terminal wealth $X_{k,1}$ and a penalty for deviations of their holdings $\theta_{k,t}$ over time from their parent  target $\ta_i$ (rebalancers) or Brownian motion $w_t$ (trackers). An investor's terminal wealth $X_{k,1}$ depends on the stock prices $S_{k,t}$ associated with $k$'s holdings $\theta_{k,t}$ over time. An exogenous continuous (deterministic) function $\kappa:[0,1]\to[0,\infty]$ models the severity of the target penalty.\footnote{Our analysis can be extended to allow for different penalty functions for the two groups of traders.} The rebalancer and tracker objectives are
\begin{align}\label{Rproblem}
\begin{split}
&\sup_{\theta_{i,t}\in\sF_{i,t}} \E\Big[ X_{i,1} - \int_0^1 \kappa(t)(\tilde{a}_i-\theta_{i,t})^2dt\Big|\,\sF_{i,0}\Big],\quad i\in\{1,...,M\},\\
&\sup_{\theta_{j,t}\in \sF_{j,t}} \E\Big[ X_{j,1} - \int_0^1 \kappa(t)(w_t-\theta_{j,t})^2dt\Big|\,\sF_{j,0}\Big],\quad j\in\{M+1,...,M+\bar{M}\},
\end{split}
\end{align}
where $\ta_i$ is the ideal holdings for rebalancer $i$ and $w_t$ is the ideal holdings for tracker $j$. However, stock-market clearing prevents $\theta_{i,t}$ and $\theta_{j,t}$ from being $\ta_i$ and $w_t$. The suprema in \eqref{Rproblem} are taken over progressively measurable holding processes  $\theta_{i,t}$ and $\theta_{j,t}$ with respect to traders' filtrations $\sF_{i,t}$ and $\sF_{j,t}$. As we shall in Sections \ref{sec:PI} and \ref{sec:eq1} below, our traders optimally use  controls given as smooth functions evaluated at a finite set of state processes (i.e., Markov controls).  The next section constructs such a set of Markovian state processes.   To rule out doubling strategies, we require square integrability
\begin{align}\label{squareint}
\E\left[ \int_0^1 \theta_{k,t}^2 dt \right]<\infty,\quad k\in\{1,...,M+\bar{M}\}. 
\end{align}Terminal wealth $X_{k,1}$ in \eqref{Rproblem} is generated by trader $k$'s perceived wealth process 
\begin{align}\label{Xit}
\begin{split}
dX_{k,t} &:=  \theta_{k,t}dS_{k,t},\quad X_{k,0} := 0,\quad k\in\{1,...,M+\bar{M}\},
\end{split}
\end{align}
which is affected by $k$’s holdings $\theta_{k,t}$ both directly and also indirectly via the impact of $k$’s holdings on an associated perceived stock-price process $S_{k,t}$.   Trader $k$’s holdings $\theta_{k,t}$ are price sensitive for two reasons: First, investors respond to market-clearing price pressure that affects price drifts. Second, price impact means that investor holdings can have a direct effect on the perceived price drift. In \eqref{Xit}, the zero initial wealth $X_{k,0}=0$ is because trader $k$'s initial endowed money market and stock holdings are normalized to zero. Given the objectives in \eqref{Rproblem}, trading reflects a combination of motives: Investors seek to have stock holdings close to their own targets $a_i$ and $w_t$, but they also seek to increase their expected terminal wealth by trading on price pressure from other investors trading on their targets. Thus, traders demand liquidity (to come close to their targets) and supply liquidity  for markets to clear (by being willing to deviate from their targets so that other traders can trade towards their targets, given the appropriate price incentives), and front-run future predictable price pressure.

Our remaining model construction involves specifying investor stock-price perceptions $S_{i,t}$ and $S_{j,t}$ and the associated investor filtrations ${\cal F}_{i,t}$ and ${\cal F}_{j,t}$. We then state conditions that these perceptions and filtrations must satisfy in equilibrium. Finally,  we give theoretical results which ensures equilibria  exist.

\subsection{State processes} \label{sec:states}
The fundamental underlying state of the market in our model depends on the aggregate parent demand imbalances $\ta_\Sigma$ and $w_t$. As already noted, there is a significant informational difference between trackers and rebalancers. Each tracker directly observes $w_t$ in \eqref{w_t} and --- as we shall see --- can therefore infer the aggregate rebalancer target $\ta_\Sigma$ in \eqref{aS} from the initial stock price. In contrast, rebalancers learn about $w_t$ and  $\ta_\Sigma$ using  dynamic filtering. Thus, the rebalancer filtrations $\sF_{i,t}$, $i\in\{1,...,M\}$, and tracker filtrations $\sF_{j,t}$, $j\in \{M+1,...,M+\bar M\}$,  are not nested. Rebalancers know prices and their individual target $\ta_i$, whereas trackers know $\ta_\Sigma$, $w_t$, and prices.

Before considering specific stock-price perceptions in Sections \ref{sec:PI} and \ref{sec:eq1} below, we describe a set of conjectured state processes $(Y_t,\eta_t,q_{i,t},w_{i,t})$ for rebalancer $i\in\{1,...,M\}$. These processes are all endogenous in the equilibria we construct. However, it is convenient to describe the state processes' informational properties first, before showing how they arise in equilibrium. The processes $(Y_t,\eta_t)$ are public in that they are adapted to $\sF_{k,t}$ for all traders $k\in\{1,...,M+\bar{M}\}$.  Furthermore, $\eta_t$ will be adapted to $\sigma(Y_u)_{u\in[0,t]}$.  The state processes $(q_{i,t},w_{i,t})$ are specific to  individual rebalancers. They are adapted to $i$'s filtration $\sF_{i,t}$, but they are not adapted to other traders' filtrations $\sF_{k,t}$ for $k\neq i$. 

Rebalancers learn by extracting information about aggregate demand imbalances from stock prices.  In the equilibria we construct, the information extracted from stock prices over time $t$ is a state process $Y_t$, which has the form
\begin{align}\label{Z}
Y_t:=w_t -  B(t)\ta_\Sigma,\quad t\in[0,1],
\end{align}
where $B:[0,1]\to\R$ is a smooth deterministic function of time that is endogenously determined in equilibrium. The function $B(t)$ controls how $\ta_\Sigma$ and $w_t$ are mixed in stock prices. The process $Y_t$ is not directly observable for the rebalancers, but Lemma \ref{lemma_infer}  below shows that $Y_t$ can be inferred from stock prices. Because rebalancer $i\in\{1,...,M\}$ also knows her own target $\ta_i$, by knowing $Y_t$ over time $t\in[0,1]$,  she equivalently knows 
\begin{align}\label{Zi}
\begin{split}
Y_{i,t}:&= Y_t + B(t)\ta_i\\
&= w_t - B(t)(\ta_\Sigma-\ta_i).
\end{split}
\end{align}
Unlike $Y_t$ in \eqref{Z}, the process $Y_{i,t}$ is independent of rebalancer $i$'s private trading target $\ta_i$ and satisfies
\begin{align}\label{RfiltrationQQQ}
\sigma(\ta_i,Y_u)_{u\in[0,t]} = \sigma(\ta_i,Y_{i,u})_{u\in[0,t]},\quad t\in[0,1].
\end{align}
Rebalancers  use  knowledge of $Y_t$ to estimate $\ta_\Sigma$ and $w_t$ from stock prices at time $t$.  For a continuously differentiable function $B:[0,1]\to\R$, we define two processes
 \begin{align}\label{dwit}
 \begin{split}
q_{i,t} &:= \E\left[\tilde{a}_\Sigma -\ta_i\,\Big|\, \sigma(Y_{i,u})_{u\in[0,t]}\right],\\
dw_{i,t}&:=dw_t-B'(t)\big(\ta_\Sigma-\ta_i - q_{i,t} \big)dt,\quad w_{i,0} := Y_{i,0},
\end{split}
\end{align}
for each rebalancer $i \in \{1,...,M\}$ and $ t\in[0,1]$. 

The expectation $q_{i,t}$ describes what rebalancer $i$ has learned up through time $t$ about the aggregate target $\ta_\Sigma -\ta_i$ of the other rebalancers and about the current value of the trackers’ target $w_t$ from the path of $Y_{i,t}$.\footnote{ Using \eqref{Zi} and \eqref{dwit}, we have $\E[w_t| \sigma(Y_{i,u})_{u\in[0,t]}] = \E[Y_{i,t} + B(t)(\ta_\Sigma-\ta_i)| \sigma(Y_{i,u})_{u\in[0,t]}] =Y_{i,t} +B(t) q_{i,t}$.}  In particular, $q_{i,t}$ is a path-dependent process because it depends on the path of $Y_{i,s}$ over time $s\in [0,t]$.  

Let the function $\Sigma(t)$ denote the remaining variance 
 \begin{align} 
\Sigma(t):= \V[\tilde{a}_\Sigma -\ta_i -q_{i,t}]=\E[(\tilde{a}_\Sigma -\ta_i -q_{i,t})^2],\quad t\in[0,1],
 \end{align}
where the second equality follows from the zero-mean assumptions for $(\ta_1,...,\ta_M)$ and $w_0$. 
Because the targets $(\ta_1,...,\ta_M)$ are assumed independent and homogeneously distributed  $\sN(0,\sigma^2_{\ta})$, the initial variance $\Sigma(0)=\E[(\tilde{a}_\Sigma -\ta_i -q_{i,0})^2]$  is identical across all rebalancers $i\in \{1,...,M\}$. This property and the formula for $\Sigma(t)$ in \eqref{Sigma} below  imply that $\Sigma(t)$ is also independent of index $i\in \{1,...,M\}$ for all $t\in[0,1]$.

Now consider the $w_{i,t}$ processes. Eq.  \eqref{dwit} gives the dynamics of $Y_{i,t}$ as
\begin{align}
\begin{split}
dY_{i,t} &= dw_t - B'(t)(\ta_\Sigma -\ta_i)dt\\
&=dw_{i,t} - B'(t)q_{i,t} dt.
\end{split}
\end{align}

The following result is a special case of the Kalman-Bucy result from filtering theory.

\begin{lemma}[Kalman-Bucy] \label{lemKB} For a continuously differentiable function $B:[0,1]\to\R$, the process $w_{i,t}$ is independent of $\ta_i$, is a Brownian motion, and satisfies  (modulo $\P$ null sets)
\begin{align}\label{RfiltrationQ}
\sigma(\tilde{a}_i,Y_{i,u})_{u\in[0,t]}=\sigma(\tilde{a}_i,w_{i,u})_{u\in[0,t]},\quad t\in[0,1].
\end{align}
Furthermore, the remaining variance at time $t$ is given by
\begin{align}\label{Sigma}
\begin{split}
\Sigma(t)&= \frac{1}{\frac{1}{\V[\tilde{a}_\Sigma -\ta_i -q_{i,0}]}+\int_0^t \big(B'(u)\big)^2du},\quad t\in[0,1].
\end{split}
\end{align}

$\endproof$
\end{lemma}
\noindent Lemma \ref{lemKB}  shows that the $w_{i,t}$ process is observable to rebalancer $i$ given $Y_{i,t}$  and $\ta_i$ and is informationally equivalent to $Y_{i,t}$. Furthermore, Lemma \ref{lemKB} shows that $w_{i,t}$ in \eqref{dwit} is rebalancer $i$'s innovations process. However, while $w_{i,t}$  on the left in \eqref{dwit} is observable by rebalancers, the individual terms $w_t$ and $\ta_\Sigma$ in $w_{i,t}$'s decomposition on the right of \eqref{dwit} are not.

Our equilibrium construction uses the stock-market clearing condition \eqref{xSigmainitial} to relate prices to the state processes driving investor demands. The sum $\sum_{i=1}^M q_{i,t}$ is an important term in this relation, so the following decomposition results are useful:

\begin{lemma}\label{lem:decomp} Let $B:[0,1]\to \R$ be a continuously differentiable  function. 
\begin{enumerate}
\item The decomposition 
\begin{align}\label{SUM1}
\sum_{i=1}^Mq_{i,t} = \eta_t + A(t) \tilde{a}_\Sigma,\quad t\in[0,1],
\end{align}
holds with the process $\eta_t$ being adapted to $\sigma(Y_u)_{u\in[0,t]}$ with $Y_t$ in \eqref{Z} and 
\begin{align}\label{dY2E}
\begin{split}
A'(t)&= - \big(B'(t)\big)^2\Sigma(t)\big(A(t) +1\big),\quad A(0)=-\tfrac{(M-1)B(0)^2\sigma^2_{\ta}}{\sigma^2_{w_0} +(M-1)B(0)^2\sigma^2_{\ta}},\\
d\eta_t & = - \big(B'(t)\big)^2\Sigma(t)\eta_tdt- MB'(t)\Sigma(t)dY_t,\quad \eta_0 =-\tfrac{M(M-1)B(0)\sigma^2_{\ta}}{\sigma^2_{w_0} +(M-1)B(0)^2\sigma^2_{\ta}}Y_0.
\end{split}
\end{align}
\item The inverse relation 
\begin{align}\label{qit_eta}
q_{i,t}&= \frac{\eta_t}{M} - F_1(t)\left( \tfrac{(M-1)B(0)^2 \sigma_a^2}{\sigma_{w_0}^2 + (M-1)B(0)^2\sigma_a^2} +F_2(t) \right) \tilde a_i
\end{align}
holds with deterministic functions $F_1(t)$ and $F_2(t)$ given by the ODEs
\begin{align}\label{F1F2}
\begin{split}
F_1'(t)&=-B'(t)^2 \Sigma(t) F_1(t), \quad F_1(0)=1,\\
F_2'(t)&=\tfrac{B'(t)^2 \Sigma(t)}{F_1(t)}, \quad F_2(0)=0.
\end{split}
\end{align}
\end{enumerate}
$\endproof$
\end{lemma}
\noindent There are two key points: First, no investor knows $\sum_{i=1}^Mq_{i,t}$, but it  can be decomposed into a public term $\eta_t$ and a term $A(t)\ta_\Sigma$ that trackers know but not the rebalancers. Second, from \eqref{dY2E}, the process $\eta_t$ depends on the path of $Y_s$ over time $s\in [0,t]$. Thus, the state process $\eta_t$ reflects common path dependence due to $w_t$. The expression \eqref{qit_eta} shows that the individual rebalancer expectation $q_{i,t}$ includes a common learning component $\frac{\eta_t}{M}$ and then the effect of $i$’s private information $\ta_i$.
In particular, it follows from \eqref{F1F2}, that $F_1(t)$ and $F_2(t)$ are both positive so that, consistent with intuition, the loading on $\ta_i$ is negative in \eqref{qit_eta}.

\section{Price-impact equilibrium}\label{sec:PI}

Investor perceptions of the impact of their trading on stock prices are a key part of the optimizations in \eqref{Rproblem} and the resulting market equilibrium.  We consider two specifications of investor stock-price perceptions. This section presents a simplified model in which perceived price impact is fully exogenous.  This approach is analogous to the exogenous price impact used in van Kerval, Kwan, and Westerholm (2020).  We then solve for the endogenous stock-price process that clears the market (and also satisfies some weak consistency conditions) and the associated optimized investor holding processes.  Section \ref{sec:eq1} presents a richer model of price impact in which investor stock-price perceptions are partially endogenized in a subgame perfect Nash financial-market equilibrium.  

Our equilibrium construction is a conjecture-and-verify analysis. Section \ref{sec31} conjectures functional forms for perceptions of investor stock-price dynamics. Section \ref{sec:PIeq} defines equilibrium and then solves for equilibrium price perception coefficients and the associated price dynamics and holdings that satisfy the definition of equilibrium.

\subsection{Stock-price perceptions}\label{sec31}
 Recall that price pressure is different from the value of future dividends.  It is a valuation adjustment  needed to clear the stock market given trading demand imbalances. This allows us to model price pressure as zero-dividend asset prices as in,  e.g., Karatzas and Shreve (1998, Chap. 4).

Rebalancers optimize \eqref{Rproblem} with respect to perceived stock-price processes of the form  
\begin{align}\label{Sit}
\begin{split}
dS^f_{i,t} &:=  \Big\{f_0(t)Y_t   +f_1(t)\tilde{a}_i +f_2(t)q_{i,t}+f_3(t)\eta_t+ \alpha\theta_{i,t}\Big\}dt+ \gamma dw_{i,t}, \\
S^f_{i,0}&:=Y_0,\quad i\in\{1,...,M\},
\end{split}
\end{align}
where $f_0,f_1,f_2,f_3:[0,1]\to\R$ are continuous (deterministic) functions of time $t\in[0,1]$ and $(\alpha,\gamma)$ are constants. The ``$f$’’ superscript indicates that the perceived price $S^f_{i,t}$ is defined with respect to a particular set of coefficient functions $f$ in \eqref{Sit}. The stock-price drift in \eqref{Sit} is perceived by rebalancer $i$ to be affine in a set of state processes.  Consistent with intuition, we will see in equilibrium that the loadings $f_0(t)$ and $f_3(t)$ on $Y_t$ and $\eta_t$ are negative. In particular, $Y_t$ with $B(t) < 0$ measures a mix of aggregate demand from rebalancers  and trackers, and $\eta_t$ reflects public expectations of aggregate private rebalancer expectations about other rebalancers’ parent demand imbalances, both of which depress price change expectations. The other coefficients describe the perceived impact of rebalancer $i$ on the stock-price drift.  The term $\alpha \theta_{i,t}$ allows for ad hoc trading frictions, and a special case sets $\alpha:=0$. Theorem \ref{thm_PI} below endogenously determines $(f_0,f_1,f_2,f_3)$ in equilibrium. The innovations in the rebalancers’ perceived stock prices $dw_{i,t}$ come from new information rebalancer $i$ learns over time about the underlying parent demand state variable $Y_t$, which has both a direct effect on the future stock-price drift and an additional indirect effect via its effect on $\eta_t$ since $\eta_t$ is adapted to $\sigma(Y_u)_{u\in[0,t]}$ from Lemma \ref{lem:decomp}. 

The zero-dividend stock valuation approach (see, e.g., Chapter 4 in Karatzas and Shreve, 1998) has several consequences: First, we model perceived and equilibrium stock-price drifts rather than price levels.  Second, in \eqref{Sit}, the stock's volatility and initial value are not determined in equilibrium but rather are model inputs. For simplicity,  we set the volatility to be a constant $\gamma> 0$ (i.e., positive demand  innovations $dw_{i,t}$ increase prices) and set the initial price to be $Y_0$ in \eqref{Sit}. However, many other choices would work equally well (e.g., $\gamma(t)$ or $g(Y_0)$). The price-impact parameter $\alpha$ is also an exogenous model input. The exogenous parameters $(\alpha,\gamma)$ can be found by calibrating model output to empirical data. A competitive market is a special case with $\alpha:=0$, whereas the empirically relevant case is  $\alpha<0$ such that buy (sell) orders decrease (increase) the future stock-price drifts.

The next result shows that $w_{i,t}$ is rebalancer $i$'s innovations process in the sense that
$w_{i,t}$ is a Brownian motion relative to $i$'s filtration defined with perceived stock prices $S^f_{i,t}$ in \eqref{Sit} and such that $S^f_{i,t}$ and $w_{i,t}$ generate the same information.

\begin{lemma}\label{lemma_infer} Let $f_0,f_1,f_2,f_3:[0,1]\to\R$  be continuous functions and let $B:[0,1]\to \R$ be a continuously differentiable  function. For a rebalancer $i\in\{1,...,M\}$, let $\theta_{i,t}$ satisfy  \eqref{squareint} and be progressively measurable with respect to $\sF_{i,t}:=\sigma(\ta_i,S^f_{i,u})_{u\in[0,t]}$ with $S^f_{i,t}$ defined in \eqref{Sit} and $Y_t$ defined in \eqref{Z}. Then, modulo $\P$-null sets, we have
\begin{align}\label{filt1}
\sigma(\ta_i, w_{i,u})_{u\in[0,t]}  = \sigma(\ta_i,S^f_{i,u})_{u\in[0,t]} ,\quad t\in[0,1],\quad  i\in\{1,...,M\}.
\end{align}
$\endproof$
\end{lemma}
\noindent Thus, given a path of perceived prices generated by a price process $S^f_{i,t}$ of the form in \eqref{Sit} and her personal target $\ta_i$, rebalancer $i$ can infer the path of $w_{i,t}$. Furthermore, given the path $w_{i,t}$,  rebalancer $i$ can infer $Y_{i,t}$ using \eqref{RfiltrationQ} and, thus, can infer $Y_t$ from \eqref{RfiltrationQQQ}. Consequently,  rebalancer $i$ can infer $(q_{i,t},\eta_t)$ where we recall from Lemma \ref{lem:decomp}  that $\eta_t$ is adapted to $\sigma(Y_t)_{t\in[0,1]}$.

Trackers optimize \eqref{Rproblem} with respect to a perceived stock-price process of the form
\begin{align}\label{New2}
\begin{split}
dS^{\bar f}_{j,t} :&=  \Big\{\bar{f}_3(t)\eta_t+\bar{f}_4(t)\ta_\Sigma+\bar{f}_5(t)w_t+ \alpha\theta_{j,t}\Big\}dt + \gamma dw_t,\\
 S^{\bar f}_{j,0}:&=Y_0,\quad j\in \{M+1,...,M+\bar{M}\},
\end{split}
\end{align}
where $\bar{f}_3,\bar{f}_4,\bar{f}_5:[0,1]\to\R$ are continuous (determinstic) functions and the  $\alpha$ is a constant.\footnote{Our model can be extended to allow for a different price-impact coefficient $\bar\alpha$ with $\bar\alpha \neq \alpha$ for the trackers.} Trackers have different information in that they observe $w_t$ directly and can infer $\ta_\Sigma$ from the initial stock price $Y_0$ using \eqref{Z} and their knowledge of $w_0$. Therefore, their perceived stock prices differ from those of the rebalancers.  Theorem \ref{thm_PI} below endogenously determines $(\bar{f}_3,\bar{f}_4,\bar{f}_5)$ in equilibrium, whereas $(\alpha,\gamma)$ are exogenous model inputs. Again, $\alpha:=0$ is the special case of a competitive market.

The motivation for these price perceptions for the trackers is as follows. First, the perceptions in \eqref{New2} allow trackers to condition their perceived price drift to take into account price pressure from target imbalances $\ta_\Sigma$ and $w_t$ that depress expected price changes.   Since trackers and rebalancers trade differently on their targets, the price-drift impacts $\bar{f}_4$ and $\bar{f}_5$ are in general different.  Second, the trackers understand that the state process $Y_t$ affects the rebalancer demand and, thus, the stock-price drift. However, $Y_t$ does not need to be included explicitly in the tracker perceived price drift in \eqref{New2} since $Y_t$ can be computed from the underlying variables $\ta_\Sigma$ and $w_t$ that are already included in the drift. Third, trackers know that rebalancers’ can infer $\eta_t$ and that this potentially affects their price perceptions in \eqref{Sit}, and, thus, is likely to affect their trading, and, thus, is likely to affect pricing.  Thus, trackers allow for the pricing effect of $\eta_t$ in their perceptions in \eqref{New2}.  Third, as already noted, $\alpha$ allows for possible exogenous price frictions, if any.

An important difference between rebalancer and tracker perceived prices in \eqref{Sit} and \eqref{New2} is that rebalancer price dynamics are based on the informational innovations $dw_{i,t}$, whereas tracker price dynamics are based on the tracker target changes  $dw_t$.  Reconciling the price perceptions of rebalancers and trackers will impose restrictions on equilibrium price perceptions and holdings and will rely on the relation between $dw_{i,t}$ and $dw_t$ in \eqref{dwit}.

Given the price perceptions in \eqref{Sit} and \eqref{New2}, we solve \eqref{Rproblem}  for optimal rebalancer and tracker holdings.
\begin{lemma} \label{PI_Le} Let $f_0,f_1,f_2,f_3,\bar{f}_3,\bar{f}_4,\bar{f}_5:[0,1]\to \R$ and  $\kappa:[0,1]\to (0,\infty]$ be  continuous functions, let $B:[0,1]\to \R$ be continuously differentiable, let $\alpha \le 0$,
and let the perceived stock-price process in the wealth dynamics \eqref{Xit} be as in \eqref{Sit} and \eqref{New2}.  Then,  for  $\sF_{i,t}:=\sigma(\ta_i,S^f_{i,u})_{u\in[0,t]}$ and $\sF_{j,t}:=\sigma(w_u,S^{\bar f}_{j,u})_{u\in[0,t]}$, and,  provided the holding processes
\begin{align}\label{Y00PI}
\begin{split}
\hat{\theta}_{i,t} &:= \frac{f_0(t)}{2 (\kappa (t)-\alpha )}Y_t+\frac{f_1(t)+2 \kappa (t)}{2(\kappa(t)- \alpha)}\ta_i+\frac{f_2(t)}{2 (\kappa (t)-\alpha )}q_{i,t}+\frac{f_3(t)}{2 (\kappa (t)-\alpha )}\eta_t,\\
\hat{\theta}_{j,t}&:= \frac{\bar{f}_3(t)}{2 (\kappa (t)-\alpha )}\eta_t+\frac{\bar{f}_5(t)+2 \kappa (t)}{2(\kappa(t)- \alpha)}w_t+\frac{\bar{f}_4(t)}{2 (\kappa (t)-\alpha )}\ta_\Sigma,
\end{split}
\end{align}
 satisfy \eqref{squareint},  the traders' maximizers for \eqref{Rproblem} are $\hat{\theta}_{i,t} $ for rebalancer $i\in\{1,...,M\}$ and $\hat{\theta}_{j,t}$ for tracker $j\in \{M+1,...,M+\bar{M}\}$.
$\endproof$
\end{lemma}

\noindent The proof of Lemma \ref{PI_Le} shows that pointwise quadratic maximization gives the maximizers for \eqref{Rproblem} for rebalancers and trackers for arbitrary $f$ and $\bar f$ functions.

Stock-price perceptions play two interconnected roles in our model.  First, rebalancers and trackers solve their optimization problems in \eqref{Rproblem} based on their perceptions in \eqref{Sit} and \eqref{New2} for how hypothetical holdings $\theta_{i,t}$ and $\theta_{j,t}$ affect price dynamics.  Second, investor stock-price perceptions affect how they learn from observed prices.  In particular, Lemma \ref{lemma_infer} shows that rebalancers use their stock-price perceptions \eqref{Sit} to infer the aggregate demand state variable $Y_t$ based on past and current stock prices.  In other words, dynamic learning by rebalancers depends critically on their stock-price perceptions.  Similarly, trackers also use their stock-price perception of $Y_0$ in \eqref{New2}  to infer the aggregate parent demand $\ta_\Sigma$ from the initial price at time $t=0$.  However, thereafter, there is no additional learning from prices by the trackers since they directly observe their target $w_t$.

\subsection{Equilibrium}\label{sec:PIeq}
This section defines our first of two equilibrium concepts and then derives price perception coefficients for the conjectured functional form in Section \ref{sec31} that satisfy the equilibrium definition along with the associated equilibrium price dynamics and holdings. The notion of equilibrium in our first construction is relatively simple, being based just on market clearing and consistency of investor price perceptions.

\begin{definition}\label{PI_eq} Deterministic functions of time $f_0,f_1,f_2,f_3,\bar{f}_3,\bar{f}_4,\bar{f}_5,B:[0,1]\to\R$ constitute a \emph{price-impact equilibrium} if:
\begin{enumerate}
\item[(i)] Maximizers $\hat{\theta}_{k,t}$ for \eqref{Rproblem} exist for traders $k \in \{1,...,M+\bar{M}\}$ given the stock-price perceptions \eqref{Sit} and \eqref{New2} for filtrations $\sF_{i,t}:=\sigma(\ta_i,S^f_{i,u})_{u\in[0,t]}$ and $\sF_{j,t}:=\sigma(w_u,S^{\bar f}_{j,u})_{u\in[0,t]}$.

\item[(ii)] Inserting trader $k$'s maximizer $\hat{\theta}_{k,t}$ into the perceived  stock-price processes \eqref{Sit} and \eqref{New2} produces identical stock-price processes across all traders   $k\in\{1,...,M+\bar{M}\}$. This common equilibrium stock-price process is denoted by $\hat{S}_t$.

\item[(iii)] The money and stock markets clear.
$\endproof$
\end{enumerate}

\end{definition}
\noindent Definition \ref{PI_eq} places only minimal restrictions on the perceived stock-price coefficient functions in \eqref{Sit} and \eqref{New2}: Markets must clear and result in consistent perceived stock-price processes when all investors use their equilibrium strategies.  Section \ref{sec:eq1} below considers a subgame perfect Nash extension of our basic model that imposes more restrictions on allowable off-equilibrium stock-price perceptions such as market clearing and various consistency requirements.

Definition \ref{PI_eq}(ii) requires that in equilibrium rebalancers and trackers perceive identical stock-price dynamics when using their equilibrium holdings. However, rebalancers and trackers have different information (i.e., rebalancers form imperfect inferences about $w_t$ and $\ta_\Sigma$, whereas trackers observe $w_t$ directly and infer $\ta_\Sigma$ at time 0).  The resolution of this  apparent paradox is investors' different information sets: Trackers and rebalancers all agree on $d\hat{S}_t$, but they disagree on how to decompose $d\hat{S}_t$ into drift and volatility components. Because the trackers observe $w_t$, they can use $dw_t$ in their decomposition of $d\hat{S}_t$. However, $w_t$ is not adapted to the rebalancers' filtrations and can therefore not be used in their $d\hat{S}_t$ decompositions. Instead, rebalancers use their innovations processes $dw_{i,t}$ when decomposing $d\hat{S}_t$ into drift and volatility. By replacing $dw_{i,t}$ in $dS^f_{i,t}$ in \eqref{Sit}  with the decomposition of $dw_{i,t}$ in terms of $dw_t$ from \eqref{dwit}, we can rewrite $dS^f_{i,t}$ in \eqref{Sit}  as
\begin{align}\label{Y32PI}
\begin{split}
dS^f_{i,t} &=  \Big\{f_0(t)Y_t  +f_1(t)\tilde{a}_i +f_2(t)q_{i,t}+f_3(t) \eta_t +\alpha \theta_{i,t}\\
&\quad -B'(t)\big(\ta_\Sigma-\ta_i - q_{i,t} \big) \gamma \Big\}dt+ \gamma dw_t,\quad i\in\{1,...,M\}.
\end{split}
\end{align}
Therefore, to ensure identical equilibrium stock-price perceptions for all traders $k\in\{1,...,M+\bar{M}\}$,   it suffices to match the drift of $dS^{\bar f}_{j,t}$ in \eqref{New2} for $j\in\{M+1,...,M+\bar{M}\}$ with the drift of $dS^f_{i,t}$ in \eqref{Y32PI}  for the equilibrium holdings  $\theta_{i,t}:= \hat{\theta}_{i,t}$, $i\in\{1,...,M\}$. This produces the following equilibrium requirement:
\begin{align}\label{driftAPI}
\begin{split}
&f_0(t)Y_t  +f_1(t)\tilde{a}_i +f_2(t)q_{i,t}+f_3(t) \eta_t +\alpha \hat{\theta}_{i,t} -B'(t)\big(\ta_\Sigma-\ta_i - q_{i,t} \big)\gamma\\
&=\bar{f}_3(t)\eta_t+\bar{f}_4(t)\ta_\Sigma+\bar{f}_5(t)w_t+ \alpha\hat{\theta}_{j,t},
\end{split}
\end{align}
for all rebalancers $i \in\{1,...,M\}$ and all trackers $j\in \{M+1,...,M+\bar{M}\}$. We note that the right-hand side of \eqref{driftAPI} does not depend on the rebalancer index $i$. Matching up coefficients in front of $(\ta_i,\ta_{\Sigma},q_{i,t},\eta_t,w_t)$ in \eqref{driftAPI} using $\hat{\theta}_{i,t}$ and $\hat{\theta}_{j,t}$ in \eqref{Y00PI} and $Y_t$ in \eqref{Z} produces five equations. In addition,  inserting $\hat{\theta}_{i,t}$ and $\hat{\theta}_{j,t}$ in \eqref{Y00PI} into  the market-clearing condition   \eqref{xSigmainitial} and using \eqref{SUM1} produce three more equations from matching $(\ta_{\Sigma},\eta_t,w_t)$ coefficients. All in all, we have eight requirements in $(f_0,f_1,f_2,f_3,\bar{f}_3,\bar{f}_4,\bar{f}_5)$ and $B'$, which give the equilibrium coefficient functions \eqref{fs} in Appendix \ref{sec:formulas} and the  ODE for $B(t)$ in \eqref{derivatives0aPI} below.

Our equilibrium existence result is based on the following technical lemma. It guarantees the existence of a solution to an autonomous system of coupled ODEs. In particular, given rebalancer stock-price perceptions of the form in \eqref{Sit} with an aggregate demand state variable $Y_t$ process of the form in \eqref{Z} (and the associated $\eta_t$ process), we must construct a deterministic function  $B(t)$ that gives an equilibrium.  

\begin{lemma}\label{PI_Lemma} 
Let $\kappa:[0,1]\to[0,\infty]$ be a continuous and integrable function (i.e., $\int_0^1 \kappa(t)dt <\infty$). For an initial constant $B(0) \in \R$, the coupled ODEs 
\begin{align}\label{derivatives0aPI}
\begin{split}
B'(t)&= \frac{2 \kappa (t) (\bar{M} B(t)+1)}{\gamma  (A(t)+\bar{M}+1)},\\
A'(t)&= - \big(B'(t)\big)^2\Sigma(t)\big(A(t) +1\big),\quad A(0)=-\tfrac{(M-1)B(0)^2\sigma^2_{\ta}}{\sigma^2_{w_0} +(M-1)B(0)^2\sigma^2_{\ta}},\\
\Sigma'(t) &= -\big(B'(t)\big)^2\Sigma(t)^2,
\quad \Sigma(0) =\tfrac{(M-1) \sigma_{\ta}^2 \sigma_{w_0}^2}{(M-1)B(0)^2  \sigma_{\ta}^2+\sigma_{w_0}^2},
\end{split}
\end{align}
have unique solutions with $\Sigma(t) \ge 0$, $\Sigma(t)$ decreasing, $A(t) \in [-1,0]$, $A(t)$ decreasing for $t\in[0,1]$, and $B(t),B'(t)<0$ when $\bar{M}B(0) +1< 0$.
$\endproof$
\end{lemma}
\noindent The ODEs for $A(t)$ and $\Sigma(t)$ in \eqref{derivatives0aPI} are consistent with the expressions in \eqref{Sigma} and \eqref{dY2E}. The exogenous price-impact coefficient $\alpha$ does appear in the ODEs \eqref{derivatives0aPI}.

The following theoretical result gives the price-impact equilibrium in terms of the ODEs \eqref{derivatives0aPI}. In this theorem, the price-impact parameter $\alpha$, volatility $\gamma$, and initial value $B(0)\in\R$ are free parameters.  The intuition for $B(0)$ being free is discussed after our equilibrium construction in Theorem \ref{thm_PI}.

\begin{theorem}\label{thm_PI} Let $\kappa:[0,1]\to (0,\infty)$ be continuous, let the functions $(B,A,\Sigma)$ be as in Lemma \ref{PI_Lemma}, and let $\alpha\le0$. Then, we have:

\begin{itemize}

\item[(i)] A price-impact equilibrium exists and is given by the price-perception functions \eqref{fs} in Appendix \ref{sec:formulas}.

\item[(ii)]  Equilibrium holdings $\hat{\theta}_{i,t}$ for rebalancer $i$ and $\hat{\theta}_{j,t}$ for tracker $j$ are 
\begin{align}\label{Y0000PI}
\begin{split}
\hat{\theta}_{i,t} &=-\tfrac{\gamma  B'(t)-2 \kappa (t)}{2 \kappa (t)-\alpha}\ta_i -\tfrac{\gamma  B'(t)}{2 \kappa (t)-\alpha }q_{i,t}
\\
&+\tfrac{\gamma  B'(t)}{(M+\bar{M}) (2 \kappa (t)-\alpha )}\eta_t-\tfrac{2 \bar{M} \kappa (t)}{(M+\bar{M}) (2 \kappa (t)-\alpha )}Y_t,\quad i\in\{1,...,M\},
\\
\hat{\theta}_{j,t} &=\tfrac{\gamma  B'(t)}{(M+\bar{M}) (2 \kappa (t)-\alpha )}\eta_t 
+\tfrac{2 M \kappa (t)}{(M+\bar{M}) (2 \kappa (t)-\alpha )}w_t
\\
&+\tfrac{\gamma  (A(t)-M+1) B'(t)-2 \kappa (t)}{(M+\bar{M}) (2 \kappa (t)-\alpha)}\ta_\Sigma,\quad j\in\{M+1,...,M+\bar{M}\}.
\end{split}
\end{align}

\item[(iii)]    There exists an equilibrium stock-price process $\hat{S}_t$ with $\hat{S}_0 := w_0 - B(0)\ta_\Sigma$ and dynamics with respect to the trackers' filtrations $\sF_{j,t}:=\sigma(w_u,S^{\bar f}_{j,u})_{u\in[0,t]}$ given by
\begin{align}\label{S_PI}
\begin{split}
d\hat{S}_t &=\Big\{\tfrac{\gamma  B'(t)}{M+\bar{M}}\eta_t-\tfrac{2 \bar{M} \kappa (t)}{M+\bar{M}}w_t +\tfrac{\gamma  (A(t)-M+1) B'(t)-2 \kappa (t)}{M+\bar{M}}\ta_\Sigma\Big\}dt + \gamma dw_t,
\end{split}
\end{align}
and dynamics with respect to the rebalancers' filtrations $\sF_{i,t}:=\sigma(\ta_i,S^f_{i,u})_{u\in[0,t]}$ given by
\begin{align}\label{rebdriftPI}
\begin{split}
d\hat{S}_t &=\Big\{\tfrac{\gamma  B'(t)}{M+\bar{M}}\eta_t-\tfrac{2 \bar{M} \kappa (t)}{M+\bar{M}}Y_t-\gamma B'(t)\big( \ta_i + q_{i,t}\big) \Big\}dt+ \gamma dw_{i,t}.
\end{split}
\end{align}
$\endproof$
\end{itemize}
\end{theorem}

Several observations follow from Theorem \ref{thm_PI}:

\begin{enumerate}
\item  Lemma \ref{lemma_infer} ensures that rebalancer $i$ can infer her innovations process $w_{i,t}$  from perceived prices $S^f_{i,t}$ and $\ta_i$, but rebalancer $i$ cannot  infer the trackers' target $w_t$ from the equilibrium prices  $\hat{S}_t$  in \eqref{S_PI}. This is because the aggregate target $\ta_\Sigma$ also appears in the drift of $d\hat{S}_t$ and $\ta_\Sigma$ is not observed by individual rebalancers.

\item The equilibrium holdings \eqref{Y0000PI} follow from inserting the $f$ and $\bar f$ functions in \eqref{fs}  in Appendix \ref{sec:formulas} into \eqref{Y00PI}. Thus, the holdings in \eqref{Y0000PI} are expressed in terms of the investors' state processes, which, in particular, are adapted to the investors' filtrations. However, these state processes are not mutually independent and so we give such representations of \eqref{Y0000PI} in \eqref{thetaiortho} and \eqref{thetajortho} in Appendix \ref{sec:formulas}. First, the price-impact equilibrium rebalancer holdings $\hat{\theta}_{i,t}$ in  \eqref{Y0000PI} can be written in terms of the independent variables $(\ta_i, \ta_\Sigma-\ta_i, w_0)$ and an  residual independent term given as a stochastic integral with respect to  $w^\circ_t$ of a deterministic function of time. Likewise, the price-impact equilibrium  tracker holdings $\hat{\theta}_{j,t}$ can be written in terms of the independent variables $(\ta_\Sigma, w_0)$ and an  residual orthogonal term given in terms of a stochastic integral with respect to  $w^\circ_t$ of a deterministic function of time. Both these residual terms are Gaussian. Section \ref{sec:num} illustrates the loading coefficients on these independent state processes. 

\item Because the exogenous price-impact coefficient $\alpha\le0$ does not appear in the ODEs \eqref{derivatives0aPI}, $\alpha$ is irrelevant for the equilibrium stock-price dynamics 
\eqref{S_PI}. However, $\alpha$ does affect the equilibrium holdings in \eqref{Y0000PI}.

\item The stock-price volatility $\gamma$ affects the stock-price drift and holdings via its impact on $B(t)$ in \eqref{derivatives0aPI} and, thus, on \eqref{fs}.

\item  It can seem paradoxical that the equilibrium stock-price process $\hat{S}_t$ has different sets of dynamics (i.e., It\^o decompositions of $d\hat S_t$ into multiple sets of drift and martingale terms). The resolution lies in the rebalancers and trackers having different filtrations: The drift and martingale terms in \eqref{rebdriftPI} are not adapted to $\sF_{j,t}$ and the drift and martingale terms in \eqref{S_PI} are not adapted to $\sF_{i,t}$ for $i\in\{1,...,M\}$ and $j\in \{M+1,...,M+\bar{M}\}$. The dynamics \eqref{S_PI} and \eqref{rebdriftPI} all produce the same process $\hat{S}_t$ because the innovations process $w_{i,t}$ in \eqref{dwit} links $dw_t$  with $dw_{i,t}$ and the drift term $B'(t)(\ta_\Sigma -\ta_i-q_{i,t})dt$.  Thus, trackers and rebalancers all perceive the same equilibrium stock-price dynamics $d\hat S_t$ but they decompose those dynamics into different perceived drifts and martingale terms.\footnote{ Rebalancers and trackers both start with private information so their filtrations are not nested. However, in equilibrium, stock-price dynamics depend on $w_t$ and $\ta_\Sigma$. Because the trackers know $w_0$ at time $t=0$, they infer $\ta_\Sigma$ from $S_{j,0}=w_0-B(0)\ta_\Sigma$, they have no need to filter at later times. On the other hand, rebalancer $i$ only has noisy dynamic predictions $\E[\ta_\Sigma|\sF_{i,t}] = q_{i,t}+\ta_i$ of the aggregate parent imbalance $\ta_\Sigma$ given her inferences based on the individual parent targets $\ta_i$ and stock-price observations.}
 
\item Investors' off-equilibrium perceived stock-price drifts differ linearly from their equilibrium drifts due to the differences $\theta_{k,t}-\hat{\theta}_{k,t}$ between their off-equilibrium and equilibrium holdings.\footnote{Eqs. \eqref{decom1} and \eqref{decom2} are similar to Eq. (3.14) in Choi, Larsen, and Seppi (2021).}  Rebalancer $i$'s perceived stock-price drift in \eqref{Sit} can be decomposed for arbitrary holdings $\theta_{i,t}$  as
\begin{align}\label{decom1}
\begin{split}
&f_0(t)Y_t   +f_1(t)\tilde{a}_i +f_2(t)q_{i,t}+f_3(t)\eta_t+ \alpha\theta_{i,t}\\
 &= -\gamma B'(t)\big( \ta_i + q_{i,t}\big) + \tfrac{\gamma  B'(t)}{M+\bar{M}}\eta_t-\tfrac{2 \bar{M} \kappa (t)}{M+\bar{M}}Y_t +\alpha(\theta_{i,t} - \hat\theta_{i,t}),
 \end{split}
\end{align}
where we have used the formulas for $(f_0,f_1,f_2,f_3)$ in \eqref{fs} in Appendix \ref{sec:formulas}. 
Likewise,  for arbitrary holdings $\theta_{j,t}$,  tracker $j$'s perceived stock-price drift in \eqref{New2} is
\begin{align}\label{decom2}
\begin{split}
&\bar{f}_3(t)\eta_t+\bar{f}_4(t)\ta_\Sigma+\bar{f}_5(t)w_t+ \alpha\theta_{j,t}\\
 &=\tfrac{\gamma  B'(t)}{M+\bar{M}}\eta_t-\tfrac{2 \bar{M} \kappa (t)}{M+\bar{M}}w_t +\tfrac{\gamma  (A(t)-M+1) B'(t)-2 \kappa (t)}{M+\bar{M}}\ta_\Sigma+\alpha(\theta_{j,t} - \hat\theta_{j,t}),
 \end{split}
\end{align}
where we have used the formulas for $(\bar{f}_3,\bar{f}_4,\bar{f}_5)$ in \eqref{fs} in Appendix \ref{sec:formulas}. 

Continuity is a reasonable property of investor perceptions. The representation of the perceived rebalancer drift in \eqref{decom1} relative to $\hat\theta_{i,t}$ from \eqref{Y0000PI} also explains the presence of the rebalancer-specific terms $(\ta_i,q_{i,t})$ in the rebalancers'  perceptions in \eqref{Sit}.

 \item Our equilibrium construction verifies that price-perception coefficients in  \eqref{Sit} and \eqref{New2} can be constructed such that an equilibrium satisfying Definition \ref{PI_eq} exists. However, as with many other rational expectation models, we do not have a proof of uniqueness.  For example, there may be other public state variables in addition to $\eta_t$ that could hypothetically be included in the perceived price drifts that might also be associated with equilibria as defined in Definition  \ref{PI_eq}.

\end{enumerate}

The function $B(t)$ from \eqref{derivatives0aPI} is key both in constructing the equilibrium and for interpreting the equilibrium price and holding processes.  First, there is the issue that the initial value $B(0)$ is a free input in Theorem \ref{thm_PI}.  The intuition is that our model determines equilibrium stock-price drifts but not price levels. As can be seen in \eqref{S_PI}, $B(0)$ controls the initial price level in our model. Second, the relation between $B(t)$ and price levels allows us to impose additional structure on $B(t)$.  In particular, $w_t$ and $\ta_\Sigma$ represent different types of demand imbalances. Thus, if $B(t) < 0$, then $Y_t$ in \eqref{Z} plays the role of an aggregate demand state variable. How the two component quantities $w_t$ and $\ta_\Sigma$ are mixed in the aggregate demand state variable $Y_t$ is different given the two components' different informational dynamics (i.e., $\ta_\Sigma$ is fixed after time 0 while $w_t$ changes randomly over time) and the different impacts on investor demands (i.e., each rebalancer only knows their personal $\ta_i$ component of $\ta_\Sigma$ where other rebalancers' targets do not affect investor $i$’s parent demand whereas $w_t$ affects both an individual tracker’s parent demands and is also information about other trackers’ parent demands).  It seems reasonable that the sign of the impact of $w_t$ and $\ta_\Sigma$ on the price level should be the same, which imposes  the additional restriction that $B(t) < 0$. From Lemma \ref{PI_Lemma}, a sufficient condition for $B(t) < 0$ for all $t \in [0,1]$ is ${\bar M}B(0) + 1<0$.\footnote{ This sufficient condition follows because the denominator in \eqref{derivatives0aPI} is positive given that $A(t) \in [-1,0]$ so that the numerator in \eqref{derivatives0aPI} determines the sign of $B’(t)$.  } 

With the economically reasonable parametric restriction that $B'(t) < 0$ and given that $\alpha \leq 0$ so that $\alpha - 2 \kappa(t) < 0$, we can sign the impact of various quantities in the model on holdings and prices, which leads to the following comparative statics:

\begin{enumerate}
\item The equilibrium holdings $\hat{\theta}_{i,t}$ of rebalancers are positively related to their parent targets $\ta_i$. This is intuitive because rebalancers want holdings close to $\ta_i$.  Rebalancer holdings $\hat{\theta}_{i,t}$ are also negatively related to the aggregate demand imbalance state variable $Y_t$.  The  fact that $\theta_{i,t}$ is decreasing in $Y_t$ is consistent with the theoretical results and empirical evidence in van Kerval, Kwan, and Westerholm (2020) that investors buy less when there is a positive parent demand imbalance for other investors in the market. However, the impact of $q_{i,t}$ on $\hat \theta_{i,t}$ is positive.  The intuition is that when rebalancer $i$ expects the other remaining rebalancers (given $i$’s ability to filter using her private target information $\ta_i$) to have a net positive parent demand imbalance $\E[\ta_\Sigma - \ta_i |\sF_{i,t}]$ from \eqref{dwit}, she buys at time $t$ to front-run the resulting anticipated  future  price pressure.

\item Tracker $j$'s holdings $\hat \theta_{j,t}$ are increasing in $w_t$ (which reflects both her own  parent demand  and also information  about the parent demands of other trackers).  Tracker holdings $\hat\theta_{j,t}$ are also decreasing in $\eta_t$, which is related to imbalances in rebalancers' aggregate parent demand expectations. The effect of $\eta_t$ is consistent with the van Kerval, Kwan, and Westerholm (2020) liquidity-provision result and empirical evidence. However, the impact of $\ta_\Sigma$ is ambiguous in \eqref{Y0000PI}, and numerical calculations in Section \ref{sec:num} show that the sign is positive.  This is again consistent with front-running future predicted price pressure due to the tracker’s superior information about aggregated latent parent demand imbalances.

\item The equilibrium stock-price drift in \eqref{S_PI} is decreasing in the tracker parent demand $w_t$. However, the impact of $\ta_\Sigma$ in the price drift is again ambiguous, which is related to  information about $\ta_\Sigma$ being useful in forecasting future price pressure.
 
\end{enumerate}

\subsection{Tractability and model structure}

This section discusses the key model components that make our model tractable. First, we assume all traders seek to maximize their individual objectives in \eqref{Rfiltration}. Linear-quadratic objectives have been used extensively in the literature because of their tractability. Such objectives have been used in, e.g., Kyle (1985), Brunnermeier and Pederson (2005), and Carlin, Lobo, and Viswanathan (2007). The linear-quadratic objectives  \eqref{Rproblem} allow us to solve for the optimal holdings in  Lemma \ref{PI_Le} using quadratic pointwise optimization.  In the price-impact equilibrium, we could equivalently use dynamic programming to produce the same optimal holdings.

Second, our stock does not pay dividends, which means that only the stock drift can be endogenously determined in equilibrium. Models with non-dividend paying stocks have been used extensively in the literature. The monograph Karatzas and Shreve (1998) gives an overview.\footnote{Similar to a money market account, a non-dividend paying stock is a \emph{financial asset} in the sense that holding one stock at time $t=1$, gives one unit of consumption at $t=1$. Likewise, being short one stock at $t=1$, means the trader provides one unit of consumption at $t=1$. Both the money market account and the non-dividend paying stock have exogenous initial prices and volatilities. It is custom for the money market account's initial price to be one and its volatility to be zero. For the non-dividend paying stock,  we set the initial price to be $Y_0$, its volatility to be a positive constant $\gamma$, and determine endogenously the drift.  }  In particular, non-dividend paying stock models have been used for short horizon models like ours where consumption only takes place at the terminal time.\footnote{There are long-lived non-dividend paying stocks too as; see, for example, Atmaz and Basak (2021) write: ``For example, Hartzmark and Solomon (2013) find that over the long-sample of 1927-2011, the average proportion of no-dividend stocks is around 35\% and accounts for 21.3\% of the aggregate US stock market capitalization. Similarly, by taking into account of rising share repurchase programs since the mid-1980ies, Boudoukh et al. (2007) report that over the 1984-2003 period, the average proportion of no-dividend stocks is 64\% and no-payout stocks, i.e., no dividends or no share repurchases, is 51\% with the relative market capitalizations of 16.4\% and 14.2\%, respectively."} The rebalancers' dynamic learning produces  forward-running filtering equations and by considering a non-dividend paying stock, we circumvent having additional backward-running equations. Equilibrium models with both forward and backward-running equations include Kyle (1985), Foster and Viswanathan (1994, 1996), Back,  Cao, and Willard (2000), and Choi, Larsen, and Seppi (2019). 

Third, standard ways price impact are modeled are as the impact of investor holdings and orders on price levels (e.g., as in Almgren (2003)) and as the impact of orders on price changes (e.g., Kyle (1985)).  However, for the sake of tractability, we follow Cuoco and Cvitani\'c (1998) and model price impact in terms of the impact of investor holdings on the price drift.  One important reason that price impact matters for the trading decisions of strategic investors is because of its effect on future expected price changes (e.g., buy orders raise prices which lowers expected future price appreciation). Our price impact specification simply assumes directly that investor holdings affect expected future price changes.  Thus, while our price impact specification is a simplification, we argue that it is a reasonable simplification that preserve the essential economics of price impact.

Fourth, instead of exogenous noise traders, we use optimizing trackers. Grossman and Stiglitz (1980) and Kyle (1985) are standard references, which use an exogenous Gaussian stock supply. Gaussian noise traders are also used in the predatory trading models in Brunnermeier and Pederson (2005) and Carlin, Lobo, and Viswanathan (2007). In our setting, we could eliminate trackers by setting $\bar{M}:=0$ and replace  the stock-market clearing condition \eqref{xSigmainitial} by using $w_t$ to model the exogenous stock supply as in 
\begin{align}\label{xSigmainitialA}
w_t=  \sum_{i=1}^{M} \theta_{i,t},\quad t\in[0,1].
\end{align}
Including noise traders as in \eqref{xSigmainitialA} in the model would be  tractable in the price-impact equilibrium. However,  surprisingly, exogenous noise-traders  complicate constructing a Nash equilibrium with dynamic learning, whereas  --- as we show in  Section \ref{sec:eq1} --- optimizing trackers and market learning in  \eqref{xSigmainitial} produce a subgame perfect Nash financial-market equilibrium in closed form. The models in Sannikov and Skrzypacz (2016) and Choi, Larsen, and Seppi (2021) have optimizing trackers but no dynamic learning.

\subsection{Numerics}\label{sec:num}

Our price-impact equilibrium is straightforward to compute numerically.  This is because equilibrium stock prices and holdings are available in closed form given the solutions to the associated coupled ODEs in \eqref{derivatives0aPI}.  We illustrate our models for several different parameterizations.  In these parameterizations, there are $M := 5$ rebalancers and $\bar M := 10$ trackers.  The penalty function  is a constant over the trading day and set to $\kappa(t):=1$.  The rebalancer target volatility is normalized to $\sigma_{\ta} := 1$ whereas we consider $\sigma_{w_0} \in\{\frac1{10}, 1\}$ to illustrate the impact of dynamic learning. Recall that $\sigma_{w_0}:=0$ gives the model with only initial learning of $\ta_\Sigma$ as developed in Choi, Larsen, and Seppi (2021).  To be consistent with our negative $B(t)$ restriction, we consider an initial value $B(0) :=-0.2$. We consider two stock-price volatility parameters $\gamma \in \{\frac12,1\}$ and a zero price-impact parameter $\alpha:=0$ (i.e., the competitive equilibrium).

\subsubsection{Equilibrium holdings}
\begin{figure}[!h]
\begin{center}
\caption{Plots of coefficient loadings over time for holdings $\hat{\theta}_{k,t}$ using the orthogonal representations in \eqref{thetajortho} and \eqref{thetaiortho} in Appendix \ref{sec:formulas}. The exogenous model parameters are $ \sigma_{\ta}:=1, M:=5, \bar{M}:=10, \;\alpha :=0$, $B(0):=-0.2$, $\kappa(t):=1$ for $t\in[0,1]$, 
$(\gamma, \sigma_{w_0}) =(\frac12,\frac1{10}) (\text{\color{blue}blue}), (\frac12,1) (\text{\color{amber}amber}), (1,\frac1{10}) (\text{\color{ao(english)}green}),$ and $ (1,1)(\text{\color{red}red}).$}\ \\
\begin{footnotesize}
$\begin{array}{cc}
\includegraphics[width=6cm, height=4.5cm]{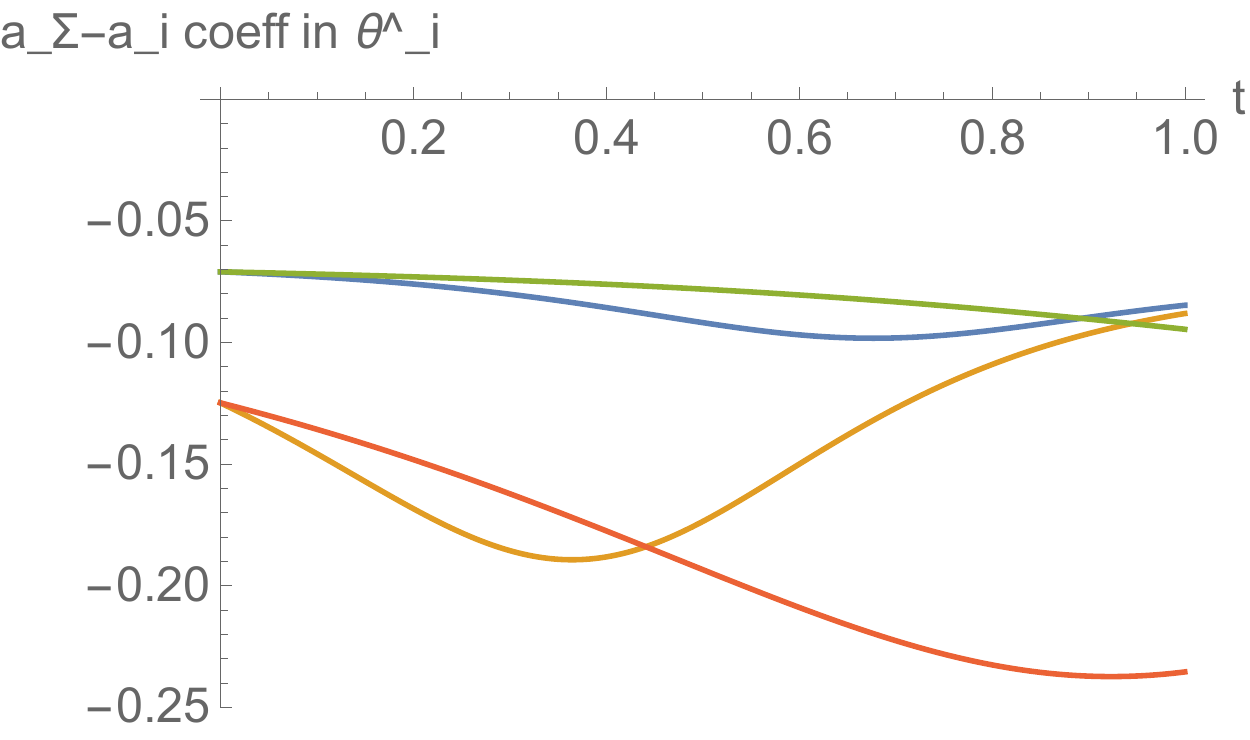} &\includegraphics[width=6cm, height=4.5cm]{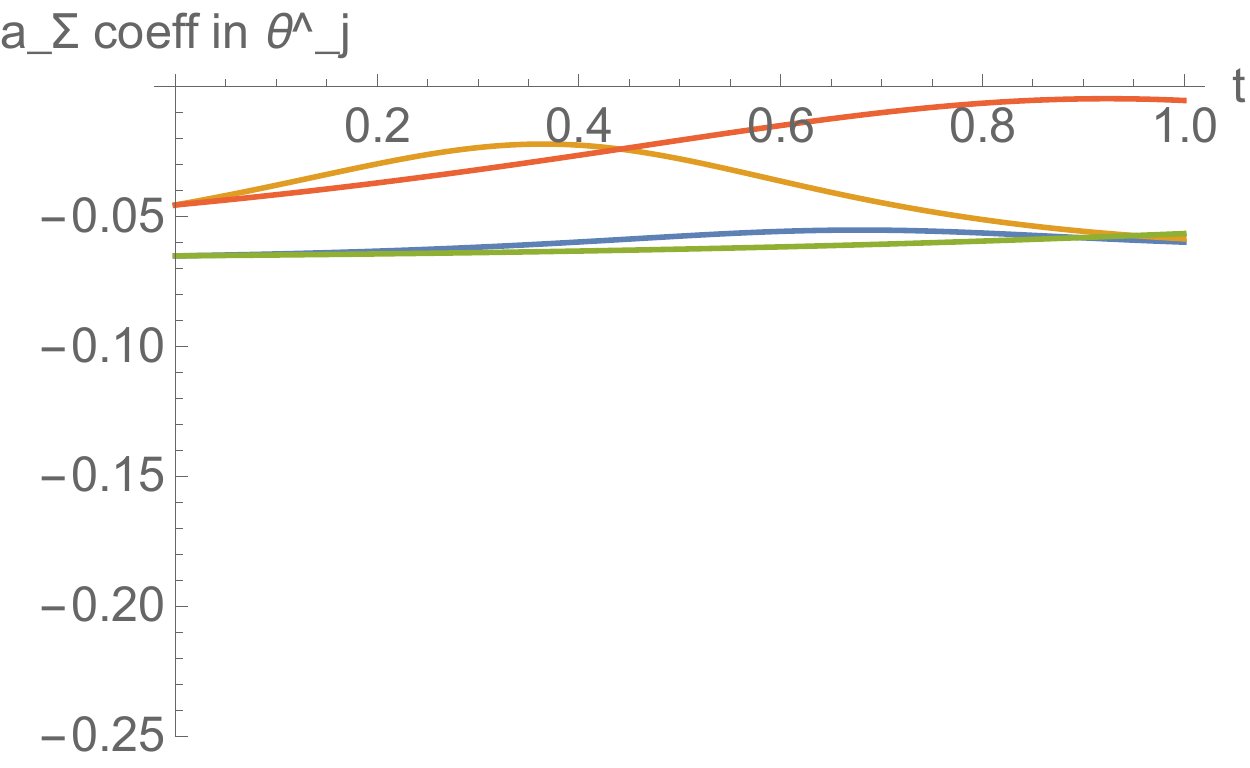} 
\\ 
\text{{\bf 1A:}  $\ta_\Sigma-\ta_i$ coefficient in } \hat{\theta}_{i,t}&\text{{\bf 1B:} $\ta_\Sigma$ coefficient in } \hat{\theta}_{j,t}\\
\\
\includegraphics[width=6cm, height=4.5cm]{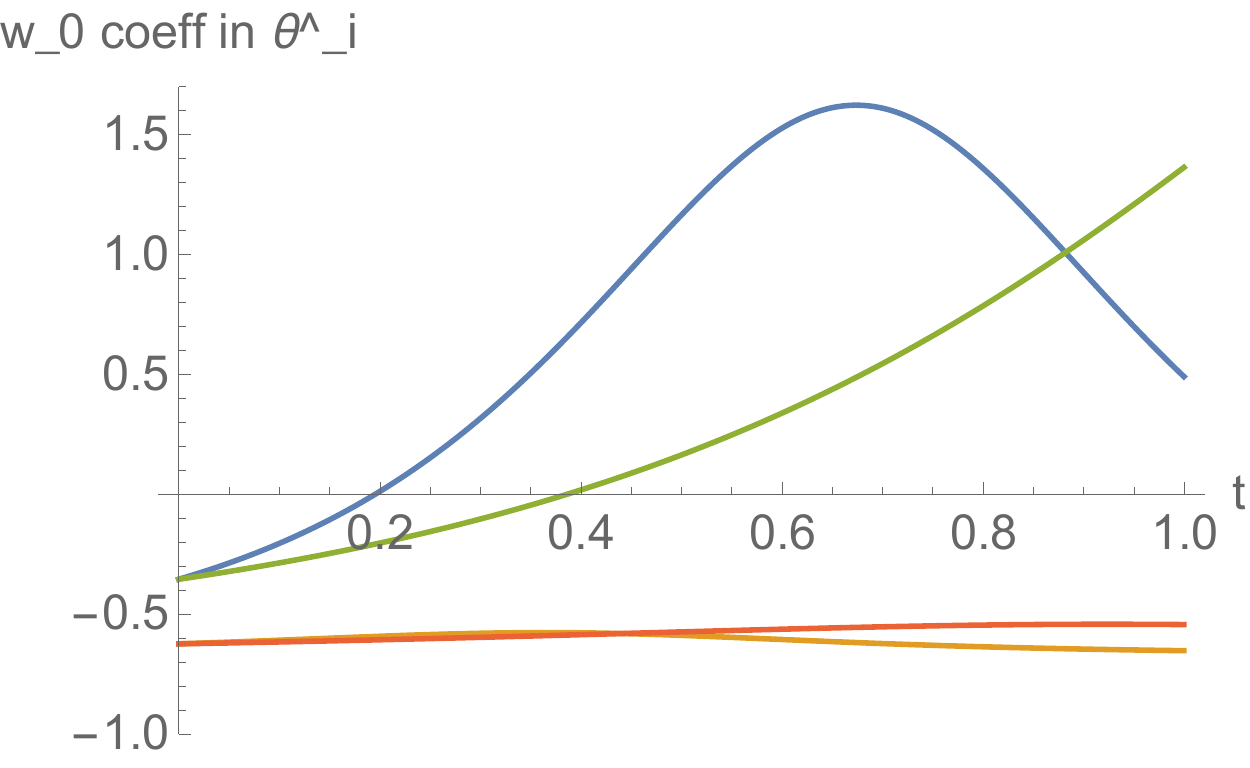} &\includegraphics[width=6cm, height=4.5cm]{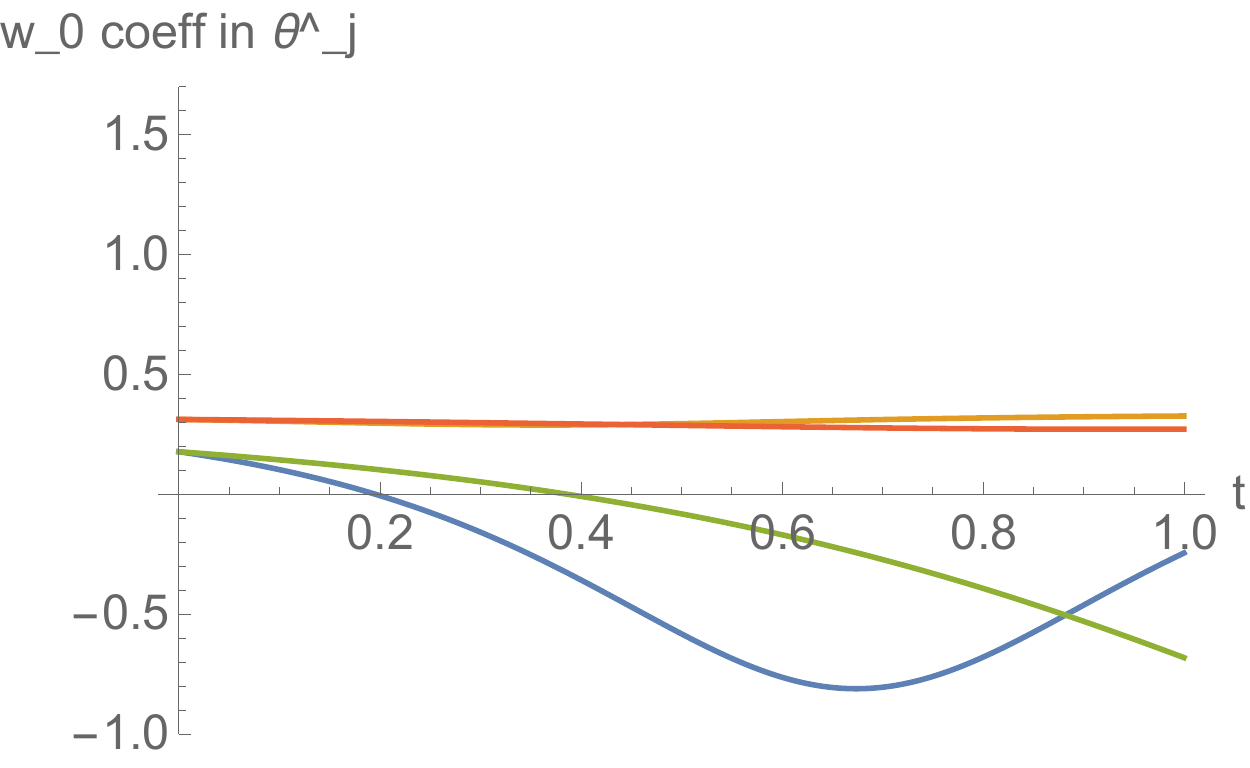} 
\\ 
\text{{\bf 1C:}  $w_0$ coefficient in } \hat{\theta}_{i,t}&\text{{\bf 1D:} $w_0$ coefficient in } \hat{\theta}_{j,t}\\
\\
\includegraphics[width=6cm, height=4.5cm]{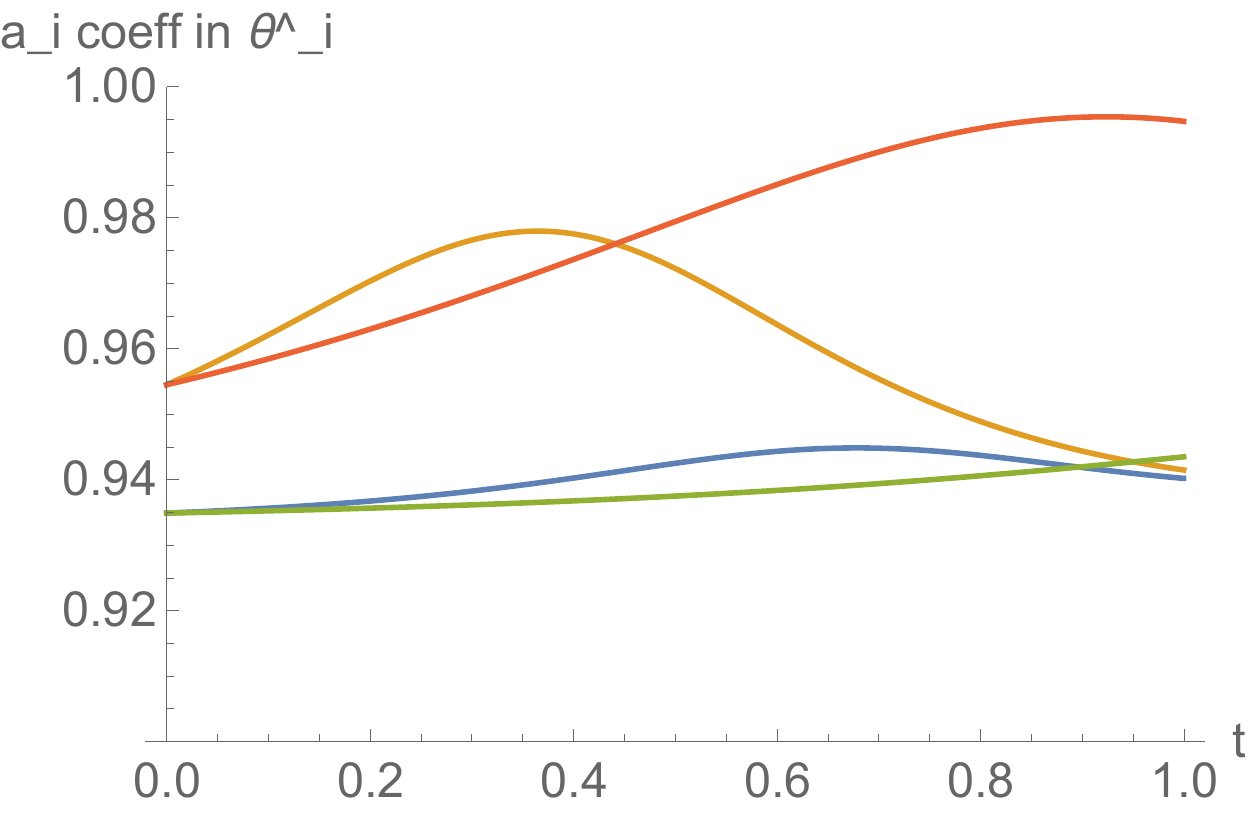} &\\ 
\text{{\bf 1E:}  $\ta_i$ coefficient in } \hat{\theta}_{i,t}&\\
\end{array}$
 \end{footnotesize}
\label{figholdings}
\end{center}
\end{figure}

First, we consider  equilibrium holdings.  Figure \ref{figholdings} shows the coefficient functions for the equilibrium stock holdings  $\hat{\theta}_{k,t}$ in \eqref{Y0000PI} for rebalancers and trackers using their orthogonal representations in  \eqref{thetaiortho} and \eqref{thetajortho} in Appendix \ref{sec:formulas}.

\newpage

Fig. \ref{figholdings}E shows rebalancer $i$'s loadings over time on her own parent target $a_i$. As expected, these loadings are close to 1, but they are less than 1 because trading towards a positive target depresses equilibrium price drifts in order for markets to clear.  The initial rebalancer loadings on $\ta_i$ of over 0.9 at time 0 indicate that rebalancers start the trading day with large block trades and then continue with more incremental trading. The negative coefficients on $\ta_\Sigma - \ta_i$ (for rebalancer $i$) and $\ta_\Sigma$ (for tracker $j$) in Fig. \ref{figholdings}A and \ref{figholdings}B  are demand accommodation.  In particular, rebalancers and trackers reduce their holdings when other rebalancers want to buy. The loadings on $w_0$ in Fig. \ref{figholdings}C and \ref{figholdings}D are more subtle.  When the initial tracker target $w_0$ has a high volatility (as in the red and amber trajectories), the tracker holdings load positively on $w_0$ over time in Fig.  \ref{figholdings}D and the negative rebalancer loadings in Fig. \ref{figholdings}C indicate demand accommodation by the rebalancers.  However, when the initial tracker target has low volatility (as in the green and blue trajectories), the initial positive tracker loadings on $w_0$ eventually flip signs as do the initial negative rebalancer loadings. At first glance, this is puzzling. The explanation is that, as noted above, the trackers and rebalancers have different stock-price drift perceptions in \eqref{S_PI} and \eqref{rebdriftPI} given their different filtrations. In particular, there is dynamic learning over time by the rebalancers based on the information $Y_t$ inferred from prices, whereas the trackers are fully informed about $\ta_\Sigma$ and $w_t$ (trackers infer $\ta_\Sigma$ at time 0). In these two low $\sigma_{w_0}$ parameterizations, the drift perceptions are quite different and illustrate how dynamic learning can have a significant impact on market dynamics. 

In addition to the effects illustrated in Fig. \ref{figholdings}, investor holdings are also affected by the realized path of $w_t = w_0+w^\circ_t$ over time.  This is because of fluctuations in the underlying tracker parent demand and also due to the effect of $w^\circ_t$ on dynamic learning by the rebalancers.  Appendix \ref{sec:formulas} shows the exact specification of this term in the tracker holdings (given as a $dw^\circ_u$ integral of a deterministic function).  Given the linearity of investor holdings and since the Brownian motion $w^\circ_t$ has zero expected increments, this random path effect disappears in ex ante expected investor holdings.

To summarize, Fig. \ref{figholdings} shows there are three main drivers of investor holdings: First, investors' holdings in most cases are drawn partially towards their own targets $\ta_i$ and $w_t$.  Second, investors provide partial accommodation to other investors' parent demands. Third, dynamic learning by the rebalancers affects their demand accommodation.  Interestingly, there is no evidence in Fig. \ref{figholdings} of predatory trading.  In particular, predatory trading differs from demand accommodation in that a predator’s holdings first load positively on another investor’s parent demand (driving up prices), and then the predator loading decreases.  In this context, the hump-shape of the blue trajectory (for low $\sigma_{w_0})$ is not predatory trading.  Indeed, the trackers eventually trade against their own initial parent target. As we shall see, the blue trajectory is explained below by price perceptions and dynamic learning rather than by predatory trading.

\begin{figure}[!h]
\begin{center}
\caption{Plots of unconditional autocorrelation  \eqref{autocorrtheta} of trading over time. The exogenous model parameters are $ \sigma_{\ta}:=1, M:=5, \bar{M}:=10, \;\alpha :=0$, $B(0):=-0.2$,  $\kappa(t):=1$ for $t\in[0,1]$, and 
$(\gamma, \sigma_{w_0}) =(\frac12,\frac1{10}) (\text{\color{blue}blue}),\,(\frac12,1) (\text{\color{amber}amber}), \,(1,\frac1{10}) (\text{\color{ao(english)}green})$, and $(1,1)(\text{\color{red}red}).$}\ \\
\begin{footnotesize}
$\begin{array}{cc}
\includegraphics[width=6cm, height=4.5cm]{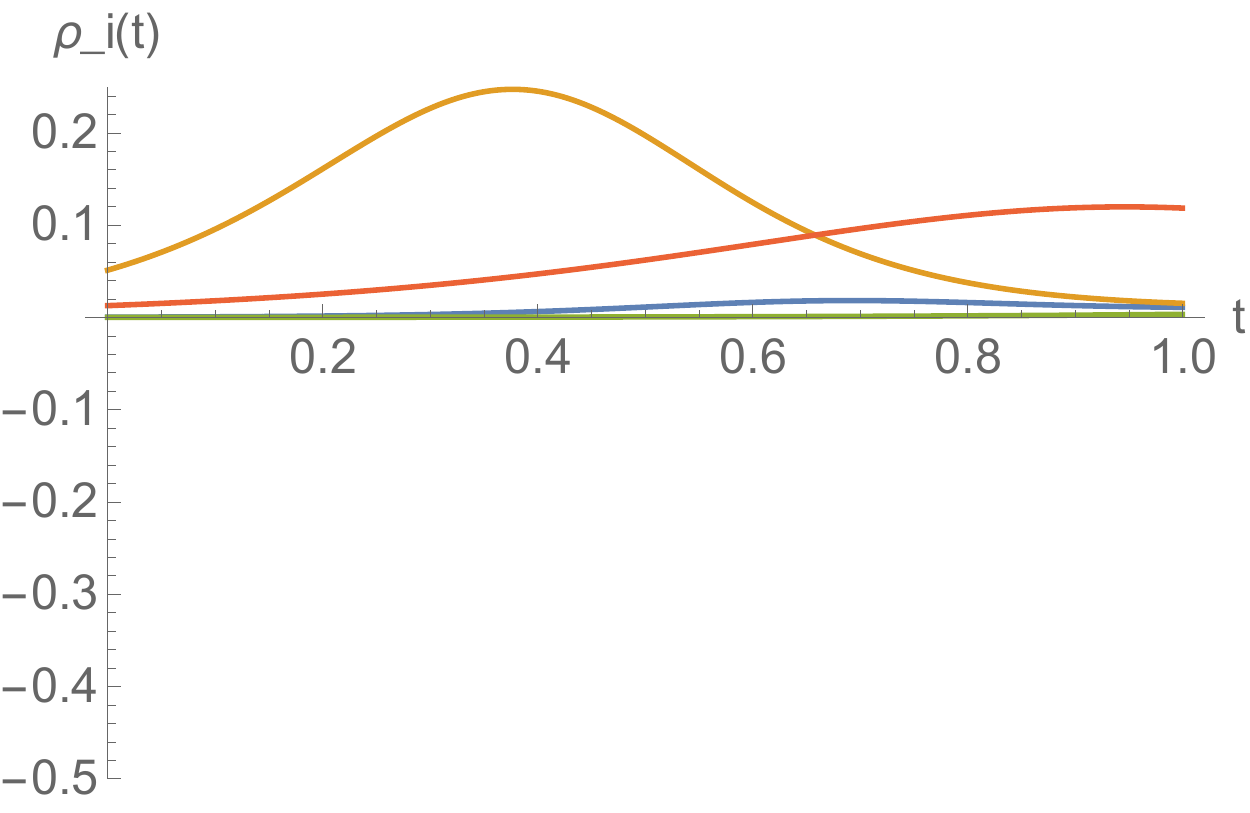} &\includegraphics[width=6cm, height=4.5cm]{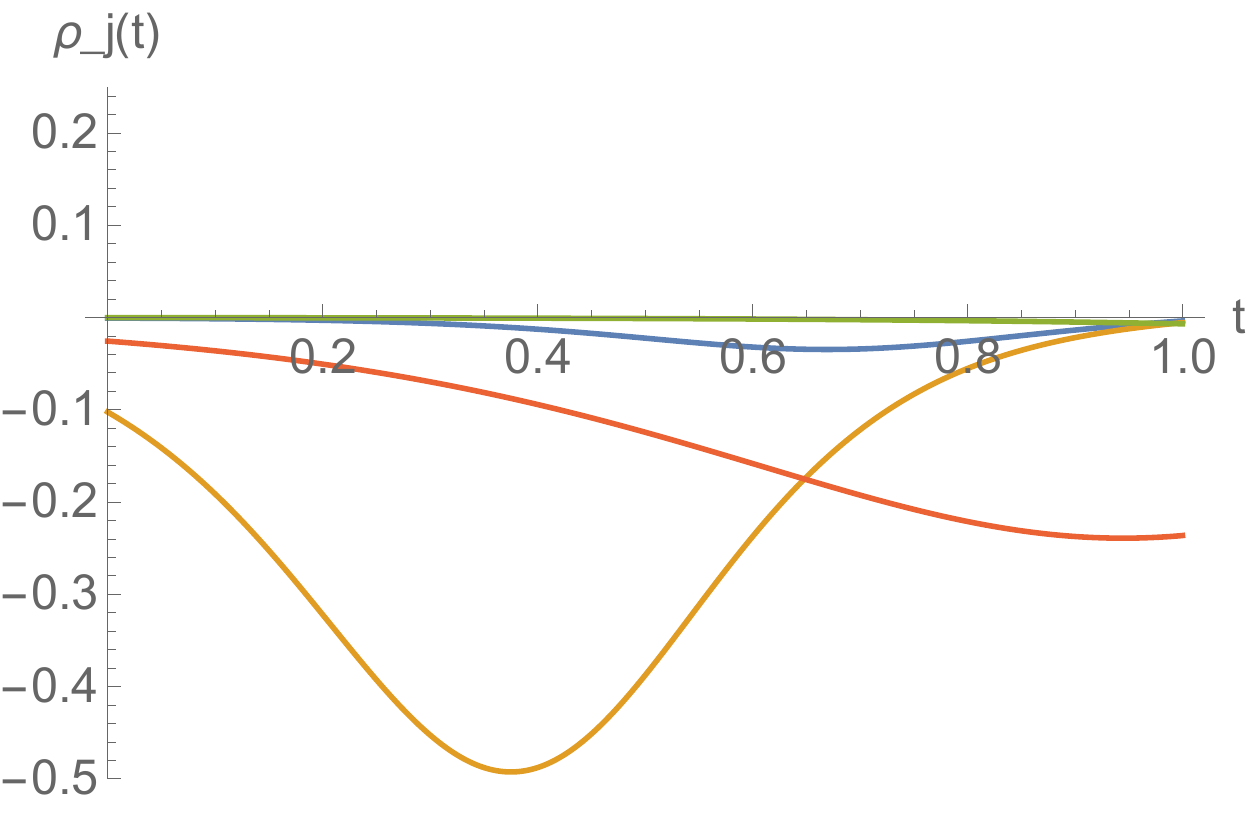} 
\\ 
\text{{\bf 2A:}  rebalancers' autocorrelation }\rho_i(t)&\text{{\bf 2B:}  trackers' autocorrelation }\rho_j(t)\\
 \end{array}$
 \end{footnotesize}
\label{figthetacorr}
\end{center}
\end{figure}

Fig. \ref{figthetacorr} plots  the  instantaneous intraday unconditional trading autocorrelations 
 \begin{align}\label{autocorrtheta}
\rho_k(t) := \lim_{h\downarrow 0}\frac{ \text{corr} (\hat{\theta}_{k,t+h}-\hat{\theta}_{k,t}, \hat{\theta}_{k,t+2h}-\hat{\theta}_{k,t+h})}{h},\quad k\in\{1,...,M+\bar{M}\},
 \end{align}
for the price-impact equilibrium holding processes for both the rebalancer and tracker in  \eqref{Y0000PI}. These autocorrelations are scaled by the time step $h>0$ (the unscaled versions converge to zero as $h\downarrow 0$).

\noindent Thus, consistent with empirical evidence, trading is autocorrelated due to order splitting.  Fig. \ref{figthetacorr} shows that rebalancers' orders are positively autocorrelated (2A) whereas trackers' orders exhibit negative autorcorrelation (2B).

Market clearing forces the intraday instantaneous unconditional cross correlation between rebalancers' and trackers'  holdings to be negatively perfectly  correlated 
 \begin{align}\label{crosscorr}
\lim_{h\downarrow 0}  \text{corr} (\hat{\theta}_{i,t+h}-\hat{\theta}_{i,t}, \hat{\theta}_{j,t+h}-\hat{\theta}_{j,t})=-1,
 \end{align}
for all $i\in\{1,...,M\}$ and $j\in \{M+1,...,M+\bar M\}$. 

\subsubsection{Equilibrium prices}

Next, we consider the price-impact equilibrium stock-price dynamics in  \eqref{S_PI}  and \eqref{rebdriftPI}.  
For the rebalancers, we  can rewrite the perceived drift in \eqref{rebdriftPI} in terms of the independent random variables $(\ta_\Sigma-\ta_i, w_0, \ta_i)$ and an  residual orthogonal term given as a stochastic integral with respect to  $w^\circ_t$ of a deterministic function of time. For the trackers, we can rewrite the drift in \eqref{S_PI} in terms of the independent random variables $(\ta_\Sigma, w_0)$ and an  residual orthogonal term given as a stochastic integral with respect to  $w^\circ_t$ of a deterministic function of time. These formulas are given in \eqref{S_PI221} and \eqref{S_PI223} in Appendix \ref{sec:formulas} and are illustrated in Fig. \ref{figstock}.

\newpage
\begin{figure}[!h]
\begin{center}
\caption{Plots of coefficient loadings  over time in stock-price drifts in  \eqref{S_PI221}  (rebalancer $i$) and \eqref{S_PI223} (tracker $j$). The exogenous model parameters are $ \sigma_{\ta}:=1, M:=5, \bar{M}:=10, \;\alpha :=0$, $B(0):=-0.2$, $\kappa(t):=1$ for $t\in[0,1]$, and 
$(\gamma, \sigma_{w_0}) =(\frac12,\frac1{10}) (\text{\color{blue}blue}), (\frac12,1) (\text{\color{amber}amber}), (1,\frac1{10}) (\text{\color{ao(english)}green}),$ and $ (1,1)(\text{\color{red}red}).$}\ \\
\begin{footnotesize}
$\begin{array}{cc}
\includegraphics[width=6cm, height=4.5cm]{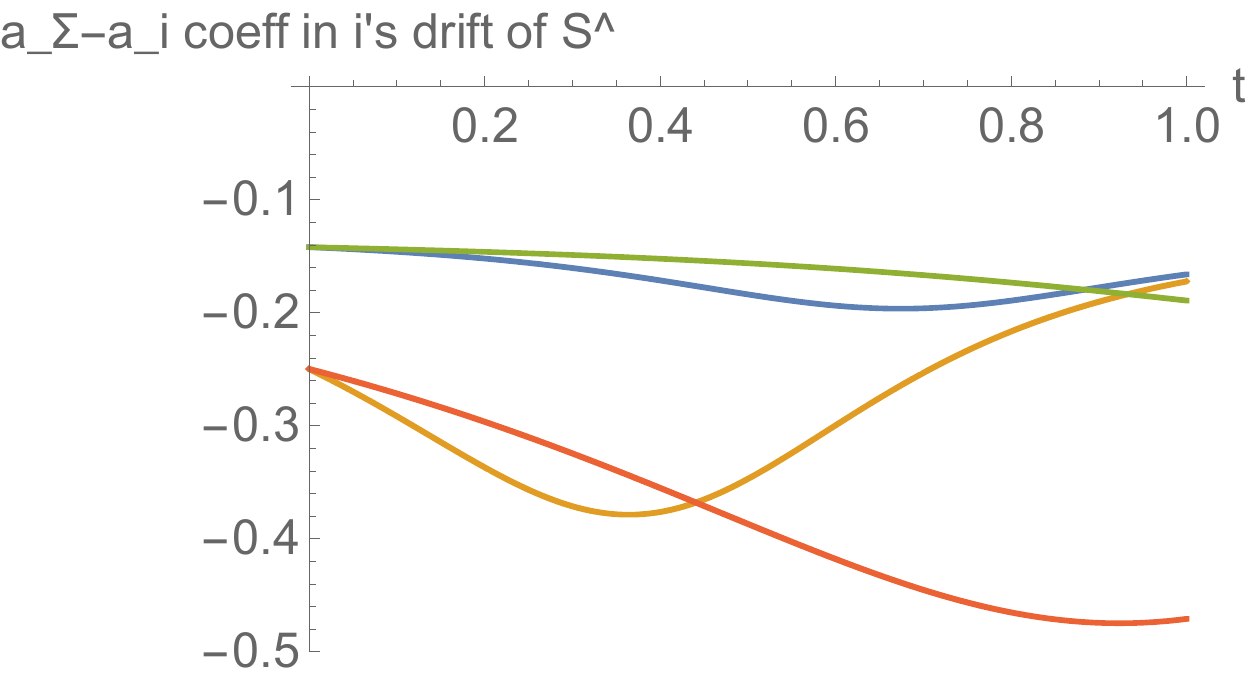} &\includegraphics[width=6cm, height=4.5cm]{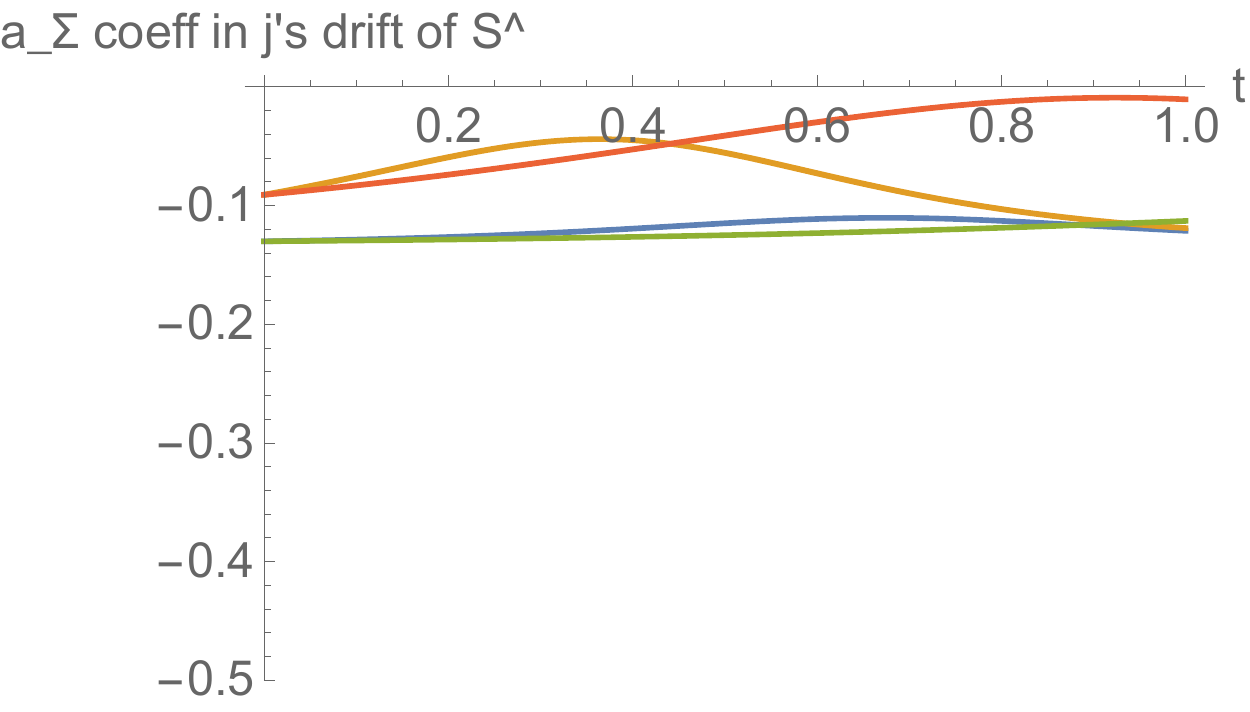} 
\\ 
\text{{\bf 3A:}  $\ta_\Sigma-\ta_i$ coefficient in rebalancers' stock drift}&\text{{\bf 3B:} $\ta_\Sigma$ coefficient  in trackers' stock drift}\\
\\
\includegraphics[width=6cm, height=4.5cm]{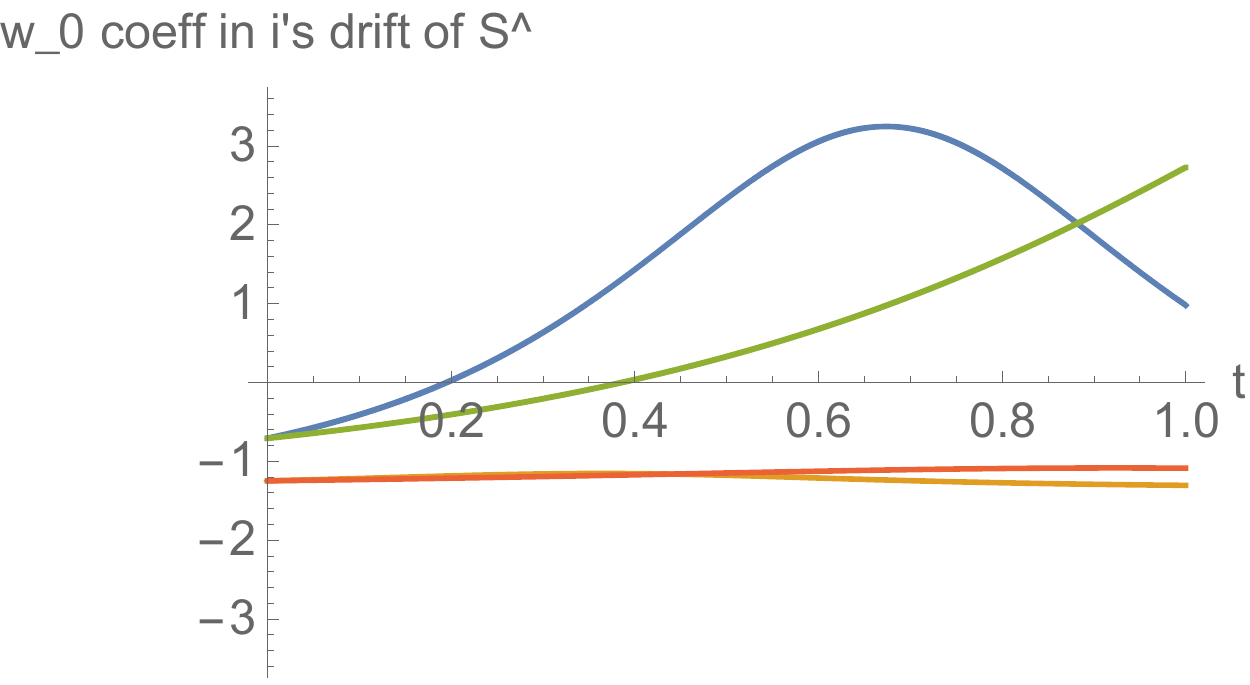} &\includegraphics[width=6cm, height=4.5cm]{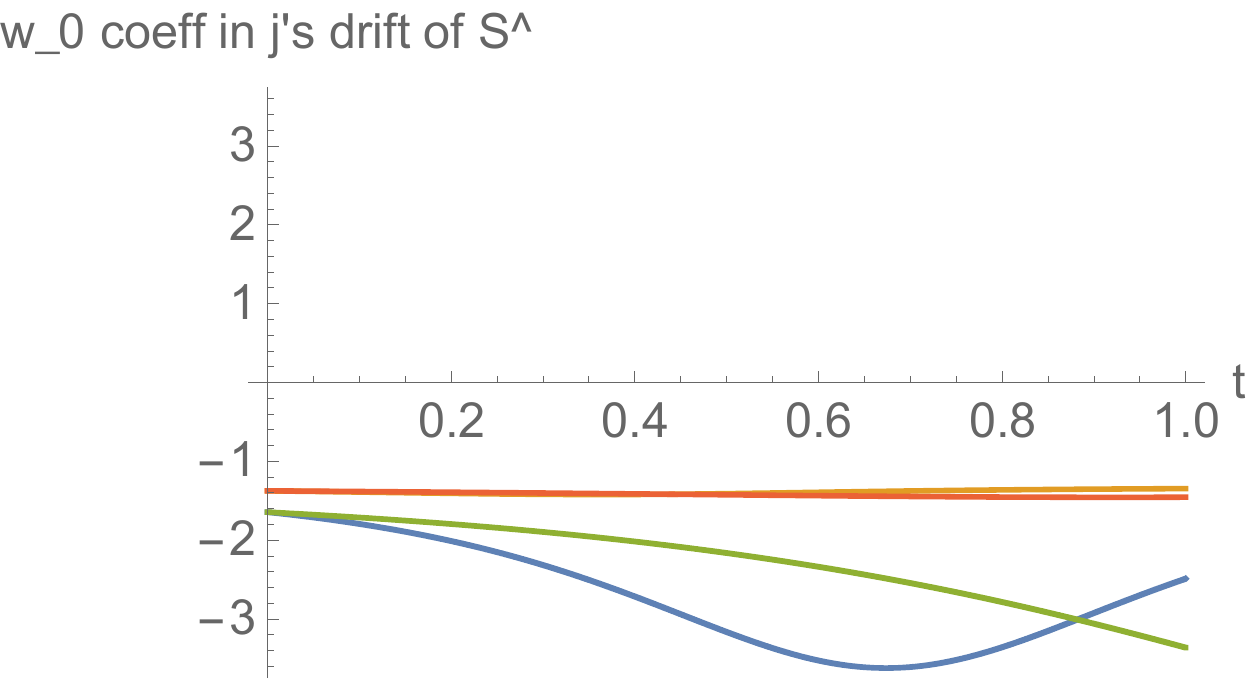} 
\\ 
\text{{\bf 3C:}  $w_0$ coefficient in rebalancers' stock drift}&\text{{\bf 3D:} $w_0$  coefficient in trackers' stock drift}\\
\\
\includegraphics[width=6cm, height=4.5cm]{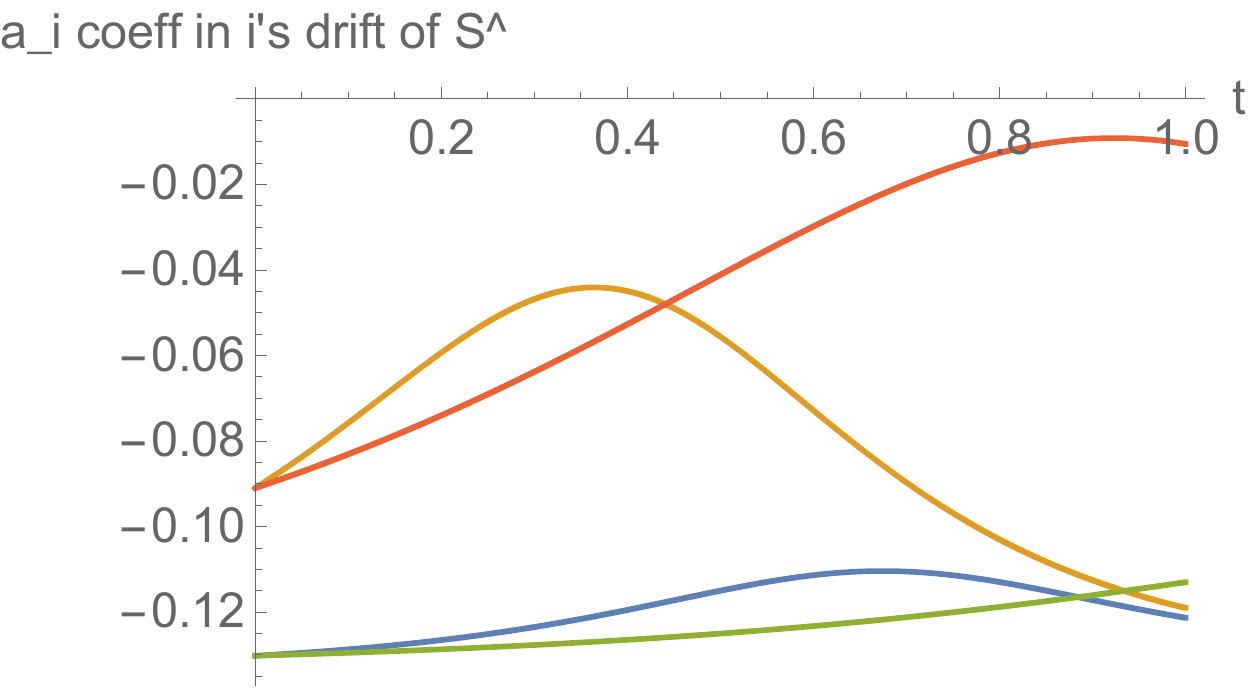} &\\ 
\text{{\bf 3E:}  $\ta_i$ coefficient in rebalancers' stock drift}&\\
\end{array}$
 \end{footnotesize}
 \label{figstock}
\end{center}
\end{figure}

Fig. \ref{figstock} shows that positive parent demands $\ta_i$, $\ta_\Sigma - \ta_i$, and $\ta_\Sigma$ all depress perceived stock-price drifts. The same is true for the tracker perceived stock-price drift loading on the initial tracker parent demand $w_0$. However, the relation between the rebalancer perceived drift and $w_0$ is more nuanced.  When the initial tracker demand volatility $\sigma_{w_0}$ is high, then rebalancers perceive that $w_0$ depresses the price drift.  However, when $\sigma_{w_0}$ is low, then the dynamic learning process --- given the inability of rebalancers to observe $w_0$ directly --- causes the rebalancer perceived stock-price drift loading on $w_0$ to change sign. For low values of $\sigma^2_{w_0}$, the trackers optimally use their superior knowledge of $w_0$ to manipulate stock-price perceptions to create gains from trade that outweigh their penalties. More specifically, the blues line in  Fig. \ref{figholdings}C and \ref{figholdings}D show that rebalancers have large positive stock holdings and trackers have large negative holdings based on a positive realization $w_0>0$. Such large negative holdings imply that trackers incur large penalties because they deviate from the target trajectory $w_t= w_0 + w^\circ_t$. Trackers find this behavior optimal because their blue line in Fig. \ref{figstock}D is negative (giving trackers large gains from trade) and rebalancers are willing to hold these large positive stock positions because their blue line in Fig. \ref{figstock}C is positive (giving also rebalancers large gains from trade).

\begin{figure}[!h]
\begin{center}
\caption{Plots of stock-price autocorrelation \eqref{autocorr} and variance of  trackers' equilibrium stock-price drift over time for the equilibrium stock-price  dynamics $d\hat{S}_t$ in \eqref{S_PI}. The exogenous model parameters are $ \sigma_{\ta}:=1, M:=5, \bar{M}:=10, \;\alpha :=0$, $B(0):=-0.2$,  $\kappa(t):=1$ for $t\in[0,1]$, and 
$(\gamma, \sigma_{w_0}) =(\frac12,\frac1{10}) (\text{\color{blue}blue}),\,(\frac12,1) (\text{\color{amber}amber}), \,(1,\frac1{10}) (\text{\color{ao(english)}green})$, and $(1,1)(\text{\color{red}red}).$}\ \\
\begin{footnotesize}
$\begin{array}{cc}
\includegraphics[width=6cm, height=4.5cm]{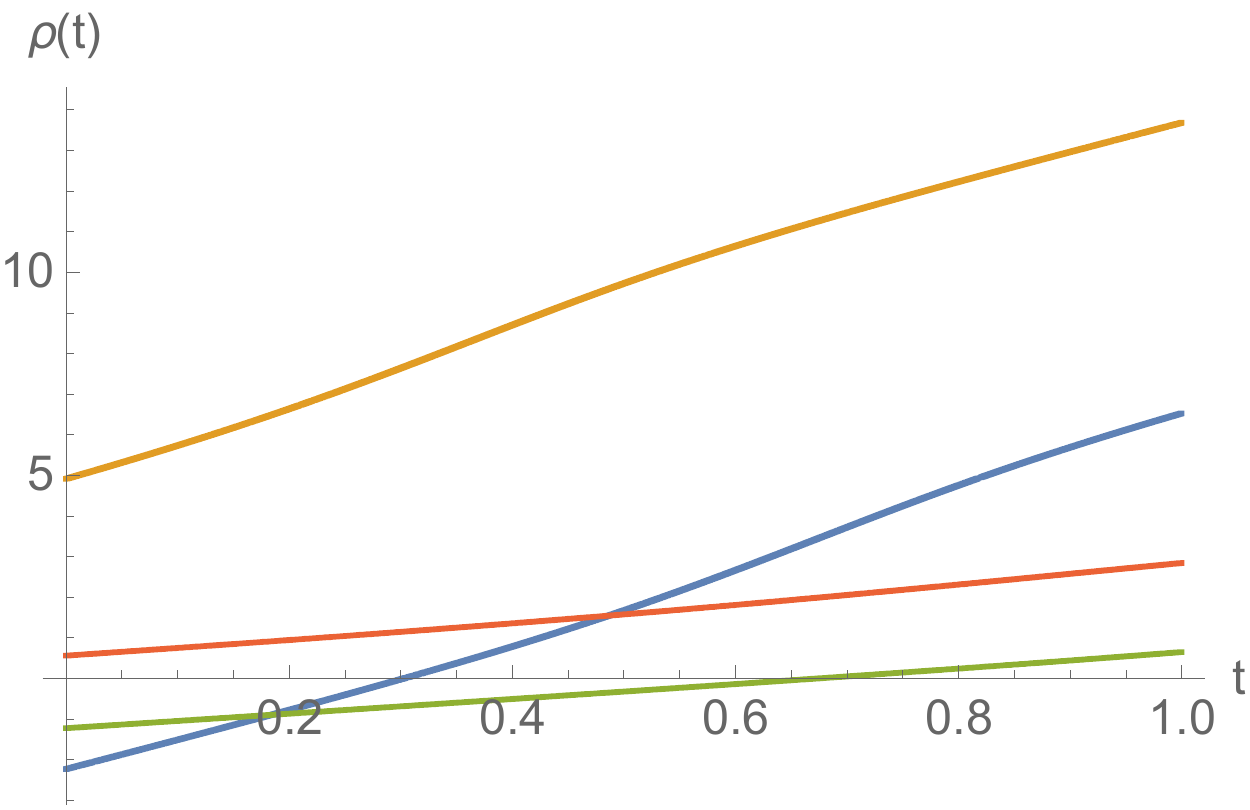} &\includegraphics[width=6cm, height=4.5cm]{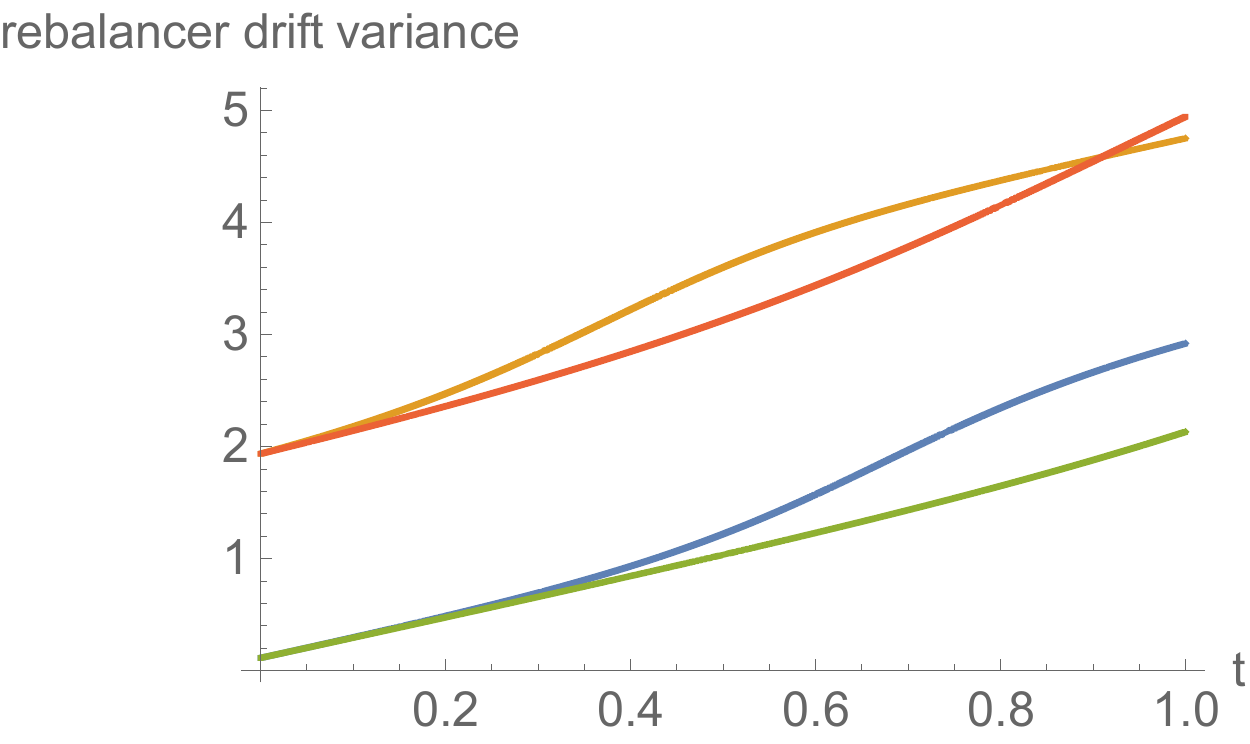} 
\\ 
\text{{\bf 4A: } stock-price autocorrelation } \rho(t)&\text{{\bf 4B: }variance of trackers' drift in \eqref{S_PI}}\\
 \end{array}$
 \end{footnotesize}
\label{figSscorr}
\end{center}
\end{figure}

Fig. \ref{figSscorr}A plots  the instantaneous intraday unconditional stock-price correlation, which is again scaled relative to $h$
\begin{align}\label{autocorr}
\rho(t) := \lim_{h\downarrow 0}\frac{ \text{corr} (\hat{S}_{t+h}-\hat{S}_{t}, \hat{S}_{t+2h}-\hat{S}_{t+h})}{h},\quad t\in[0,1),
\end{align}
for the equilibrium stock-price process $\hat{S}_t$. We see that price pressure from persistent parent demands lead to rising intraday price autocorrelation over the trading day.  Fig. \ref{figSscorr}B plots the time trajectory of the unconditional variance of intraday price drifts over the trading day  based on the trackers' equilibrium perceptions in  \eqref{S_PI}.  Predictable price drifts are important in actual markets as incentives for intraday liquidity provision by HFT market makers (represented in our model by rebalancers with realizations  $\ta_i = 0$.) We see that price-drift variability due to price pressure increases over the trading day.

\section{Subgame perfect Nash equilibrium}\label{sec:eq1}

This section builds on the analysis in Section \ref{sec:PI} by endogenizing stock-price perceptions and price impact. In particular, we partially endogenize the impact of hypothetical off-equilibrium investor holdings on market-clearing stock prices based on her perceptions of how other investors perceive prices and on other investors' resulting optimal response functions to her off-equilibrium holdings. More specifically, a subgame perfect Nash equilibrium involves describing how each trader $k_0$ (who might be a rebalancer $i_0$ or a tracker $j_0$ with their different filtrations) perceives market-clearing stock prices given $k_0$'s stock-price perceptions about other traders $k \neq k_0$ (where $k$ can be rebalancers $i$ or trackers $j$).

In our subgame perfect Nash model, a generic trader $k_0$ perceives that other rebalancers and trackers have stock-price perceptions of the form
\begin{align}\label{Sit3a}
\begin{split}
dS^Z_{i,t} &:=  \Big\{Z_t   +\mu_1(t)\tilde{a}_i+\mu_2(t)q_{i,t}+\mu_3(t)\eta_t + \alpha\theta_{i,t}\Big\}dt+ \gamma dW_{i,t}, \\
 S^Z_{i,0}&:=Z_0,\quad i\in\{1,...,M\},\\
dS^Z_{j,t} &:=  \Big\{Z_t  +\bar{\mu}_4(t)\ta_\Sigma+\bar{\mu}_5(t)w_t+ \alpha\theta_{j,t}\Big\}dt + \gamma dW_{j,t},\\
  S^Z_{j,0}&:=Z_0,\quad j\in\{M+1,...,M+\bar{M}\},
\end{split}
\end{align}
where $W_{k,t}$ is a Brownian motion for each trader $k\in\{1,...,M+\bar{M}\}$ and $Z_t$ is an arbitrary It\^o process.  The ``$Z$’’ superscript in \eqref{Sit3a} indicates that the perceived stock prices $S_{i,t}^Z$ and $S_{j,t}^Z$ are defined with respect to a particular It\^o process $Z_t$ (i.e., $Z_t$ is a sum of drift and volatility). We use the market-clearing condition \eqref{xSigmainitial} to construct  two such It\^o processes in \eqref{Y02i} and \eqref{Y0jj} below. These $Z_t$ processes differ from $Y_t$ in  \eqref{Sit} and \eqref{New2} in that we use $Z_t$ to capture the effect of arbitrary off-equilibrium stock holdings by trader $k_0$ on market-clearing prices given optimal responses by other investors $k$, $k\neq k_0$. We then go on to determine endogenously the deterministic functions $(\mu_1,\mu_2,\mu_3,\bar{\mu}_4, \bar{\mu}_5)$ in equilibrium in Theorem \ref{thm_Main} below.

The major difference between the price-impact equilibrium in Section \ref{sec:PI} and the following subgame perfect Nash equilibrium analysis lies in the traders' stock-price perceptions. In the price-impact equilibrium, the forms of the stock-price perceptions \eqref{Sit} and \eqref{New2} were conjectured with no additional justification beyond them leading to equilibrium existence in Theorem \ref{thm_PI}. In contrast, for a subgame perfect Nash equilibrium, investor stock-price perceptions must be such that:

\begin{itemize}
\item[(i)] Trader $k_0$'s own stock-price perceptions must be consistent with market-clearing for any off-equilibrium holdings $\theta_{k_0,t}$ used by $k_0$, when other traders' holding responses are optimal given the stock-price dynamics $k_0$ perceives other traders $k\neq k_0$ to have. This off-equilibrium market-clearing requirement can be found in, e.g., Vayanos (1999).

\item [(ii)] Trader $k_0$'s equilibrium holdings are found by solving her optimization problem using her own market-clearing stock-price dynamics from (i). 

\item[(iii)] All optimizers from (i) must be consistent with traders' equilibrium holdings in (ii).
\end{itemize}
\noindent  Definition \ref{Nash_eq} below makes properties (i)-(iii) operational. We refer to the last property (iii) as a consistency requirement between off and on-equilibrium holdings.

\subsection{Optimal off-equilibrium responses}\label{sec_offeq}

Lemma \ref{response}  gives trader $k$'s optimal response to an arbitrary It\^o process $Z_t$ and is the Nash equilibrium analogue of   Lemma \ref{PI_Le}.

\begin{lemma}[Optimal responses to $Z_t$] \label{response}Let $\mu_1,\mu_2,\mu_3,\bar{\mu}_4,\bar{\mu}_5:[0,1]\to \R$ and  $\kappa:[0,1]\to (0,\infty]$ be  continuous functions, let $\alpha\le0$, let $(Z_t)_{t\in[0,1]}$ be an It\^o process, and let the perceived stock-price process in the wealth dynamics \eqref{Xit} be as in \eqref{Sit3a}.  Then, $Z_t$ is adapted to both $\sF_{i,t}:=\sigma(\ta_i,Y_u,W_{i,u},S^Z_{i,u})_{u\in[0,t]}$ and $\sF_{j,t}:=\sigma(\ta_\Sigma,w_u,Y_u,W_{j,u},S^Z_{j,u})_{u\in[0,t]}$ and, provided 
\begin{align}\label{Y00fct}
\begin{split}
\theta^Z_{i,t} &:= \tfrac{1}{2 (\kappa (t)-\alpha ) }Z_t+\tfrac{2 \kappa (t)+\mu_1(t)}{2 (\kappa (t)-\alpha )}\ta_i+\tfrac{\mu_2(t)}{2 (\kappa (t)-\alpha )}q_{i,t}+\tfrac{\mu_3(t)}{2 (\kappa (t)-\alpha )}\eta_t,\\
\theta^Z_{j,t}&:= \tfrac{1}{2 (\kappa (t)-\alpha )}Z_t+\tfrac{2 \kappa(t)+\bar{\mu}_5(t)}{2 (\kappa (t)-\alpha )}w_t +\tfrac{\bar{\mu}_4(t)}{2 (\kappa (t)-\alpha )}\ta_\Sigma,
\end{split}
\end{align}
satisfy \eqref{squareint},  the traders' maximizers for \eqref{Rproblem} are $\theta^Z_{i,t} $ for rebalancer $i\in\{1,...,M\}$ and $\theta^Z_{j,t}$ for tracker $j\in \{M+1,...,M+\bar{M}\}$.
$\endproof$
\end{lemma}
\noindent Similar to Lemma \ref{PI_Le}, Lemma \ref{response} is proven using  pointwise quadratic maximization. Unlike $Y_t$ in Lemma \ref{PI_Le},  there is no Markov structure imposed on $Z_t$ in  Lemma \ref{response}, which makes dynamical programming inapplicable. Therefore,  the simplicity of the linear-quadratic objectives in  \eqref{Rproblem} is crucial for the proof of the optimality of $\theta^Z_{i,t}$ and $\theta^Z_{j,t}$ in \eqref{Y00fct}.

\subsection{Market-clearing stock-price perceptions}\label{sec_priceperceptions}

Investor $k_0$'s   perceptions about other investors' stock-price perceptions ensure that the stock market clears for any choice of $k_0$'s holdings. Thus, when solving for trader $k_0$'s individual equilibrium holdings, we require  $k_0$'s perceived stock-price process $S^\nu_{k_0,t}$ to clear the stock market for arbitrary hypothetical holdings $\theta_{k_0,t}$.   We assume that a given trader $k_0\in\{1,...,M+\bar{M}\}$ perceives that other  traders $k\neq k_0$ perceive the stock-price processes in \eqref{Sit3a}. Hence, trader $k_0$ perceives that other traders $k$, $k\neq k_0$, optimally hold $\theta^Z_{k,t}$ in \eqref{Y00fct} shares of stock. Given this, we then  find market-clearing $Z_{k_0,t}$ processes associated with arbitrary  hypothetical holdings $\theta_{k_0,t}$ for trader $k_0$.

First, consider a rebalancer $i_0\in\{1,...,M\}$. We construct a process $Z_{i_0,t}$ such that the stock market clears in the sense
\begin{align}\label{Y0}
\begin{split}
0&= \underbrace{\theta_{i_0,t}}_{\text{rebalancer $i_0$}}+\underbrace{ \sum_{i=1, i\neq i_0}^M \theta^{Z_{i_0}}_{i,t}}_{\text{other rebalancers}}+\underbrace{\sum_{j=M+1}^{\bar{M}}\theta^{Z_{i_0}}_{j,t}}_{\text{trackers}},\quad t\in[0,1],
\end{split}
\end{align}
where $\theta_{i_0,t}$ denotes an arbitrary stock-holdings process for rebalancer $i_0$ and other investors' responses $\theta^{Z_{i_0}}_{k,t}$ are from \eqref{Y00fct} for $Z_t := Z_{i_0,t}$. Clearly, any   solution $Z_{i_0,t}$  of  \eqref{Y0} is specific for rebalancer $i_0$. To describe one particular solution, we consider a specific continuously differentiable function $B:[0,1]\to\R$ satisfying
\begin{align}\label{B}
B(t)=-\frac{A(t) \mu_2(t)+\bar{M} \bar{\mu}_4(t)+2 \kappa (t)+\mu_1(t)}{2 \bar{M} \kappa (t)+\bar{M} \bar{\mu}_5(t)},
\end{align}
where $A(t)$ is as in \eqref{dY2E}. Because $A(t)$ in \eqref{dY2E} depends on $B(t)$, Eq. \eqref{B} is a fixed point requirement for $B(t)$. Below, we show that the coupled ODEs in \eqref{derivatives0a} characterize $(A,B)$ in \eqref{B}, and we give conditions ensuring that \eqref{derivatives0a} has a solution. Given a solution $B(t)$ to  \eqref{B}, we use $Y_t:=w_t - B(t)\ta_\Sigma$ from  \eqref{Z}  to express a solution of \eqref{Y0} as\footnote{The specific $B(t)$ function in \eqref{B} lets us combine $w_t$ and $\ta_\Sigma$ terms from \eqref{Y0} into the $Y_t$ term in \eqref{Y02i} using $Y_t = w_t - B(t)\ta_\Sigma$ from \eqref{Z}.}
\begin{align}\label{Y02i}
\begin{split}
Z_{i_0,t}&:=\tfrac{2 (\alpha -\kappa (t))}{M+\bar{M}-1}\theta_{i_0,t} +\tfrac{2\kappa(t)+\mu_1(t)}{M+\bar{M}-1}\ta_{i_0}+\tfrac{\mu_2(t)}{M+\bar{M}-1}q_{i_0,t}\\
&-\tfrac{(M-1) \mu_3(t)+\mu_2(t)}{M+\bar{M}-1}\eta_t 
-\tfrac{\bar{M} (2 \kappa (t)+\bar{\mu}_5(t))}{M+\bar{M}-1}Y_t,\quad t\in[0,1].
\end{split}
\end{align} 
The process $Z_{i_0,t}$ in \eqref{Y02i} captures the impact of arbitrary holdings $\theta_{i_0,t}$ by rebalancer $i_0$ on market-clearing stock prices given $i_0$'s perceptions of how other traders optimally respond using $\theta_{k,t}^{Z_{i_0}}$.

We then describe rebalancer $i_0$'s stock-price perceptions for $i_0\in\{1,...,M\}$. Rebalancer $i_0$ filters based on her  own target $\ta_i$ and on observations of past and current perceived market-clearing  stock prices $S^\nu_{i_0,u}$ defined by
\begin{align}\label{New22i}
\begin{split}
dS^\nu_{i_0,t} &:=  \Big\{\nu_0(t)Z_{i_0,t}   +\nu_1(t)\tilde{a}_{i_0} +\nu_2(t)q_{i_0,t}+\nu_3(t)\eta_t+ \alpha\theta_{i_0,t}\Big\}dt + \gamma dw_{i_0,t},\\
S^\nu_{i_0,0}&:=Y_0,\quad i_0\in\{1,...,M\},
\end{split}
\end{align}
where $(\ta_{i_0},\theta_{i_0,t})$ are known and $(Z_{i_0,t} ,q_{i_0,t},\eta_{t_0})$ are inferred by rebalancer $i_0$. The ``$\nu$’’ superscript in \eqref{New22i} indicates that the perceived stock prices are defined with respect to a particular set of deterministic functions $(\nu_0,\nu_1,\nu_2,\nu_3)$, which we endogenously determine in Theorem \ref{thm_Main} below.  More specifically, by observing $\ta_{i_0}$ and $(S^\nu_{i_0,u})_{u\in[0,t]}$ defined  in \eqref{New22i}, rebalancer $i_0$  infers  $Y_t:=w_t - B(t)\tilde{a}_\Sigma$ from \eqref{Z} using the Volterra argument behind Lemma \ref{lemma_infer}. To see this, we insert \eqref{Y02i} into \eqref{New22i} to produce rebalancer $i_0$'s perceived market-clearing stock-price dynamics
\begin{align}\label{New22ia}
\begin{split}
dS^\nu_{i_0,t} &=  \Big\{\Big(\tfrac{\nu_0(t) (2 \kappa (t)+\mu_1(t))}{M+\bar{M}-1}+\nu_1(t)\Big)\tilde{a}_{i_0} +\Big(\tfrac{\mu_2(t) \nu_0(t)}{M+\bar{M}-1}+\nu_2(t)\Big)q_{i_0,t}\\
&+\Big(\nu_3(t)-\tfrac{\nu_0(t) ((M-1) \text{$\mu $3}(t)+\mu_2(t))}{M+\bar{M}-1}\Big)\eta_t-\tfrac{\bar{M} \nu_0(t) (2 \kappa (t)+\bar{\mu}_5(t))}{M+\bar{M}-1}Y_t\\
&+\Big(\alpha -\tfrac{2 \nu_0(t) (\kappa (t)-\alpha )}{M+\bar{M}-1}\Big)\theta_{i_0,t}\Big\}dt + \gamma dw_{i_0,t}.
\end{split}
\end{align}
Because the expressions multiplying $(\ta_{i_0},q_{i_0,t},\eta_t,Y_t,\theta_{i_0,t})$ in  \eqref{New22ia} are continuous (deterministic) functions of time $t\in[0,1]$,  Lemma \ref{lemma_infer} applies and shows that by observing $\ta_{i_0}$ and $(S^\nu_{i_0,u})_{u\in[0,t]}$ in \eqref{New22ia} over time $t\in[0,1]$, rebalancer $i_0$ can infer $w_{i_0,t}$. Subsequently, rebalancer $i_0$ can use \eqref{RfiltrationQQQ} and \eqref{RfiltrationQ} to also infer $Y_t$ over time $t\in[0,1]$.

Next, consider a tracker $j_0\in\{M+1,...,M+\bar{M}\}$. For arbitrary off-equilibrium holdings $\theta_{j_0,t}$, the market-clearing solution $Z_{j_0,t}$ from 
\begin{align}\label{Y0jj}
\begin{split}
0&= \underbrace{ \theta_{j_0,t}}_{\text{tracker }j_0}+\underbrace{\sum_{j=M+1, j\neq j_0}^{\bar{M}}\theta^{Z_{j_0}}_{j,t}}_{\text{other trackers}} +\underbrace{\sum_{i=1}^M \theta^{Z_{j_0}}_{i,t}}_{\text{rebalancers}},\quad t\in[0,1],
\end{split}
\end{align}
is given by
\begin{align}\label{Y02}
\begin{split}
Z_{j_0,t}&:=\tfrac{2 (\alpha -\kappa (t))}{M+\bar{M}-1}\theta_{j_0,t}-\tfrac{M \mu_3(t)+\mu_2(t)}{M+\bar{M}-1}\eta_t -\tfrac{(\bar{M}-1) (2 \kappa (t)+\bar{\mu}_5(t))}{M+\bar{M}-1}w_t\\
&\;-\tfrac{A(t) \mu_2(t)+(\bar{M}-1) \bar{\mu}_4(t)+2 \kappa (t)+\mu_1(t)}{M+\bar{M}-1}\ta_\Sigma.
\end{split}
\end{align}
Once again, $Z_{j_0,t}$ captures tracker $j_0$'s perceptions of the impact of her holdings $\theta_{j_0,t}$ on market-clearing stock prices given $j_0$'s perceptions of other investors' responses $\theta_{k,t}^{Z_{j_0}}$ to $\theta_{j_0,t}$.

Tracker $j_0$'s  perceived market-clearing stock-price process is defined as
\begin{align}\label{New22}
\begin{split}
dS^{\bar\nu}_{j_0,t} &:=  \Big\{Z_{j_0,t} +\bar{\nu}_3(t)\eta_t+\bar{\nu}_4(t)\ta_\Sigma+\bar{\nu}_5(t)w_t+ \alpha\theta_{j_0,t}\Big\}dt + \gamma dw_t,\\
S^{\bar\nu}_{j_0,0}&:=Y_0,\quad j\in \{M+1,...,M+\bar{M}\},
\end{split}
\end{align}
where $\bar{\nu}_3,\bar{\nu}_4,\bar{\nu}_5:[0,1]\to \R$ are deterministic functions of time (endogenously determined Theorem \ref{thm_Main} below). Inserting \eqref{Y02} into \eqref{New22} gives tracker $j_0$'s perceived market-clearing stock-price dynamics
\begin{align}\label{New22a}
\begin{split}
dS^{\bar\nu}_{j_0,t} &=  \Big\{\Big(\bar{\nu}_3(t)-\tfrac{M \text{$\mu $3}(t)+\mu_2(t)}{M+\bar{M}-1}\Big)\eta_t\\
&+\Big(\bar{\nu}_5(t)-\tfrac{(\bar{M}-1) (2 \kappa (t)+\bar{\mu}_5(t))}{M+\bar{M}-1}\Big)w_t \\
&+\Big(\bar{\nu}_4(t)-\tfrac{A(t) \mu_2(t)+(\bar{M}-1) \bar{\mu}_4(t)+2 \kappa (t)+\mu_1(t)}{M+\bar{M}-1}\Big) \ta_\Sigma\\
&+\tfrac{\alpha  (M+\bar{M}+1)-2 \kappa (t)}{M+\bar{M}-1}\theta_{j_0,t}\Big\}dt + \gamma dw_t.
\end{split}
\end{align}
We note that tracker $j_0$'s perceived market-clearing stock-price dynamics $dS^{\bar\nu}_{j_0,t}$ in \eqref{New22a} are driven by the exogenous Brownian motion $w_t$ from \eqref{w_t} whereas rebalancer $i_0$'s stock prices $dS^\nu_{i_0,t}$ in \eqref{New22ia} are driven by $i_0$'s innovations process $dw_{i_0,t}$ from \eqref{dwit}. This is due to the different information  sets of rebalancers and trackers.
 
Unlike the price-impact equilibrium in Theorem \ref{thm_PI},  we see from \eqref{New22ia} and \eqref{New22a} that, even with no direct price impact in the sense $\alpha := 0$ in \eqref{New22i} and \eqref{New22}, the remaining net price impacts $-\frac{2 \nu_0(t) \kappa (t)}{M+\bar{M}-1}$ and $-\frac{2 \kappa (t)}{M+\bar{M}-1}$ of $\theta_{i,t}$ and $\theta_{j,t}$ are nonzero. This is because price pressure in \eqref{New22ia} and \eqref{New22a} clears the stock market for arbitrary holdings $\theta_{i,t}$ and $\theta_{j,t}$.

The next result gives the optimal holdings $\theta^*_{k,t}$ for all traders $k_0:=k\in\{1,...,M+\bar{M}\}$ given their perceptions of market-clearing stock prices in  \eqref{New22ia} and \eqref{New22a}. While both  $\theta^*_{k,t}$ and the optimal response holdings $\theta^{Z}_{k,t}$ in \eqref{Y00fct} maximize \eqref{Rproblem}, they differ because they are based on different perceived stock-price processes.  On one hand, the optimal responses $\theta^{Z}_{k,t}$ in \eqref{Y00fct} are based on the stock-price perceptions  in \eqref{Sit3a}. On the other hand, the optimizer $\theta^*_{k,t}$ is based on the market-clearing stock-price perceptions in \eqref{New22ia} and \eqref{New22a}.

\begin{lemma}[Trader $k$'s maximizer for market-clearing stock-price perceptions] \label{lemma_eqholdings}Let $\nu_0,\nu_1$, $\nu_2,\nu_3,\bar{\nu}_3,\bar{\nu}_4,\bar{\nu}_5:[0,1]\to \R$ and  $\kappa:[0,1]\to (0,\infty]$ be  continuous functions  with $\nu_0>0$ and  assume  $\alpha \le0$. Let the perceived market-clearing stock-price processes in the wealth dynamics \eqref{Xit} be given by \eqref{New22ia} and \eqref{New22a} with corresponding filtrations $\sF_{i,t}:= \sigma(\ta_i,S^\nu_{i,u})_{u\in[0,t]}$ and $\sF_{j,t}:= \sigma(w_u,S^{\bar\nu}_{j,u})_{u\in[0,t]}$ for  $ i\in \{1,...,M\}$ and $j\in \{M+1,...,M+\bar{M}\}$. Then, provided the holding processes
\begin{align}\label{Y000}
\begin{split}
\theta_{i,t}^* :&=\tfrac{2 \kappa (t) (M+\bar{M}+\nu_0(t)-1)+(M+\bar{M}-1) \nu_1(t)+\mu_1(t) \nu_0(t)}{2 (\kappa (t)-\alpha ) (M+\bar{M}+2 \nu_0(t)-1)}\ta_i\\
&+\tfrac{(M+\bar{M}-1) \nu_2(t)+\mu_2(t) \nu_0(t)}{2 (\kappa (t)-\alpha ) (M+\bar{M}+2 \nu_0(t)-1)}q_{i,t}\\
&- \tfrac{\nu_0(t) ((M-1) \mu_3(t)+\mu_2(t))-(M+\bar{M}-1) \nu_3(t)}{2 (\kappa (t)-\alpha ) (M+\bar{M}+2 \nu_0(t)-1)}\eta_t\\
&-\tfrac{\bar{M} \nu_0(t) (2 \kappa (t)+\bar{\mu}_5(t))}{2 (\kappa (t)-\alpha ) (M+\bar{M}+2 \nu_0(t)-1)}Y_t,\\
\theta_{j,t}^* :&= \tfrac{(M+\bar{M}-1) \bar{\nu}_3(t)-M \mu_3(t)-\mu_2(t)}{2 (M+\bar{M}+1) (\kappa (t)-\alpha)}\eta_t \\
&+\tfrac{(M+\bar{M}-1) \bar{\nu}_5(t)+2 M \kappa (t)-(\bar{M}-1) \bar{\mu}_5(t)}{2 (M+\bar{M}+1) (\kappa (t)-\alpha )}w_t\\
&-\tfrac{A(t) \mu_2(t)-(M+\bar{M}-1) \bar{\nu}_4(t)+(\bar{M}-1) \bar{\mu}_4(t)+2 \kappa (t)+\mu_1(t)}{2 (M+\bar{M}+1) (\kappa (t)-\alpha )}\ta_\Sigma,
\end{split}
\end{align}
satisfy \eqref{squareint},  the traders' maximizers for \eqref{Rproblem} are $\theta_{i,t}^*$ for rebalancer $i\in\{1,...,M\}$ and $\theta_{j,t}^*$ for tracker $j\in \{M+1,...,M+\bar{M}\}$.
$\endproof$
\end{lemma}

 From Lemma \ref{lemma_eqholdings}, we note that a generic rebalancer $i_0$ has filtration   $\sigma(\ta_{i_0},S^\nu_{i_0,u})_{u\in[0,t]}$ whereas she perceives that other rebalancers $i\neq i_0$ have filtrations $\sigma(\ta_i,Y_u,W_{i,u},S^Z_{i,u})_{u\in[0,t]}$ as in Lemma \ref{response}. Because these are $i_0$'s off-equilibrium perceptions, this is allowable as long as they are consistent with $i$'s equilibrium holdings. We require this consistency in Definition \ref{Nash_eq}(iii) below. We also note from Lemma \ref{response} that rebalancer $i$ can infer $Z_{i_0,t}$ in  \eqref{Y02i}. In turn, this allows rebalancer $i$, $i\neq i_0$, to also know the process
\begin{align}\label{extrainfer}
\tfrac{2 (\alpha -\kappa (t))}{M+\bar{M}-1}\theta_{i_0,t} +\tfrac{2\kappa(t)+\mu_1(t)}{M+\bar{M}-1}\ta_{i_0}+\tfrac{\mu_2(t)}{M+\bar{M}-1}q_{i_0,t}.
\end{align}
However, knowing \eqref{extrainfer} is insufficient for rebalancer $i$, $i\neq i_0$, to infer rebalancer $i_0$'s private target $\ta_{i_0}$.

\subsection{Equilibrium}\label{sec:eq}
\begin{definition}\label{Nash_eq} Deterministic functions of time $\mu_1,\mu_2,\mu_3,\bar{\mu}_4,\bar{\mu}_5,\nu_0,\nu_1,\nu_2,\nu_3,\bar{\nu}_4,\bar{\nu}_5:[0,1]\to\R$ constitute a \emph{subgame perfect Nash financial-market equilibrium} if:
\begin{enumerate}
\item[(i)] For $k \in \{1,...,M+\bar{M}\}$, trader $k$'s  maximizer $\theta^*_{k,t}$ for \eqref{Rproblem} exists given the market-clearing stock-price perceptions \eqref{New22ia} and \eqref{New22a}.

\item[(ii)] For  $k\in\{1,...,M+\bar{M}\}$, inserting trader $k$'s maximizer $\theta^*_{k,t}$ into the perceived  market-clearing  stock-price processes \eqref{New22ia} and \eqref{New22a} produces identical stock-price processes across all traders.  This common equilibrium stock-price process is denoted by $S^*_t$.

\item[(iii)] Optimizers and equilibrium holdings must be consistent in the sense that trader $k$'s perceived response to trader $k_0$'s maximizer $\theta^*_{k_0,t}$ is trader $k$'s maximizer $\theta^*_{k,t}$.

\item[(iv)] The money and stock markets clear.
$\endproof$
\end{enumerate}
\end{definition}

The identical stock-price requirement in Definition \ref{Nash_eq}(ii) is similar to the one in Definition \ref{PI_eq}(ii). We see from the rebalancers' perceptions  \eqref{New22i} that both the drifts and the martingale terms have $i$ dependence. Similar to \eqref{Y32PI}, we replace $dw_{i,t}$ in $dS^\nu_{i,t}$ in \eqref{New22i} with the decomposition of $dw_{i,t}$ in terms of $dw_t$ in \eqref{dwit} and rewrite $dS^\nu_{i,t}$ in \eqref{New22i}  as 
\begin{align}\label{Y32}
\begin{split}
dS^\nu_{i,t} &=  \Big\{\nu_0(t)Z_{i,t}  +\nu_1(t)\tilde{a}_i +\nu_2(t)q_{i,t}+\nu_3(t) \eta_t +\alpha \theta_{i,t}\\
&\quad -B'(t)\big(\ta_\Sigma-\ta_i - q_{i,t} \big) \gamma \Big\}dt+ \gamma dw_t,\quad i\in\{1,...,M\}.
\end{split}
\end{align}
Therefore, to ensure identical equilibrium stock-price perceptions for all traders $k\in\{1,...,M+\bar{M}\}$,   it suffices to match the drift of $dS^{\bar\nu}_{j,t}$ in \eqref{New22} for $j\in\{M+1,...,M+\bar{M}\}$ with the drift of $dS^\nu_{i,t}$ in \eqref{Y32} for  the optimal holdings $\theta_{i,t}:= \theta^*_{i,t}$ for $i\in\{1,...,M\}$ . This produces the requirement 
\begin{align}\label{driftA}
\begin{split}
&\nu_0(t)Z^*_{i,t}  +\nu_1(t)\tilde{a}_i +\nu_2(t)q_{i,t}+\nu_3(t) \eta_t +\alpha \theta^*_{i,t} -B'(t)\big(\ta_\Sigma-\ta_i - q_{i,t} \big)\gamma\\
&=\bar{\nu}_3(t)\eta_t+\bar{\nu}_4(t)\ta_\Sigma+\bar{\nu}_5(t)w_t+ \alpha\theta^*_{j,t},
\end{split}
\end{align}
for all rebalancers $i \in\{1,...,M\}$ and all trackers $j\in \{M+1,...,M+\bar{M}\}$. The right-hand side of \eqref{driftA} does not depend on the rebalancer index $i$. In \eqref{driftA}, the process $Z_{i,t}^*$ is \eqref{Y02i} evaluated at  $\theta_{i,t}:= \theta^*_{i,t}$,  and $Z_{j,t}^*$ is  \eqref{Y02} evaluated at  $\theta_{j,t}:= \theta^*_{j,t}$ so that:
\begin{align}\label{Y02star}
\begin{split}
Z^*_{i,t}&:=\tfrac{2 (\alpha -\kappa (t))}{M+\bar{M}-1}\theta^*_{i,t} + \tfrac{2\kappa(t)+\mu_1(t)}{M+\bar{M}-1}\ta_{i}+\tfrac{\mu_2(t)}{M+\bar{M}-1}q_{i,t}\\
&\;-\tfrac{(M-1) \mu_3(t)+\mu_2(t)}{M+\bar{M}-1}\eta_t 
-\tfrac{\bar{M} (2 \kappa (t)+\bar{\mu}_5(t))}{M+\bar{M}-1}Y_t,\\
Z^*_{j,t}&:=\tfrac{2 (\alpha -\kappa (t))}{M+\bar{M}-1}\theta^*_{j,t}-\tfrac{M \mu_3(t)+\mu_2(t)}{M+\bar{M}-1}\eta_t \\
&\;-\tfrac{(\bar{M}-1) (2 \kappa (t)+\bar{\mu}_5(t))}{M+\bar{M}-1}w_t-\tfrac{A(t) \mu_2(t)+(\bar{M}-1) \bar{\mu}_4(t)+2 \kappa (t)+\mu_1(t)}{M+\bar{M}-1}\ta_\Sigma,
\end{split}
\end{align}
for rebalancers $ i\in\{1,...,M\}$ and trackers $j\in\{M+1,...,M+\bar{M}\}$. 

As for the consistency requirement in Definition \ref{Nash_eq}(iii), we first fix a rebalancer $i_0\in \{1,...,M\}$. We require that the response holdings in \eqref{Y00fct} are consistent with $\theta^*_{i_0,t}$ in the sense that
\begin{align}\label{Y00fctbb}
\begin{split}
\theta^*_{i,t} &=\tfrac{1}{2( \kappa (t)- \alpha)}Z^*_{i_0,t}+\tfrac{2 \kappa (t)+\mu_1(t)}{2( \kappa (t)- \alpha)}\ta_i+\tfrac{\mu_2(t)}{2( \kappa (t)- \alpha)}q_{i,t}+\tfrac{\mu_3(t)}{2( \kappa (t)- \alpha)}\eta_t,\\
\theta^*_{j,t}&= \tfrac{1}{2( \kappa (t)- \alpha)}Z^*_{i_0,t}+\tfrac{2 \kappa(t)+\bar{\mu}_5(t)}{2( \kappa (t)- \alpha)}w_t +\tfrac{\bar{\mu}_4(t)}{2( \kappa (t)- \alpha)}\ta_\Sigma,
\end{split}
\end{align}
for rebalancers $i\in\{1,...,M\}\setminus \{i_0\}$ and  trackers $j\in \{M+1,...,M+\bar{M}\}$.  Second, we fix a tracker $j_0\in \{M+1,...,M+\bar{M}\}$ and require that the response holdings in \eqref{Y00fct} must be  consistent with $\theta^*_{j_0,t}$ in the sense that
\begin{align}\label{Y00fctbb2}
\begin{split}
\theta^*_{i,t} &=\tfrac{1}{2( \kappa (t)- \alpha)}Z^*_{j_0,t}+\tfrac{2 \kappa (t)+\mu_1(t)}{2 \alpha -2 \kappa (t)}\ta_i+\tfrac{\mu_2(t)}{2( \kappa (t)- \alpha)}q_{i,t}+\tfrac{\mu_3(t)}{2( \kappa (t)- \alpha)}\eta_t,\\
\theta^*_{j,t}&=\tfrac{1}{2( \kappa (t)- \alpha)}Z^*_{j_0,t}+\tfrac{2 \kappa(t)+\bar{\mu}_5(t)}{2( \kappa (t)- \alpha)}w_t +\tfrac{\bar{\mu}_4(t)}{2( \kappa (t)- \alpha)}\ta_\Sigma,
\end{split}
\end{align}
for rebalancers $i\in\{1,...,M\}$ and trackers $j\in \{M+1,...,M+\bar{M}\}\setminus \{j_0\}$. 

Similar to the price-impact equilibrium, our Nash equilibrium existence result is based on a  technical lemma, which guarantees the existence of a solution to an autonomous system of coupled ODEs. 

\begin{lemma}\label{main_Lemma} 
Let $\kappa:[0,1]\to(0,\infty]$ be a continuous and integrable function (i.e., $\int_0^1 \kappa(t)dt <\infty$), let $M+\bar{M}>2$, and let $\alpha \le0$. For a constant $B(0) \in \R$, the coupled ODEs 
\begin{footnotesize}
\begin{align}\label{derivatives0a}
\begin{split}
B'(t)
&=\frac{\begin{array}{l}\Big\{2 \kappa (t) \Big(\bar{M} B(t) (M+\bar{M}-1) \big(\alpha  (M+\bar{M})-2 (M+\bar{M}-1) \kappa (t)\big)\\
+(M+\bar{M}-2)\big(\alpha  (M+\bar{M}+1)-2 (M+\bar{M}) \kappa (t)\big)\Big)\Big\}\\
 \end{array}}{\begin{array}{l}
\Big\{ \gamma  \Big(A(t) (M+\bar{M}-2) \big(\alpha  (M+\bar{M}+1)-2 (M+\bar{M}) \kappa (t)\big)\\
+\alpha  \big((M^2+M-1) \bar{M}+M^2+2 M \bar{M}^2-M+\bar{M}^3-2\big)\\
-2 \left((M^2-1) \bar{M}+(2 M-1) \bar{M}^2+(M-2) M+\bar{M}^3\right) \kappa (t)\Big)\Big\},\\
 \end{array}
},\\
A'(t)&= - \big(B'(t)\big)^2\Sigma(t)\big(A(t)+1\big),\quad A(0)=-\frac{(M-1)B(0)^2\sigma^2_{\ta}}{\sigma^2_{w_0} +(M-1)B(0)^2\sigma^2_{\ta}},\\
\Sigma'(t) &= -\big(B'(t)\big)^2\Sigma(t)^2,
\quad \Sigma(0) =\frac{(M-1) \sigma_{\ta}^2 \sigma_{w_0}^2}{(M-1)B(0)^2 \sigma_{\ta}^2+\sigma_{w_0}^2},
\end{split}
\end{align}
\end{footnotesize}have unique solutions with $\Sigma(t) \ge 0$, $\Sigma(t)$ decreasing, $A(t) \in [-1,0]$, and $A(t)$ decreasing for $t\in[0,1]$.
$\endproof$
\end{lemma}
\noindent The affine ODE for $B(t)$ in  \eqref{derivatives0a} is more complicated than the corresponding affine ODE in \eqref{derivatives0aPI} because the Nash equilibrium has the additional fixed point requirement in \eqref{B} that is absent in the price-impact equilibrium. However, both ODEs for $B(t)$ are affine.

Our main theoretical result gives a Nash equilibrium in terms of the ODEs \eqref{derivatives0a}. In this theorem, the price-impact parameter $\alpha\le0 $, volatility $\gamma>0$, and initial value $B(0)\in \R$ are free parameters.

\begin{theorem}\label{thm_Main}  Let $\kappa:[0,1]\to (0,\infty)$ be continuous, let the functions $(B,A,\Sigma)$ be as in Lemma \ref{main_Lemma}, let $M+\bar{M}>2$, and let $\alpha\le0$. Then, we have:

\begin{itemize}

\item[(i)] A subgame perfect Nash financial-market equilibrium exists and is given by the functions in \eqref{nus} in Appendix \ref{sec:formulas}.
\item[(ii)]  Equilibrium holdings  are 
\begin{footnotesize}
\begin{align}
\theta_{i,t}^* &:=
-\frac{(M+\bar{M}-2) \left(2 \kappa (t)-\gamma  B'(t)\right)}{\alpha  (M+\bar{M})-2 (M+\bar{M}-1) \kappa (t)}\ta_i\nonumber\\
& +\frac{\gamma  (M+\bar{M}-2) B'(t)}{\alpha  (M+\bar{M})-2 (M+\bar{M}-1) \kappa (t)}q_{i,t}\nonumber
\\
&-\frac{\begin{array}{l}\Big\{\gamma  (M+\bar{M}-2)^2 B'(t) (\alpha  (M+\bar{M}+1)-2 (M+\bar{M}) \kappa (t))\Big\}\nonumber\\
 \end{array}}{\begin{array}{l}
\Big\{(\alpha  (M+\bar{M})-2 (M+\bar{M}-1) \kappa (t)) \big(\alpha  \big((3 M-1) \bar{M}^2+M (3 M-2) \bar{M}\nonumber\\
+(M-2) M (M+1)+\bar{M}^3\big)-2 \left((M+\bar{M}-2) (M+\bar{M})^2+\bar{M}\right) \kappa (t)\big)\Big\}
 \end{array}
}\eta_t,\nonumber\\
&+\frac{\begin{array}{l}\Big\{2 \bar{M} (M+\bar{M}-2) (M+\bar{M}-1) \kappa (t)\Big\}\\
 \end{array}}{\begin{array}{l}
\Big\{\alpha  \left((3 M-1) \bar{M}^2+M (3 M-2) \bar{M}+(M-2) M (M+1)+\bar{M}^3\right)\\
-2 \left((M+\bar{M}-2) (M+\bar{M})^2+\bar{M}\right) \kappa (t)\Big\}
 \end{array}
}
Y_t,\label{Y0000}
\end{align}
\end{footnotesize}
\vspace{-0.5cm}
\begin{align*}
\begin{split}
\theta_{j,t}^* :&= 
-\tfrac{\gamma  (M+\bar{M}-2) (M+\bar{M}-1) B'(t)}{\alpha  \left((3 M-1) \bar{M}^2+M (3 M-2) \bar{M}+(M-2) M (M+1)+\bar{M}^3\right)-2 \left((M+\bar{M}-2) (M+\bar{M})^2+\bar{M}\right) \kappa (t)}\eta_t 
\\
&-\tfrac{2 M (M+\bar{M}-2) (M+\bar{M}-1) \kappa (t)}{\alpha  \left((3 M-1) \bar{M}^2+M (3 M-2) \bar{M}+(M-2) M (M+1)+\bar{M}^3\right)-2 \left((M+\bar{M}-2) (M+\bar{M})^2+\bar{M}\right) \kappa (t)}w_t
\\
&+\tfrac{(M+\bar{M}-2) (M+\bar{M}-1) \left(\gamma  (-A(t)+M-1) B'(t)+2 \kappa (t)\right)}{\alpha  \left((3 M-1) \bar{M}^2+M (3 M-2) \bar{M}+(M-2) M (M+1)+\bar{M}^3\right)-2 \left((M+\bar{M}-2) (M+\bar{M})^2+\bar{M}\right) \kappa (t)}
\ta_\Sigma,
\end{split}
\end{align*}

for rebalancers $i\in \{1,...,M\}$ and  trackers $ j\in\{M+1,...,M+\bar{M}\}$.

\item[(iii)] There exists a Nash equilibrium stock-price process $S^*_t$ with $S^*_0 := w_0 - B(0)\ta_\Sigma$ and dynamics with respect to the trackers' filtrations $\sF_{j,t}:=\sigma(w_u,S^{\bar\nu}_{j,u})_{u\in[0,t]}$ given by
\begin{align}\label{dhatS}
\begin{split}
dS^*_t &:=\Big\{\tfrac{\gamma  (M+\bar{M}-2) B'(t) (\alpha  (M+\bar{M}+1)-2 (M+\bar{M}) \kappa (t))}{\alpha  \left((3 M-1) \bar{M}^2+M (3 M-2) \bar{M}+(M-2) M (M+1)+\bar{M}^3\right)-2 \left((M+\bar{M}-2) (M+\bar{M})^2+\bar{M}\right) \kappa (t)}\eta_t\\
&-\tfrac{2 \bar{M} (M+\bar{M}-1) \kappa (t) (\alpha  (M+\bar{M})-2 (M+\bar{M}-1) \kappa (t))}{\alpha  \left((3 M-1) \bar{M}^2+M (3 M-2) \bar{M}+(M-2) M (M+1)+\bar{M}^3\right)-2 \left((M+\bar{M}-2) (M+\bar{M})^2+\bar{M}\right) \kappa (t)}w_t \\
&-\tfrac{(M+\bar{M}-2) (\alpha  (M+\bar{M}+1)-2 (M+\bar{M}) \kappa (t)) \left(\gamma  (-A(t)+M-1) B'(t)+2 \kappa (t)\right)}{\alpha  \left((3 M-1) \bar{M}^2+M (3 M-2) \bar{M}+(M-2) M (M+1)+\bar{M}^3\right)-2 \left((M+\bar{M}-2) (M+\bar{M})^2+\bar{M}\right) \kappa (t)}\ta_\Sigma\Big\}dt \\
&+ \gamma dw_t,
\end{split}
\end{align}
and dynamics with respect to the rebalancers' filtrations $\sF_{i,t}:=\sigma(\ta_i,S^\nu_{i,u})_{u\in[0,t]}$ given by
\begin{align}\label{dhatS34}
\begin{split}
dS^*_t &:=\Big\{\tfrac{\gamma  (M+\bar{M}-2) B'(t) (\alpha  (M+\bar{M}+1)-2 (M+\bar{M}) \kappa (t))}{\alpha  \left((3 M-1) \bar{M}^2+M (3 M-2) \bar{M}+(M-2) M (M+1)+\bar{M}^3\right)-2 \left((M+\bar{M}-2) (M+\bar{M})^2+\bar{M}\right) \kappa (t)}\eta_t\\
&-\tfrac{2 \bar{M} (M+\bar{M}-1) \kappa (t) (\alpha  (M+\bar{M})-2 (M+\bar{M}-1) \kappa (t))}{\alpha  \left((3 M-1) \bar{M}^2+M (3 M-2) \bar{M}+(M-2) M (M+1)+\bar{M}^3\right)-2 \left((M+\bar{M}-2) (M+\bar{M})^2+\bar{M}\right) \kappa (t)}Y_t \\
&-\gamma B'(t)(\ta_i+q_{i,t})\Big\}dt + \gamma dw_{i,t}.
\end{split}
\end{align}
$\endproof$
\end{itemize}
\end{theorem}

The following observations follow from Theorem \ref{thm_Main}:

\begin{enumerate}

\item The logic for the initial value $B(0)$ being a free input parameter is the same as in the price-impact equilibrium.

\item The price-impact parameter $\alpha$ and stock-price volatility $\gamma$ affect the stock-price drift and holdings via its impact on $B(t)$ in \eqref{derivatives0a}. The dependence on $\alpha$ is different from the price-impact equilibrium where the corresponding $B(t)$ in \eqref{derivatives0aPI} is independent of $\alpha$. The reason is that $\alpha$ affects the perceived optimal responses in \eqref{Y00fct}.

\item Similar to \eqref{decom1} and  \eqref{decom2}, for an arbitrary trader $k_0 \in \{1,...,M+\bar{M}\}$ and her arbitrary  holdings $\theta_{k_0,t}$,  the optimal responses in \eqref{Y00fct} can be decomposed as
\begin{align}\label{decomp3}
\begin{split}
\theta^{Z_{k_0}}_{i,t} &= \theta^*_{i,t} -\frac{1}{M+\bar{M}-1} (\theta_{k_0,t}-\theta^*_{k_0,t}),\quad i \in \{1,...,M\},\\
\theta^{Z_{k_0}}_{j,t}&=\theta^*_{j,t} -\frac{1}{M+\bar{M}-1} (\theta_{k_0,t}-\theta^*_{k_0,t}),\quad j\in \{M+1,...,M+\bar{M}\},
\end{split}
\end{align}
where the equilibrium holdings $(\theta^*_{i,t}, \theta^*_{j,t},\theta^*_{k_0,t})$  are in \eqref{Y0000}.\footnote{This is similar to Eq. (2.16) in Chen, Choi, Larsen, and Seppi (2021).}

\item The subgame perfect Nash financial-market  equilibrium is attractive because of its reasonable off-equilibrium market-clearing perceptions.  However, although much of the mathematic structure is similar, the expressions for the equilibrium stock price and holding coefficients are algebraically more complex. Nonetheless, our numerical results in Section \ref{sec:num} below show that the differences between the price-impact and the subgame perfect Nash financial-market  equilibria are quantitatively small. This, in turn, suggests that the economic logic from the price-impact equilibrium carries over to the Nash equilibrium.
\end{enumerate}

\subsection{Numerics} 

We have experimented extensively with the subgame perfect Nash model's numerics, and its numerics  are very similar to the numerics of the price-impact equilibrium in Section \ref{sec:PI}. The numerical similarity of the two equilibria suggests that the intuitions for the signs of the various coefficients in the price-impact equilibrium carry over to the subgame perfect Nash financial-market equilibrium. Because the two equilibria produce similar numerics, it appears that the  in-equilibrium market-clearing requirement (common in both equilibria) has a much larger effect on equilibrium prices relative to the off-equilibrium market-clearing  requirement (only present in the subgame perfect Nash equilibrium).  

\section{Empirical predictions}

The primary contribution of our paper’s analysis is theoretical. The Kyle model has provided a tractable framework for a large body of theoretical research on price discovery and dynamic order splitting given long-lived asymmetric information about stock cash flows. However, no corresponding tractable framework exists for modeling price discovery and dynamic order splitting with private trading targets (e.g., by large index funds). Our model provides such a framework.  While our zero-dividend modeling approach precludes statements about the impact of order on price levels, our analysis does have empirical implications for intraday price drifts:

First, intraday price predictability is an important empirical driver of high-frequency liquidity provision.  Our model’s equilibrium price dynamics in \eqref{S_PI} and \eqref{dhatS} suggest that intraday price drifts are path dependent (via the $\eta_t$ term) and also that learning about parent demands imbalances early in the trading day is associated with predictable price drifts later in the day.

Second, our analysis provides insights about the determinants of price impact as it relates to imbalance-related parent trading demands and toxic cumulative order flow.  In particular, the holdings $\theta_{k,t}$ are cumulative trading up through time $t$, and large parent targets $\ta_i$ lead to toxic streams of orders. Our subgame perfect Nash model endogenizes the price drift impact of investor holdings (i.e., cumulative trading).  The Nash model's price-impact coefficient in the rebalancer's perceived stock-price dynamics \eqref{New22ia} is given by
\begin{align}\label{emp1}
-\frac{2 \nu_0(t) (\kappa (t)-\alpha )}{M+\bar{M}-1} = -2\frac{  \kappa (t)-\alpha }{M+\bar{M}-2},
\end{align}
where we have inserted $\nu_0(t)$ from \eqref{nus}. An implication of \eqref{emp1} is that if, as is widely believed, investor target penalties become stronger as time passes (i.e., if $\kappa(t)$ increases with time), then our Nash model predicts that the total price impact in \eqref{emp1} should increase.  On its face, this is contrary to evidence in Barardehi and Bernhardt (2021) that price impact declines over the trading day.  We conjecture, however, that a richer model can be reconciled with these stylized facts if the number of investors (and, thus, the available inventory bearing capacity to absorb aggregate parent demand imbalances) is also allowed to grow as the market approaches the end of the trading day. Increased investor participation toward the end of the trading day is also empirically common.

\section{Measuring execution costs}
As an application, this section gives a measure of a rebalancer's costs of rebalancing from zero endowed shares at time $t=0$ to a given target $\ta_i$. We present the measure in the price-impact equilibrium in Section \ref{sec:PI} (the Nash analogue is logically similar and produces similar numerics). In the price-impact equilibrium, rebalancer $i$'s value function is 
\begin{align}\label{Rproblema}
\begin{split}
J(\ta_i,0,\eta_0,Y_0,q_{i,0}):=& \E\Big[ \int_0^1 \hat \theta_{i,t}d\hat S_t - \int_0^1 \kappa(t)(\tilde{a}_i-\hat \theta_{i,t})^2dt\Big|\,\sF_{i,0}\Big],
\end{split}
\end{align}
where $\hat \theta_{i,t}$ denotes rebalancer $i$'s equilibrium stock holdings in \eqref{Y0000PI}  and $\sF_{i,t}:=\sigma(\ta_i,S^f_{i,u})_{u\in[0,t]}$ where the $f$ coefficient functions are as in \eqref{fs} in Appendix \ref{sec:formulas} for $i\in\{1,...,M\}$. We seek a value function $J= J(\ta_i,s,q,Y,q_i)$ such that the process
\begin{align}\label{Rproblemc}
\begin{split}
J(\ta_i,s,\eta_s,Y_s,q_{i,s})&+ \int_0^s \Big\{\hat\theta_{i,t}\Big(f_0(t)Y_t  +f_1(t)\tilde{a}_i +f_2(t)q_{i,t}+f_3(t)\eta_t+\alpha\hat\theta_{i,t}\Big)\\
& - \kappa(t)(\tilde{a}_i-\hat\theta_{i,t})^2\Big\}dt,\quad s\in[0,1],
\end{split}
\end{align}
is a martingale with respect to $\sF_{i,t}$. Because rebalancer $i$'s objective in \eqref{Rproblem} is linear-quadratic, the value function $J$ is again linear-quadratic in the state processes. Thus, $J$ can be written as
\begin{align}\label{Rproblemd}
\begin{split}
J(\ta_i,s,\eta,Y,q_i)&= J_0(s) + J_{\eta}(s) \eta + J_Y(s) Y+ J_{q_i}(s) q_i+ J_{\eta\eta}(s)\eta^2 \\
& + J_{\eta Y}(s) \eta Y + J_{YY}(s) Y^2+J_{q_iq_i}(s) q^2_i+ J_{q_i\eta}(s)q_i\eta+J_{q_iY}(s) q_iY,
\end{split}
\end{align}
for deterministic functions of time $(J_0, J_\eta, J_Y, J_{q_i},J_{\eta\eta},J_{\eta Y}, J_{YY},J_{q_iq_i},J_{q_i\eta},J_{q_iY})$. These functions are given by a coupled set of ODEs with zero terminal conditions (we omit the ODEs for brevity). In \eqref{Rproblemd}, the dummy variables $(\eta,Y,q_i)$ are real numbers and $s\in[0,1]$.
 
To quantify the costs associated with rebalancer $i$'s trading target $\ta_i$,  the quadratic mapping RC (Rebalancing Costs) defined by
\begin{align}\label{def_RC}
\begin{split}
\text{RC}(\ta_i):=J(0,0,\eta,Y,q_{i})-J(\ta_i,0,\eta,Y,q_{i}),
\end{split}
\end{align}
measures the dependence the change in profit (i.e., change in value function) associated with a non-zero target $\ta_i$. 

\begin{figure}[!h]
\begin{center}
\caption{Plots of the rebalancers' value function $J$ for various values of $(\gamma,\sigma_{w_0})$.  The exogenous model parameters are $ \sigma_{\ta}:=1, M:=\bar{M}:=10, \;\alpha :=-0.1,\;  B(0):=-1,\;\kappa(t):=1$ for $t\in[0,1]$, and $w_0:= B(0)(\ta_\Sigma-\ta_i)$.  }\ \\
\begin{footnotesize}
$\begin{array}{c c}
\includegraphics[width=6cm, height=4.5cm]{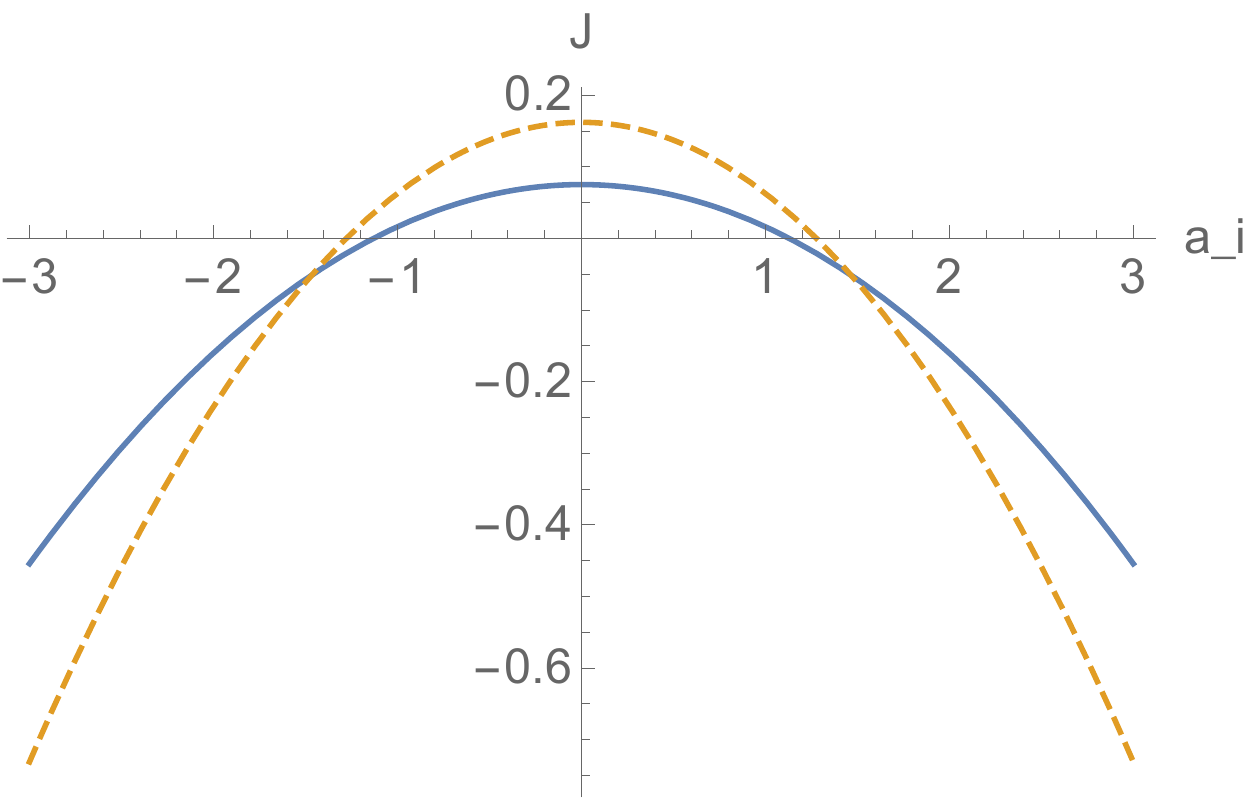} &\includegraphics[width=6cm, height=4.5cm]{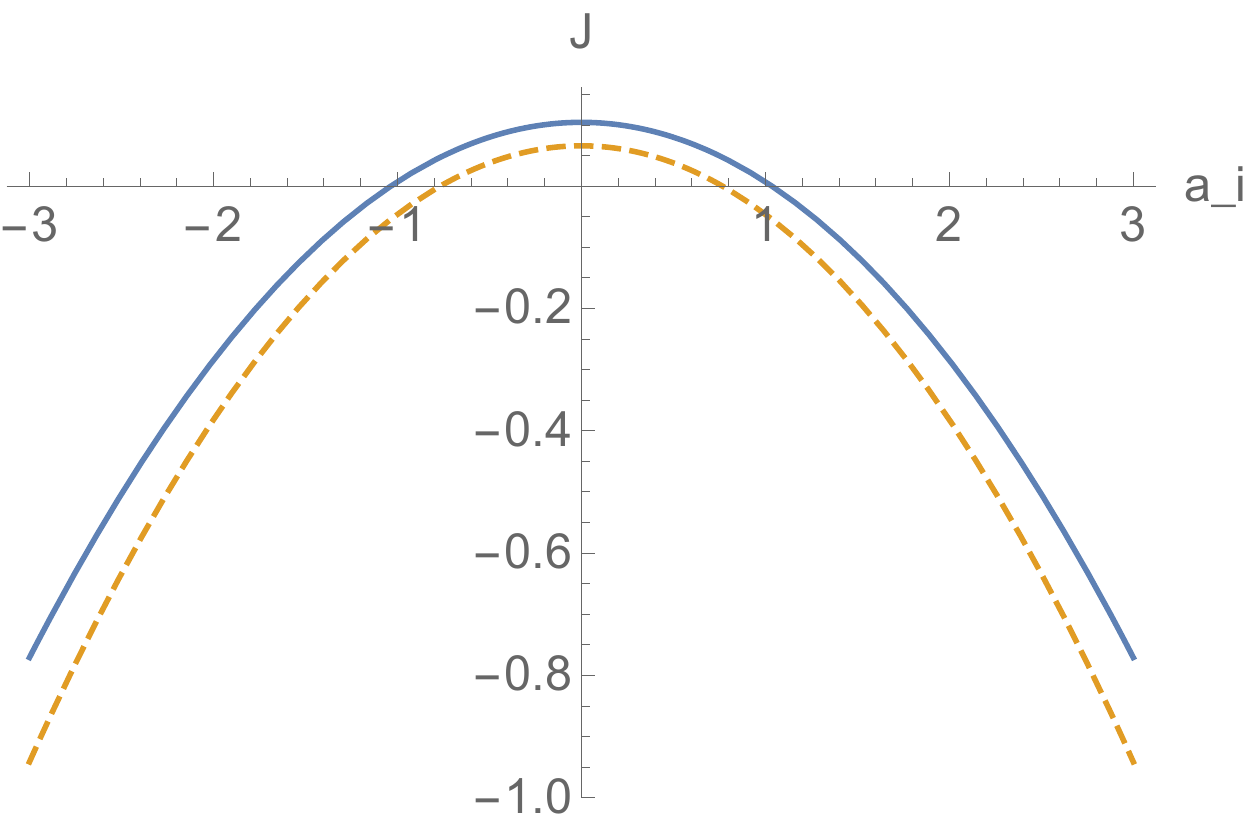} 
\\ 
\text{{\bf 6A:}   $\gamma:=1\;\text{(-----)},\; \gamma:=0.5\; (- -),\;\sigma_{w_0} :=1 $}&\text{{\bf 6B:}  $\gamma:=1\;\text{(-----)},\; \gamma:=0.5\; (- -),\;\sigma_{w_0} :=0.1 $}\\
\end{array}$
 \end{footnotesize}
\label{fig23}
\end{center}
\end{figure}

Figure \ref{fig23}  plots the rebalancer’s value function $J$ for different target values $\ta_i$ for different model parameterizations.  When the  target $\ta_i$ is close to zero, the rebalancers become high-frequency liquidity providers.  Their value function is positive due expected profit from liquidity provision and price-pressure front-running.  As the target moves away from zero, the rebalancer starts to have larger stock-holding penalties that eventually drive the rebalancer's value function negative. Interestingly, the impact of the stock-price volatility parameter $\gamma$ on the rebalancer’s value function can be positive or negative.  Liquidity providing rebalancers are better off with a small $\gamma$ whereas rebalancers with large rebalancing targets are better off when $\gamma$ is large.

The rebalancing cost RC in \eqref{def_RC} for a target $\ta_i$ is computed as the difference between the value function evaluated at $\ta_i$ and the function evaluated at $\ta_i = 0$.  Since the value function $J$ is highest at $\ta_i = 0$, the measure RC is positive.

\section{Conclusion}

This paper presents the first analytically tractable model of dynamic learning about parent trading demand imbalances with optimized order-splitting. In particular, we provide closed-form expressions prices and stock holdings in terms of solutions to systems of coupled ODEs in both price-impact and Nash equilibria. We then show that trading in our models reflects a combination of reaching investor’s own trading targets, liquidity provision so that markets can clear, and front-running based on predictions of future price pressure.  

There are many interesting directions for future research based on our analysis. First, replacing the zero-dividend stock approach with valuation based on a terminal payoff would be a significant technical step. Second, the model could be enriched by allowing for investor heterogeneity in the form of different penalty functions $\kappa(t)$ and by having multiple tracker targets (which would weaken the trackers’ informational advantage).   Third, it would be interesting to investigate if other off-equilibrium refinements have larger equilibrium effects. Fourth, incorporating risk-aversion into the investors' objectives would be interesting too. For example,   how can  Lemma \ref{response} be extended if the objectives in \eqref{Rproblem} are changed to exponential utilities?

\newpage

\appendix

\section{Formulas}\label{sec:formulas}

\subsection{Price-perception coefficients for the price-impact equilibrium}

\begin{align}\label{fs}
\begin{split}
f_0(t)&:=\frac{4 \bar{M} \kappa (t) (\kappa (t)-\alpha )}{(M+\bar{M}) (\alpha -2 \kappa (t))},\\
f_1(t)&:=\frac{2 \gamma  B'(t) (\kappa (t)-\alpha )+2 \alpha  \kappa (t)}{\alpha -2 \kappa (t)},\\
f_2(t)&:=\frac{2 \gamma  B'(t) (\kappa (t)-\alpha )}{\alpha -2 \kappa (t)},\\
f_3(t)&:=\frac{2 \gamma  B'(t) (\alpha -\kappa (t))}{(M+\bar{M}) (\alpha -2 \kappa (t))},\\
\bar{f}_3(t) &:= \frac{2 \gamma  B'(t) (\alpha -\kappa (t))}{(M+\bar{M}) (\alpha -2 \kappa (t))},\\
\bar{f}_4(t) &:= \frac{2 (\alpha -\kappa (t)) \left(\gamma  (A(t)-M+1) B'(t)-2 \kappa (t)\right)}{(M+\bar{M}) (\alpha -2 \kappa (t))},\\
\bar{f}_5(t) &:=  \frac{2 \kappa (t) (\alpha  (M-\bar{M})+2 \bar{M} \kappa (t))}{(M+\bar{M}) (\alpha -2 \kappa (t))}. 
\end{split}
\end{align}

\subsection{Orthogonal representations for the price-impact equilbrium}
Let the deterministic functions $F_1(t)$ and $F_2(t)$ be as in \eqref{F1F2}. 

\subsubsection{Price-impact equilibrium holdings}
The price-impact equilibrium holdings $\hat\theta_{i,t}$ in \eqref{Y0000PI} for rebalancer $ i\in\{1,...,M\}$ has an orthogonal representation given by
\begin{footnotesize}
\begin{align}\label{thetaiortho}
\begin{split}
\hat{\theta}_{i,t} &
=-\frac{\begin{array}{l}\Big\{2 \kappa (t) \Big(A(t) (\bar{M} B(t)+M+\bar{M}) \left((M-1)B(0)^2 \sigma_{\ta}^2+\sigma_{w_0}^2\right)\\
+M F_1(t) F_2(t) (\bar{M} B(t)+1) \left((M-1)B(0)^2 \sigma_{\ta}^2+\sigma_{w_0}^2\right)\\
+(M-1) \left(B(0)^2 M \sigma_{\ta}^2 F_1(t) (\bar{M} B(t)+1)-\bar{M} B(t) \left((M-1)B(0)^2 \sigma_{\ta}^2+\sigma_{w_0}^2\right)\right)\\
+\bar{M} (M+\bar{M}) \left((M-1)B(0)^2 \sigma_{\ta}^2+\sigma_{w_0}^2\right)\Big)\Big\}
 \end{array}}{\begin{array}{l}
\Big\{(M+\bar{M}) (A(t)+\bar{M}+1) (\alpha -2 \kappa (t)) \left((M-1)B(0)^2 \sigma_{\ta}^2+\sigma_{w_0}^2\right)\Big\}
 \end{array}
}
\ta_i
\\&
+ \frac{\begin{array}{l}\Big\{2 \bar{M} \kappa (t) \Big(B(t) \big(-A(t) \left((M-1)B(0)^2 \sigma_{\ta}^2+\sigma_{w_0}^2\right)\\
+\bar{M} F_1(t) \left(F_2(t) \left((M-1)B(0)^2 \sigma_{\ta}^2+\sigma_{w_0}^2\right)+(M-1)B(0)^2 \sigma_{\ta}^2\right)\\
-(\bar{M}+1) \left((M-1)B(0)^2 \sigma_{\ta}^2+\sigma_{w_0}^2\right)\big)\\
+F_1(t) \left(F_2(t) \left((M-1)B(0)^2 \sigma_{\ta}^2+\sigma_{w_0}^2\right)+(M-1)B(0)^2 \sigma_{\ta}^2\right)\Big)\Big\}
 \end{array}}{\begin{array}{l}
\Big\{(M+\bar{M}) (A(t)+\bar{M}+1) (\alpha -2 \kappa (t)) \left((M-1)B(0)^2 \sigma_{\ta}^2+\sigma_{w_0}^2\right)\Big\}
 \end{array}
}
(\ta_\Sigma -\ta_i)
\\
&+\frac{2 \bar{M} \kappa (t) \left(1-\frac{B(0) (M-1) \sigma_{\ta}^2 F_1(t) (\bar{M} B(t)+1)}{(A(t)+\bar{M}+1) \left((M-1)B(0)^2 \sigma_{\ta}^2+\sigma_{w_0}^2\right)}\right)}{(M+\bar{M}) (\alpha -2 \kappa (t))}w_0
\\
&+\frac{2 \bar{M} \kappa (t)}{(M+\bar{M}) (\alpha -2 \kappa (t))} w^\circ_t -\frac{2 \bar{M} F_1(t) \kappa (t) (\bar{M} B(t)+1)}{(M+\bar{M}) (A(t)+\bar{M}+1) (\alpha -2 \kappa (t))}\int_0^t \frac{B'(s)\Sigma(s)}{F_1(s)} dw^\circ_s.
\end{split}
\end{align}
\end{footnotesize}The price-impact equilibrium holdings $\hat\theta_{j,t}$ in \eqref{Y0000PI} for tracker $ j\in\{M+1,...,M+\bar M\}$ has an orthogonal representation given by
\begin{footnotesize}
\begin{align}\label{thetajortho}
\begin{split}
\hat{\theta}_{j,t} &=-\frac{\begin{array}{l}\Big\{2 \kappa (t) \Big(\bar{M} B(t) \Big(A(t) \left((M-1)B(0)^2 \sigma_{\ta}^2+\sigma_{w_0}^2\right)\\
+M F_1(t) F_2(t) \left((M-1)B(0)^2 \sigma_{\ta}^2+\sigma_{w_0}^2\right)\\
+(M-1) \left(B(0)^2 M \sigma_{\ta}^2 F_1(t)-(M-1)B(0)^2 \sigma_{\ta}^2-\sigma_{w_0}^2\right)\Big)\\
+M F_1(t) \big(F_2(t) \left((M-1)B(0)^2 \sigma_{\ta}^2+\sigma_{w_0}^2\right)\\
+(M-1)B(0)^2 \sigma_{\ta}^2\big)-(M+\bar{M}) \left((M-1)B(0)^2 \sigma_{\ta}^2+\sigma_{w_0}^2\right)\Big)\Big\}
 \end{array}}{\begin{array}{l}
\Big\{(M+\bar{M}) (A(t)+\bar{M}+1) (\alpha -2 \kappa (t)) \left((M-1)B(0)^2 \sigma_{\ta}^2+\sigma_{w_0}^2\right)\Big\}
 \end{array}
}\ta_\Sigma\\
&+\frac{2 M \kappa (t) \left(\frac{B(0) (M-1) \sigma_{\ta}^2 F_1(t) (\bar{M} B(t)+1)}{(A(t)+\bar{M}+1) \left((M-1)B(0)^2 \sigma_{\ta}^2+\sigma_{w_0}^2\right)}-1\right)}{(M+\bar{M}) (\alpha -2 \kappa (t))}w_0
\\
&-\frac{2 M \kappa (t)}{(M+\bar{M}) (\alpha -2 \kappa (t))}w^\circ_t 
+\frac{2 M F_1(t) \kappa (t) (\bar{M} B(t)+1)}{(M+\bar{M}) (A(t)+\bar{M}+1) (\alpha -2 \kappa (t))}
\int_0^t \frac{B'(s)\Sigma(s)}{F_1(s)} dw^\circ_s.
\end{split}
\end{align}
\end{footnotesize}

\subsubsection{Price-impact equilibrium stock dynamics}

For the trackers, we can rewrite the drift in \eqref{S_PI} in terms of  $(\ta_\Sigma, w_0)$ and an  residual orthogonal  term as
\begin{align}\label{S_PI223}
\begin{split}
&\tfrac{\gamma  B'(t)}{M+\bar{M}}\eta_t-\tfrac{2 \bar{M} \kappa (t)}{M+\bar{M}}w_t +\tfrac{\gamma  (A(t)-M+1) B'(t)-2 \kappa (t)}{M+\bar{M}}\ta_\Sigma\\
&=-\tfrac{\frac{B(0) \gamma  (M-1) M\sigma_{\ta}^2 F_1(t) B'(t)}{(M-1)B(0)^2\sigma_{\ta}^2+\sigma_{w_0}^2}+2 \bar M \kappa (t)}{M+\bar M}w_0\\
&+\tfrac{\gamma  B'(t) \left(A(t)+M F_1(t) \left(\frac{(M-1)B(0)^2\sigma_{\ta}^2}{(M-1)B(0)^2\sigma_{\ta}^2+\sigma_{w_0}^2}+F_2(t)\right)-M+1\right)-2 \kappa (t)}{M+\bar M} \tilde a_\Sigma\\
&-\tfrac{2 \bar M \kappa (t)}{M+\bar M}w^\circ_t-\tfrac{\gamma  M F_1(t) B'(t)}{M+\bar M} \int_0^t \tfrac{B'(s)\Sigma(s)}{F_1(s)}dw^\circ_s.
\end{split}
\end{align}

For the rebalancers, we  can rewrite the drift in \eqref{rebdriftPI} in terms of $(\ta_\Sigma-\ta_i, w_0, \ta_i)$ and an residual orthogonal term as
\begin{align}\label{S_PI221}
\begin{split}
&-\gamma B'(t)\big( \ta_i + q_{i,t}\big) + \tfrac{\gamma  B'(t)}{M+\bar{M}}\eta_t-\tfrac{2 \bar{M} \kappa (t)}{M+\bar{M}}Y_t\\
&=\tfrac{\bar M \left(\frac{B(0) \gamma  (M-1)\sigma_{\ta}^2 F_1(t) B'(t)}{(M-1)B(0)^2\sigma_{\ta}^2+\sigma_{w_0}^2}-2 \kappa (t)\right)}{M+\bar M}w_0\\
&+\tfrac{\gamma  B'(t) \left(M F_1(t) \left(\frac{(M-1)B(0)^2\sigma_{\ta}^2}{(M-1)B(0)^2\sigma_{\ta}^2+\sigma_{w_0}^2}+F_2(t)\right)-M-\bar M\right)+2 \bar M B(t) \kappa (t)}{M+\bar M}\ta_i \\
&+\tfrac{\bar M \left(\gamma  F_1(t) B'(t) \left(-\frac{(M-1)B(0)^2\sigma_{\ta}^2}{(M-1)B(0)^2\sigma_{\ta}^2+\sigma_{w_0}^2}-F_2(t)\right)+2 B(t) \kappa (t)\right)}{M+\bar M}(\ta_\Sigma -\ta_i)\\
&-\tfrac{2 \bar M \kappa (t)}{M+\bar M}w_t^\circ + \tfrac{\gamma  \bar M F_1(t) B'(t)}{M+\bar M}\int_0^t \tfrac{B'(s)\Sigma(s)}{F_1(s)}dw^\circ_s.
\end{split}
\end{align}

\subsection{Price-perception coefficients for the Nash equilibrium}
\begin{footnotesize}
\begin{align}\label{nus}
\begin{split}
\mu_1(t)&:=\frac{2 \gamma  (M+\bar{M}-2) B'(t) (\kappa (t)-\alpha )+2 \kappa (t) (\alpha  (M+\bar{M}-4)+2 \kappa (t))}{\alpha  (M+\bar{M})-2 (M+\bar{M}-1) \kappa (t)},\\
\mu_2(t)&:=-\frac{2 \gamma  (M+\bar{M}-2) B'(t) (\alpha -\kappa (t))}{\alpha  (M+\bar{M})-2 (M+\bar{M}-1) \kappa (t)},\\
\mu_3(t)&:=\frac{\begin{array}{l}\Big\{-\big(4 \gamma  (M+\bar{M}-2) B'(t) (\alpha -\kappa (t))^2\big)\Big\}
 \end{array}}{\begin{array}{l}
\Big\{\big((\alpha  (M+\bar{M})-2 (M+\bar{M}-1) \kappa (t)) \big(\alpha  ((3 M-1) \bar{M}^2+M (3 M-2) \bar{M}\\
+(M-2) M (M+1)+\bar{M}^3)
-2 \left((M+\bar{M}-2) (M+\bar{M})^2+\bar{M}\right) \kappa (t)\big)\big)\Big\}
 \end{array}
},
\\
\bar{\mu}_4(t)&:=\frac{\begin{array}{l}\Big\{-\Big(2 (M+\bar{M}-2) (\alpha -\kappa (t)) \big(\gamma  B'(t) \big(\alpha  \left(-2 A(t)+(M+\bar{M}-1) (M+\bar{M})^2-2\right)\\
-2 \kappa (t) \left(-A(t)+(3 M-2) \bar{M}^2+M (3 M-4) \bar{M}+(M-1)^2 M+\bar{M}^3+\bar{M}-1\right)\big)+4 \kappa (t) (\alpha -\kappa (t))\big)\Big)\Big\}
 \end{array}}{\begin{array}{l}
\Big\{\Big((\alpha  (M+\bar{M})-2 (M+\bar{M}-1) \kappa (t))\big(\alpha  ((3 M-1) \bar{M}^2+M (3 M-2) \bar{M}\\
+(M-2) M (M+1)+\bar{M}^3)-2 \left((M+\bar{M}-2) (M+\bar{M})^2+\bar{M}\right) \kappa (t)\big)\Big)\Big\}
 \end{array}
},
\\
\bar{\mu}_5(t)&:=
\frac{\begin{array}{l}\Big\{\Big(2 \kappa (t) \big(\alpha  \big(M^2 (3 \bar{M}-5)+M^3+M (\bar{M} (3 \bar{M}-10)+6)\\
+(\bar{M}-4) (\bar{M}-1) \bar{M}\big)+2 \left(M^2+2 M (\bar{M}-1)+(\bar{M}-1) \bar{M}\right) \kappa (t)\big)\Big)\Big\}
 \end{array}}{\begin{array}{l}
\Big\{\Big(\alpha  \left((3 M-1) \bar{M}^2+M (3 M-2) \bar{M}+(M-2) M (M+1)+\bar{M}^3\right)\\
-2 \left((M+\bar{M}-2) (M+\bar{M})^2+\bar{M}\right) \kappa (t)\Big)\Big\}
 \end{array}
},
\\
\nu_0(t) &:=\frac{1}{M+\bar{M}-2}+1,\\
\nu_1(t) &:= \frac{2 \alpha  (M+\bar{M}-2) \kappa (t)-2 \gamma  (M+\bar{M}-1) B'(t) (\alpha -\kappa (t))}{\alpha  (M+\bar{M})-2 (M+\bar{M}-1) \kappa (t)},
\\
\nu_2(t) &:= -\frac{2 \gamma  (M+\bar{M}-1) B'(t) (\alpha -\kappa (t))}{\alpha  (M+\bar{M})-2 (M+\bar{M}-1) \kappa (t)},
\\
\nu_3(t) &:=\frac{\begin{array}{l}\Big\{-(M+\bar{M}-1)4 \gamma  B'(t) (\alpha -\kappa (t))^2\Big\}
 \end{array}}{\begin{array}{l}
\Big\{\Big((\alpha  (M+\bar{M})-2 (M+\bar{M}-1) \kappa (t)) \big(\alpha  \big((3 M-1) \bar{M}^2+M (3 M-2) \bar{M}\\
+(M-2) M (M+1)+\bar{M}^3\big)-2 \left((M+\bar{M}-2) (M+\bar{M})^2+\bar{M}\right) \kappa (t)\big)\Big)\Big\}
 \end{array}
},\\
\bar{\nu}_3(t)&:=\frac{2 \gamma  (M+\bar{M}-2) B'(t) (\alpha -\kappa (t))}{\alpha  \left((3 M-1) \bar{M}^2+M (3 M-2) \bar{M}+(M-2) M (M+1)+\bar{M}^3\right)-2 \left((M+\bar{M}-2) (M+\bar{M})^2+\bar{M}\right) \kappa (t)},\\
\bar{\nu}_4(t)&:=\frac{\begin{array}{l}\Big\{-\Big(2 (M+\bar{M}-2) (\alpha -\kappa (t)) \big(\gamma  B'(t) \big(\alpha  \big(-A(t) (M+\bar{M}+2)+(M+\bar{M}-1) \left(M^2+2 M \bar{M}+M+\bar{M}^2\right)\\
-\bar{M}-2\big)+2 \kappa (t) \left(A(t) (M+\bar{M})+M^2 (1-3 \bar{M})-M^3-3 M (\bar{M}-1) \bar{M}+M-(\bar{M}-2) \bar{M}^2\right)\big)\\
+2 \kappa (t) (\alpha  (M+\bar{M}+2)-2 (M+\bar{M}) \kappa (t))\big)\Big)\Big\}
 \end{array}}{\begin{array}{l}
\Big\{\Big((\alpha  (M+\bar{M})-2 (M+\bar{M}-1) \kappa (t)) \big(\alpha  \left((3 M-1) \bar{M}^2+M (3 M-2) \bar{M}+(M-2) M (M+1)+\bar{M}^3\right)\\
-2 \left((M+\bar{M}-2) (M+\bar{M})^2+\bar{M}\right) \kappa (t)\big)\Big)
\Big\}
 \end{array}
},
\\
\bar{\nu}_5(t)&:=\frac{2 (M+\bar{M}-1) \kappa (t) \left(\alpha  \left(M^2+2 M (\bar{M}-1)+(\bar{M}-4) \bar{M}\right)+2 \bar{M} \kappa (t)\right)}{\alpha  \left((3 M-1) \bar{M}^2+M (3 M-2) \bar{M}+(M-2) M (M+1)+\bar{M}^3\right)-2 \left((M+\bar{M}-2) (M+\bar{M})^2+\bar{M}\right) \kappa (t)}.
\end{split}
\end{align}
\end{footnotesize}

\newpage
\section{Kalman-Bucy filtering}

The proof of Lemma \ref{lemKB} follows from the well-known Kalman-Bucy result in filtering theory and can be found in, e.g., Lipster and Shiryaev (Chapter 8, 2001). We note that the solution to the Riccati equation \eqref{SigmaKB3} below is given by  \eqref{Sigma}.

\begin{theorem}[Kalman-Bucy]\label{thm:KB} Let $B:[0,1]\to \R$ be a continuously differentiable function and consider the Gaussian  observation process $Y_{i,t}:= w_t - B(t)(\ta_\Sigma-\ta_i)$ from \eqref{Zi}  with dynamics
\begin{align} \label{KBdYi}
dY_{i,t} =   dw_t -  B'(t)\big(\ta_\Sigma-\ta_i\big)dt,\quad Y_{i,0}= w_0 - B(0)(\ta_\Sigma-\ta_i)
\end{align}
and corresponding innovations process $w_{i,t}$ in \eqref{dwit}. Then, \eqref{RfiltrationQ} holds and the filtering property in \eqref{dwit} holds if $q_{i,t}$ has dynamics given by
\begin{align}\label{SigmaKB2}
\begin{split}
d q_{i,t} &=   -B'(t)\Sigma(t)dY_{i,t}- \big(B'(t)\big)^2\Sigma(t) q_{i,t}dt\\
&=   -B'(t)\Sigma(t)dw_{i,t},\\
 q_{i,0} &= \E[\tilde{a}_\Sigma -\ta_i |\sigma(Y_{i,0})]\\
 &= \E[\tilde{a}_\Sigma -\ta_i |\sigma\big(w_0-B(0)(\ta_\Sigma -\ta_i)\big)]\\
 &= -\frac{B(0)\V[\ta_\Sigma -\ta_i]}{\V[w_0] +B(0)^2\V[\ta_\Sigma -\ta_i]}\big(w_0-B(0)(\ta_\Sigma -\ta_i)\big)\\
& = -\frac{(M-1)B(0)\sigma^2_{\ta}}{\sigma^2_{w_0}+(M-1)B(0)^2\sigma^2_{\ta}}\big(w_0-B(0)(\ta_\Sigma -\ta_i)\big),
\end{split}
\end{align}
and the remaining variance is given by
\begin{align}\label{SigmaKB3}
\begin{split}
\Sigma'(t)&= -\big(B'(t)\big)^2\Sigma(t)^2,
\end{split}
\end{align}
with initial value
\begin{align}\label{SigmaKB4}
\begin{split}
\Sigma(0) &= \V[\tilde{a}_\Sigma -\ta_i -q_{i,0}]\\
&= \E[(\tilde{a}_\Sigma -\ta_i -q_{i,0})^2]\\
&= \E\left[\left(\tilde{a}_\Sigma -\ta_i +\tfrac{(M-1)B(0)\sigma^2_{\ta}}{\sigma^2_{w_0}+(M-1)B(0)^2\sigma^2_{\ta}}\big(w_0-B(0)(\ta_\Sigma -\ta_i)\big)\right)^2\right]\\
&= \left(\tfrac{(M-1)B(0)\sigma^2_{\ta}}{\sigma^2_{w_0}+(M-1)B(0)^2\sigma^2_{\ta}}\right)^2\sigma_{w_0}^2+ \left(1-B(0)\tfrac{(M-1)B(0)\sigma^2_{\ta}}{\sigma^2_{w_0}+(M-1)B(0)^2\sigma^2_{\ta}}\right)^2(M-1)\sigma_{\tilde{a}}^2\\
&=\frac{(M-1) \sigma_{\ta}^2 \sigma_{w_0}^2}{(M-1)B(0)^2 \sigma_{\ta}^2+\sigma_{w_0}^2}.
\end{split}
\end{align}

\end{theorem}

\section{Remaining proofs}

\proof[Proof of Lemma \ref{lem:decomp}] To see that \eqref{SUM1} holds, we use the Kalman-Bucy filter \eqref{SigmaKB2} to write 
\begin{align}\label{dY22}
\begin{split}
q_{i,t} &=  q_{i,0}  - \int_0^t B'(u)\Sigma(u)dw_{i,u},\quad t\in[0,1].
\end{split}
\end{align}
Then, 
\begin{align}\label{dY222}
\begin{split}
\sum_{i=1}^M q_{i,0}&=-\frac{(M-1)B(0)\sigma^2_{\ta}}{\sigma^2_{w_0}+(M-1)B(0)^2\sigma^2_{\ta}}\big(Mw_0-B(0)(M\ta_\Sigma -\ta_\Sigma)\big)\\
&=-\frac{(M-1)B(0)\sigma^2_{\ta}}{\sigma^2_{w_0} +(M-1)B(0)^2\sigma^2_{\ta}}\big(MY_0+B(0)\ta_\Sigma\big),\\
\sum_{i=1}^M B'(t)\Sigma(t)dw_{i,t}&=  B'(t)\Sigma(t)\Big(Mdw_t+B'(t)\Big\{\ta_\Sigma + \sum_{i=1}^Mq_{i,t}-M\ta_\Sigma \Big\}dt\Big)\\
&=  B'(t)\Sigma(t)\Big(MdY_t+B'(t)\Big\{\ta_\Sigma + \sum_{i=1}^Mq_{i,t} \Big\}dt\Big).
\end{split}
\end{align}
To explicitly solve for $\sum_{i=1}^M q_{i,t}$, we note
\begin{align}\label{dY2B}
\begin{split}
d e^{\int_0^t (B'(u))^2\Sigma(u)du} \sum_{i=1}^M q_{i,t}&=e^{\int_0^t (B'(u))^2\Sigma(u)du}\Big\{(B'(t))^2\Sigma(t)\sum_{i=1}^M q_{i,t}dt-\sum_{i=1}^M B'(t)\Sigma(t)dw_{i,t}\Big\}\\
&= -e^{\int_0^t (B'(u))^2\Sigma(u)du} B'(t)\Sigma(t)\Big(MdY_t+B'(t)\ta_\Sigma dt\Big).
\end{split}
\end{align}
We get the solution $\sum_{i=1}^M q_{i,t}$ by integrating 
\begin{align}\label{dY2C}
\begin{split}
 \sum_{i=1}^M q_{i,t}&=  e^{-\int_0^t (B'(u))^2\Sigma(u)du}\sum_{i=1}^M q_{i,0} \\
 &\quad - \int_0^t e^{-\int_s^t (B'(u))^2\Sigma(u)du}B'(s)\Sigma(s)\Big(MdY_s+B'(s)\ta_\Sigma ds\Big).
\end{split}
\end{align}
Thus, the decomposition \eqref{SUM1} holds with 
\begin{align}\label{dY2D}
\begin{split}
A(t)&:= -  e^{-\int_0^t (B'(u))^2\Sigma(u)du}\tfrac{(M-1)B(0)^2\sigma^2_{\ta}}{\sigma^2_{w_0} +(M-1)B(0)^2\sigma^2_{\ta}} - \int_0^t e^{-\int_s^t(B'(u))^2\Sigma(u)du}\big(B'(s)\big)^2\Sigma(s)ds,\\ 
\eta_t&:=  - e^{-\int_0^t (B'(u))^2\Sigma(u)du}\tfrac{M(M-1)B(0)\sigma^2_{\ta}}{\sigma^2_{w_0} +(M-1)B(0)^2\sigma^2_{\ta}}Y_0- M\int_0^t e^{-\int_s^t (B'(u))^2\Sigma(u)du}B'(s)\Sigma(s)dY_s.
\end{split}
\end{align}

For the second part, we write the solution to the Ornstein-Uhlenbeck SDE for $d\eta_t$ in \eqref{dY2E} as
\begin{align}\label{eta_explicit}
\eta_t&=F_1(t)\left( \eta_0 + M F_2(t) \tilde a_\Sigma  - M  \int_0^t \tfrac{B'(s)\Sigma(s)}{F_1(s)}dw^\circ_s \right),
\end{align}
where the deterministic functions $F_1(t)$ and $F_2(t)$ are given by the ODEs in \eqref{F1F2}.
Similarly, the the Ornstein-Uhlenbeck SDE for $dq_{i,t}$ in \eqref{SigmaKB2} has solution
\begin{align}\label{qit_explicit}
q_{i,t}&=F_1(t)\left( q_{i,0}+ F_2(t)(\tilde a_\Sigma -\tilde a_i) - \int_0^t \tfrac{B'(s)\Sigma(s)}{F_1(s)}dw^\circ_s \right).
\end{align}
By comparing \eqref{eta_explicit} and \eqref{qit_explicit}, we get \eqref{qit_eta}.

$\endproof$

\proof[Proof of Lemma \ref{lemma_infer}] The inclusion ``$\supseteq$" in \eqref{filt1} follows from \eqref{Rfiltration}, \eqref{RfiltrationQQQ}, and \eqref{RfiltrationQ}. To see the inclusion ``$\subseteq$", we use $Y_t$ in \eqref{Z}, $\eta_t$ in \eqref{dY2D}, and $q_{i,t}$ in \eqref{SigmaKB2} to find deterministic functions $h_0,h$, and $H$ such that
\begin{align}
\begin{split}
dS^f_{i,t}-\alpha\theta_{i,t}dt&= \Big\{f_0(t)Y_t   +f_1(t)\tilde{a}_i +f_2(t)q_{i,t}+f_3(t)\eta_t\Big\}dt + \gamma dw_{i,t}\\
&=
 \Big\{h_0(t)Y_0+ h(t) \ta_i + \int_0^t H(u,t)dw_{i,u}\Big\}dt + \gamma dw_{i,t}.
\end{split}
\end{align}
We define
\begin{align}\label{Sitbb}
\begin{split}
dZ_{i,t}&:=  dS^f_{i,t}-\Big\{\alpha\theta_{i,t} +h_0(t)Y_0+  h(t)\ta_i\Big\}dt, \quad Z_{i,0} :=w_{i,0}, 
\end{split}
\end{align}

The inclusion ``$\subseteq$" in \eqref{filt1} will follow from the inclusion
\begin{align}\label{filt2}
 \sigma(w_{i,u})_{u\in[0,t]}\subseteq \sigma(Z_{i,u})_{u\in[0,t]}.
\end{align}
To see \eqref{filt2}, let $t_0 \in [0,t]$ be arbitrary and let $f(s)$, $s\in[0,t]$, solve the following Volterra integral equation of the second kind (such $f$ exists by Lemma 4.3.3 in Davis (1977) because $\gamma\neq0$):
\begin{align}\label{Sitbb222}
\begin{split}
\int_r^tf(s) H(r,s)ds+f(r) \gamma = 1_{[0,t_0]}(r),\quad r\in[0,t].
\end{split}
\end{align}
This gives us
\begin{align}\label{Sitbb22}
\begin{split}
\int_0^ t f(s) dZ_{i,s} & =  \int_0^t f(s) \int_0^s H(r,s)dw_{i,r} ds +  \int_0^t f(s) \gamma dw_{i,s}\\
& =  \int_0^t  \int_r^tf(s) H(r,s)dsdw_{i,r} +  \int_0^t f(s) \gamma dw_{i,s}\\
& =  \int_0^t \Big( \int_r^tf(s) H(r,s)ds+f(r) \sigma\Big)dw_{i,r}\\
& = \int_0^t 1_{[0,t_0]}(r) dw_{i,r}\\
& = w_{i,t_0}-w_{i,0}.
\end{split}
\end{align}

$\endproof$

\proof[Proof of Lemma \ref{PI_Le}] Consider a rebalancer $i\in\{1,...,M\}$. For arbitrary holdings $\theta_{i,t}$, the expectation in the $i$'th objective in \eqref{Rproblem} is
\begin{align}\label{pp1}
\begin{split}
& \E\Big[\int_0^1 \theta_{i,t} dS^f_{i,t} - \int_0^1 \kappa(t)(\ta_i - \theta_{i,t})^2 dt \Big]\\
& =\E\Big[\int_0^1 \theta_{i,t} \Big\{f_0(t)Y_t   +f_1(t)\tilde{a}_i +f_2(t)q_{i,t}+f_3(t)\eta_t+ \alpha\theta_{i,t}\Big\}dt- \int_0^1 \kappa(t)(\ta_i - \theta_{i,t})^2 dt \Big].
\end{split}
\end{align}
The equality in \eqref{pp1} follows from the square integrability condition \eqref{squareint}, which ensures that the stochastic integral $\int_0^s \theta_{i,t} dw_{i,t}$  is a martingale with zero expectation. We can maximize the integrand in \eqref{pp1} pointwise because the second-order condition  $\alpha <\kappa(t)$ holds. This gives the first formula in \eqref{Y00PI}. 

The second  formula for a tracker $j$ in \eqref{Y00PI} is proved similarly.
$\endproof$

\proof[Proof of Lemma \ref{PI_Lemma}]The local Lipschitz property of the ODEs  \eqref{derivatives0aPI} ensures that there exists a maximal interval of existence $[0, \tau)$ with $\tau\in (0,\infty]$ by the Picard-Lindel\"of theorem (see, e.g., Theorem II.1.1 in Hartman 2002). We assume that $\tau<1$ and construct a contradiction. To this end, we set 
\begin{equation}\label{kappaint}
K:= \int_0^1\kappa(s)ds<\infty.
\end{equation}

First, the Riccati ODE for $\Sigma(t)$ has the explicit solution in \eqref{Sigma},  which cannot explode as $t\uparrow \tau$ (even if $B(t)$ should explode as $t\uparrow \tau$). 

Second, the initial value $A(0)$ in \eqref{derivatives0aPI} ensures $A(0)\ge-1$ and to see that implies $A(t)\geq-1$ for all  $t\in [0,\tau)$, we note
\begin{align}
\frac{\partial}{\partial t}\big(A(t)+1\big)=-\big(B'(t)\big)^2\Sigma(t)\big(A(t)+1\big),
\end{align}
which implies 
\begin{align}
\begin{split}
A(t)+1&=\big(A(0)+1\big)e^{-\int_0^t(B'(s))^2\Sigma(s)ds}\\
&\ge0.
\end{split}
\end{align}
This shows that $A(t)$ cannot explode as $t\uparrow \tau$ (even if $B(t)$ should explode as $t\uparrow \tau$). 

Third, we show $B(t)$ is uniformly bounded for $t\in [0,\tau)$; hence, also $B(t)$ cannot explode as $t\uparrow \tau$. This then gives the desired contradiction because of Theorem II.3.1 in Hartman (2002). The affine ODE for $B(t)$ in 
\eqref{derivatives0a} has the explicit solution 
\begin{align}\label{Bexplicit}
    B(t)=e^{\int_0^t\frac{2\bar M \kappa(s)}{\gamma(A(s)+1+\bar M)}ds}\Big(B(0)+\int_0^t\frac{2  \kappa(s)}{\gamma(A(s)+1+\bar M)}e^{-\int_0^s\frac{2\bar M \kappa(u)}{\gamma(A(u)+1+\bar M)}du}ds\Big).
\end{align}
We can use $K$ in \eqref{kappaint} to produce the upper bound
\begin{align}\label{bound1}
\begin{split}
\int_0^t\frac{2\bar{M} \kappa(s)}{\gamma\big(A(s)+1+\bar{M}\big)}ds \le \int_0^t\frac{2\bar{M} \kappa(s)}{\gamma \bar{M}}ds\leq\frac{2K}\gamma,\quad t\in[0,\tau).
\end{split}
\end{align}
In turn, the bound \eqref{bound1} and \eqref{Bexplicit} imply
\begin{align}\label{Btbounded}
\begin{split}
|B(t)|&\leq e^{\frac{2K}\gamma}\Big(|B(0)|+\int_0^t\frac{2 \kappa(s)}{\gamma(A(s)+1+\bar{M})}ds\Big)\\
&\le e^{\frac{2K}\gamma}\Big(|B(0)|+\frac{2K}{\gamma \bar M}\Big),
\end{split}
\end{align}
for $t\in[0,\tau)$. Because the upper bound in \eqref{Btbounded} is uniform over $t\in[0,\tau)$, $B(t)$ cannot explode as $t\uparrow \tau$.  
$\endproof$

\proof[Proof of Theorem \ref{thm_PI}] To see that the holdings in \eqref{Y0000PI} satisfy the square integrability condition \eqref{squareint}, we insert $B'(t)$ from  \eqref{derivatives0aPI} to get
\begin{align}\label{pf1}
\begin{split}
\hat{\theta}_{i,t} &= -\tfrac{2 \kappa (t) (A(t)+\bar{M} (1-B(t)))}{(A(t)+\bar{M}+1) (\alpha -2 \kappa (t))}\ta_i + \tfrac{2 \kappa (t) (\bar{M} B(t)+1)}{(A(t)+\bar{M}+1) (\alpha -2 \kappa (t))}q_{i,t}
\\
&-\tfrac{2 \kappa (t) (\bar{M} B(t)+1)}{(M+\bar{M}) (A(t)+\bar{M}+1) (\alpha -2 \kappa (t))}\eta_t+\tfrac{2 \bar{M} \kappa (t)}{(M+\bar{M}) (\alpha -2 \kappa (t))}Y_t,\\
\hat{\theta}_{j,t} &=-\tfrac{2 \kappa (t) (\bar{M} B(t)+1)}{(M+\bar{M}) (A(t)+\bar{M}+1) (\alpha -2 \kappa (t))}\eta_t-\tfrac{2 M \kappa (t)}{(M+\bar{M}) (\alpha -2 \kappa (t))}w_t\\
&+\tfrac{2 \kappa (t) (\bar{M} B(t) (-A(t)+M-1)+M+\bar{M})}{(M+\bar{M}) (A(t)+\bar{M}+1) (\alpha -2 \kappa (t))}\ta_\Sigma.
\end{split}
\end{align}
Because $\kappa:[0,1]\to(0,\infty)$ is continuous, $\kappa(t)$ is uniformly bounded. This gives us that
$B'(t)$ in \eqref{derivatives0aPI} is also uniformly bounded. As a consequence, the variances $\V[q_{i,t}], \V[\eta_t]$, and $\V[Y_t]$ are also uniformly bounded functions of $t\in[0,1]$. Therefore, the holding processes in \eqref{pf1} satisfy  \eqref{squareint} if the coefficient functions for $(\ta_i, q_{i,t},\eta_t,Y_t, w_t,\ta_\Sigma)$ are square integrable over $t\in [0,1]$. For example, the coefficient function for $\ta_i$ in $\hat{\theta}_{i,t}$ is bounded because
\begin{align}
|\tfrac{2 \kappa (t) (A(t)+\bar{M} (1-B(t)))}{(A(t)+\bar{M}+1) (\alpha -2 \kappa (t))}|\le \tfrac{2  |A(t)+\bar{M} (1-B(t))|}{\bar{M}},
\end{align}
which is continuous for $t\in[0,1]$. Similarly, the remaining coefficients functions can be seen to be bounded too. The optimality in Definition \ref{PI_eq}(i) then follows from Lemma \ref{PI_Le} and the fact that the holdings \eqref{Y0000PI} are those in \eqref{Y00PI} with the $f$ functions in  \eqref{fs}  inserted.

Definition \ref{PI_eq}(ii)+(iii) are ensured by the specific $f$ functions in  \eqref{fs}.

$\endproof$

\proof[Proof of Lemma \ref{response}]  Lemma A.1 in Choi, Larsen, and Seppi (2021) and the continuity of $Z_t$'s paths imply that $Z_t$ is  adapted to both $\sF_{i,t}$ and $\sF_{j,t}$. The rest of this proof is similar to the proof of Lemma \ref{PI_Le} given above and is therefore omitted. $\endproof$

\proof[Proof of Lemma \ref{lemma_eqholdings}] The rebalancers' second-order condition is
\begin{align}\label{pf2}
\begin{split}
\frac{\big(\alpha -\kappa (t)\big) \big(M+\bar{M}+2 \nu_0(t)-1\big)}{M+\bar{M}-1}<0,
\end{split}
\end{align}
whereas  the trackers' second-order condition is $\alpha <\kappa(t)$. Inequality \eqref{pf2} holds because $\nu_0(t)\ge0$ and $\alpha <\kappa(t)$. The rest of this proof is similar to the proof of Lemma \ref{PI_Le} given above and is therefore omitted.
$\endproof$

\proof[Proof of Lemma \ref{main_Lemma}] The proof only requires minor changes to the proof of Lemma \ref{PI_Lemma}. As before, we let  $[0, \tau)$ be the maximal interval of existence with $\tau\in (0,\infty]$ and assume that $\tau<1$ to construct a contradiction. As in the proof of Lemma \ref{PI_Lemma}, $\Sigma(t)=\frac{1}{\frac1{\Sigma(0)}+\int_0^t(B'(t))^2dt}$ and $A(t)\geq-1$. Next, to show $B(t)$ is bounded on  $[0,\tau)$, we rewrite the ODE for $B(t)$ in \eqref{derivatives0a} as
\begin{equation}\label{ppp3}
  B'(t)=\frac{2\kappa(t)\big(B(t)(c(t)+\bar{M})+1\big)}{\gamma \big(A(t)+1+\bar{M}+c(t)\big)}.
\end{equation}
where the deterministic function $c(t)$ is  defined as
\begin{equation}
c(t):=\frac{2\bar{M}\big(\kappa(t)-\alpha\big)}{(M+\bar{M}-2)\big(2(M+\bar{M})\kappa(t)-\alpha(M+\bar{M}+1)\big)},\quad t\in[0,1].
\end{equation}
Because $\alpha \le 0$ and $\kappa(t)>0$, we have $c(t) >0$.  Furthermore, $c(t)$ is bounded because
\begin{align}\label{c0}
\begin{split}
 c(t) &\leq\frac{2\bar{M}(\kappa(t)-\alpha)}{(M+\bar{M}-2)(M+\bar{M}+1)(\kappa(t)-\alpha)}\\
 &=\frac{2\bar{M}}{(M+\bar{M}-2)(M+\bar{M}+1)}\\
 &=:c_0,
\end{split}
\end{align}
where the inequality follows from $2(M+\bar M) > (M+\bar M +1)$ and the positivity of $\kappa(t)$. Because $A(t) + 1\ge0$ and $c(t)>0$ we get the two estimates
\begin{align}\label{pp2222}
\begin{split}
    \int_0^t\frac{2\kappa(s)(c(s)+\bar{M})}{\gamma(A(s)+1+\bar{M}+c(s))}ds
    &\le \frac{2(c_0+\bar{M})}{\gamma\bar{M}}K,\\
    \int_0^t \frac{2\kappa(s)}{\gamma(A(s)+1+\bar{M}+c(s))}ds&\le\frac{2}{\gamma\bar{M}}K,
\end{split}
\end{align}
where $K$ is as in \eqref{kappaint}. Similar to \eqref{Bexplicit}, the explicit solution of \eqref{ppp3} is
\begin{align}\label{pp22}
\begin{split}
    B(t)&=e^{\int_0^t\frac{2\kappa(s)(c(s)+\bar{M})}{\gamma(A(s)+1+\bar{M}+c(s))}ds}B(0)+\int_0^te^{\int_s^t\frac{2\kappa(u)(c(u)+\bar{M})}{\gamma(A(u)+1+\bar{M}+c(u))}du}\tfrac{2\kappa(s)}{\gamma(A(s)+1+\bar{M}+c(s))}ds.
\end{split}
\end{align}
Combing this expression for $B(t)$ with the bounds \eqref{pp2222} produces
\begin{equation}\label{Btboundeddd}
  \begin{aligned}
    B(t)&\leq e^{\frac{2(c_0+\bar{M})K}{\gamma\bar{M}}}|B(0)|+\int_0^te^{\frac{2(c_0+\bar{M})K}{\gamma\bar{M}}}\tfrac{2\kappa(s)}{\gamma(A(s)+1+\bar{M}+c(s))}ds\\
    &\leq e^{\frac{2(c_0+\bar{M})K}{\gamma\bar{M}}}(|B(0)|+\frac{2K}{\gamma\bar{M}}).
  \end{aligned}
\end{equation}
 $\endproof$

\proof[Proof of Theorem \ref{thm_Main}] From \eqref{ppp3} we see that
\begin{align}\label{ppp35}
\begin{split}
  |B'(t)|&=\frac{2\kappa(t)\big(|B(t)|(c(t)+\bar{M})+1\big)}{\gamma \big(A(t)+1+\bar{M}+c(t)\big)}\\
&\le \frac{2\kappa(t)\big(|B(t)|(c_0+\bar{M})+1\big)}{\gamma\bar{M}},
\end{split}
\end{align}
where $c_0$ is defined in \eqref{c0}. Because $\kappa(t)$ is continuous on $t\in[0,1]$, $\kappa(t)$ is bounded and  from \eqref{Btbounded} we know that $B(t)$ is bounded too. Therefore, from \eqref{ppp35}, we see that $B'(t)$ is also uniformly bounded. Consequently,  the variances $\V[q_{i,t}], \V[\eta_t]$, and $\V[Y_t]$ are also uniformly bounded functions of $t\in[0,1]$.

As before, the coefficient functions for $(\ta_i, q_{i,t},\eta_t,Y_t, w_t,\ta_\Sigma)$ in  \eqref{Y0000}  are all uniformly bounded for $t\in [0,1]$. Therefore, the square-integrability condition   \eqref{squareint} holds.

The requirements in Definition \ref{Nash_eq} follow from the definition of the functions in \eqref{nus}.

$\endproof$

\end{document}